\documentclass[a4paper,fleqn,usenatbib]{mnras}
\usepackage[T1]{fontenc}
\usepackage{ae,aecompl}
\pdfoutput =1
\usepackage{graphicx,soul}	
\usepackage{amsmath}	
\usepackage{amssymb}	
\usepackage{cases}
\usepackage{color}
\usepackage{bm}
\title[disc central surface brightness]{Galaxy disc central surface brightness distribution in the optical and near-infrared bands}
\author[L. Zhou et al.]{
L. Zhou$^{1,2,3}$\thanks{E-mail: lizhou@bao.ac.cn},
W. Du$^{2}$\thanks{E-mail:wdu@bao.ac.cn},
H. Wu$^{2}$\thanks{E-mail:hwu@bao.ac.cn},
Y. C. Liang$^{2}$,
M. He$^{2,3}$,
P. S. Zhao$^{2,3}$,
\newauthor{
L. C. Deng$^{2}$,
W. B. Shi$^{1,2}$\thanks{E-mail:swb@sdu.edu.cn},
Y. H. Wang$^{2,3}$}
\\
$^1$Shandong Provincial Key Laboratory of Optical Astronomy and Solar-Terrestrial Environment, \\
School of Space Science and Physics, Shandong University at Weihai, Weihai 264209, China \\
$^2$Key Laboratory of Optical Astronomy, National Astronomical Observatories, Chinese Academy of Sciences, \\
20A Datun Road, Chaoyang District, Beijing 100012, China \\
$^3$School of Astronomy and Space Science, University of the Chinese Academy of Sciences, Beijing 101408, China
}

\date{Accepted XXX. Received YYY; in original form ZZZ}

\pubyear{2018}

\begin{document}
\label{firstpage}
\pagerange{\pageref{firstpage}--\pageref{lastpage}}
\maketitle

\begin{abstract}

To study the disc central surface brightness ($\mu_0$) distribution in optical and near-infrared bands, we select 708 disc-dominated galaxies within a fixed distance of 57 Mpc from SDSS DR7 and UKIDSS DR10.   Then we fit $\mu_0$ distribution by using single and double Gaussian profiles with an optimal bin size for the final sample of 538 galaxies in optical $griz$ bands and near-infrared $YJHK$ bands.

Among the 8 bands, we find that $\mu_{0}$ distribution in optical bands can not be much better fitted with double Gaussian profiles. However, for all the near-infrared bands, the evidence of being better fitted by using double Gaussian profiles is positive.  
Especially for $K$ band, the evidence of a double Gaussian profile being better than a single Gaussian profile for $\mu_{0}$ distribution is very strong, the reliability of which can be approved by 1000 times test for our sample. 
No dust extinction correction is applied. The difference of $\mu_{0}$ distribution between optical and near-infrared bands could be caused by the effect of dust extinction in optical bands. 
Due to the sample selection criteria, our sample is not absolutely complete. However, the sample incompleteness does not change the double Gaussian distribution of $\mu_{0}$ in $K$ band.
Furthermore, we discuss some possible reasons for the fitting results of $\mu_{0}$ distribution in $K$ band. Conclusively, the double Gaussian distribution of $\mu_{0}$ in $K$ band for our sample may depend on bulge-to-disk ratio, color and disk scalelength, rather than the inclination of sample galaxies, bin size and statistical fluctuations. 
\end{abstract}
\begin{keywords}
Galaxies: fundamental parameters -- Galaxies: photometry -- Galaxies: structure -- Galaxies: evolution
\end{keywords}

%
\section{Introduction}
The distributions of different galaxy parameters in different environments 
could provide important constraints for setting up theoretical
models of galaxy formation and evolution (e.g. \citealp{2003ApSS.284..353T}). There are
some works which found bimodal distributions of galaxy parameters, such as color (
e.g. \citealp{2004ApJ...600..681B}, \citealp{2006MNRAS.372..253M}, \citealp{2009ApJ...706L.173B}, \citealp{2011ApJ...735...86W}), star formation rate 
(\citealp{2012MNRAS.424..232W}), and disc central surface brightness ($\mu_0$; e.g. \citealp{1997ApJ...484..145T}, \citealp{2009MNRAS.393..628M}, \citealp{2009MNRAS.394.2022M}, \citealp{2013MNRAS.433..751S} etc.). 
\citealp{1948AnAp...11..247D}, \citealp{1959HDP....53..311D}, \citealp{1968adga.book.....S}, and \citealp{1970ApJ...160..811F} studied the light profiles of galaxies. 

The first work which provided convincing evidence for the bimodal distribution of $\mu_0$ is \citealp{1997ApJ...484..145T}, which selected 79 sample galaxies from Ursa Major 
cluster and revealed that there is an apparent lack of intermediate surface brightness galaxies (ISB) in $BRI$ and $K'$ bands. 
The separate $\mu_0$ distribution of
high (HSB) and low surface brightness (LSB) suggested
that there are discrete but stable radial configurations. 
The authors also pointed that the bimodal distribution of $\mu_0$ could be resulted from the errors induced during 
fitting galaxy disc process, because the shallow $K'$ band could lead to the disc premature truncations and 
result in the mixture of bulge components. As raised by \citealp{bell2000}, there is a probability for incorrect inclination corrections and small-number statistics to result in the bimodality of $\mu_0$ distribution. 

To overcome the error induced by small-number statistics, \citealp{2009MNRAS.394.2022M} selected a larger number of sample galaxies (286 galaxies) from Virgo cluster and also found a $\mu_0$ bimodal distribution. However, they claimed that only Virgo cluster 
has been studied, so they inferred that the bimodality 
of $\mu_0$ could not be intrinsic and the $\mu_0$ distribution could be different in different environment.
\citealp{2013MNRAS.433..751S} selected 438 galaxies from Spitzer Survey of Stellar Structure in Galaxies ($S^4G$) by limiting with distance and morphology type. 
They demonstrated that there is a bimodality in $\mu_0$ distribution, which implies that there is a gap between LSB galaxies and HSB galaxies. They investigated some possible reasons that would lead to the bimodality, such as small-number statistics, environment influences, low signal-to-noise (S/N) 
and galaxy inclination correction, 
and found the bimodality of $\mu_0$ cannot be due to any of these biases or statistical fluctuations. 
In \citealp{2016MNRAS.455.2644S}, they showed that there is no bimodality of the central surface brightness for galaxies in sheets, while there is bimodality in voids and filaments.
\citealp{2013MNRAS.433..751S} extended the study outside of clusters and found that there is still bimodality for the overall sample composed of galaxies inside and outside of clusters. Therefore, bimodality found in clusters is not only a cluster feature. \citealp{2016MNRAS.455.2644S} pushed the limit further by splitting sample using the cosmic web and found that indeed bimodality is not only a cluster but also a filament and void feature, while it is not a sheet feature.
They suggested that there may be two stable states of galaxy being: 
LSB galaxies are dominated by dark matter component and HSB galaxies are dominated by 
baryonic matter in the center. The reason for the low number of ISB galaxies could be that the co-dominated situation of 
baryonic and dark matter in the center is not stable.   

According to previous studies, we would like to study $\mu_0$ distribution of a sample of galaxies using data from SDSS and UKIDSS. This paper is
organized as follows: In section 2, we describe the sample selections. The photometry and geometric fitting are shown in Section 3. The distributions of surface brightness in 
different bands are shown in section 4. Finally, we discuss our results in section 5 and give  conclusions in section 6. Throughout this paper, we adopt 
a cosmological model with $H_0 $=$ 70 km s^{-1} Mpc ^{-1}$, $\Omega_M $=$ 0.3$, $\Omega_{\Lambda} $=$ 0.7$. 
We use AB magnitudes for Sloan Digital Sky Survey (SDSS) and Vega magnitudes for United Kingdom Infra-Red Telescope (UKIRT) Infrared Deep Sky Survey (UKIDSS).
\section{Data}
\subsection{SDSS DR7}
Using a dedicated wide-field 2.5-meter telescope at Apache Point Observatory in New Mexico,
SDSS(\citealp{1998AJ....116.3040G}, \citealp{2000AJ....120.1579Y}, \citealp{lu01}, \citealp{2002AJ....124.1810S}, \citealp{2002AJ....123..485S}) 
is intended to map one quarter of the whole sky($\sim 10,000 deg^2$) with CCD imaging in five bands ($u, g, r, i, z$) 
and spectroscopy ranging from 3800 to 9200 {\AA} for millions of galaxies, quasars and stars. 
All the sample galaxies used in this work are drawn from the main galaxy sample of SDSS Data Release 7 (DR7) (\citealp{2009ApJS..182..543A}).
The main galaxy sample is comprised of galaxies with $r-band$ Petrosian magnitude brighter than 17.77 $mag$ (\citealp{2002AJ....124.1810S}).
In the SDSS DR7 image catalogue, there are 11,663 $deg^2$ of imaging data.
For $u, g, r, i$ and $z$ bands, the 95\% completeness magnitude limits are 22.0, 22.2, 22.2, 21.3 and 20.5 $mag$, respectively (\citealp{ab04}).

\subsection{UKIDSS LAS DR10}
The UKIRT UKIDSS has been carried out using the Wide Field Camera (WFCAM;\citealp{2007AA...467..777C}), which has a field of view of 0.21 $deg^2$ and a pixel size of 0.4 $arcsecond$ on
the 3.8-meter UKIRT. It can be considered as the near-infrared counterpart of the SDSS (\citealp{2000AJ....120.1579Y}).
There are several surveys in UKIDSS, including the Large Area Survey (LAS), the Galactic Clusters Survey(GCS), the Galactic Plane Survey (GPS), 
Deep Extragalactic Survey (DXS) and Ultra Deep Survey (UDS), which cover various combinations of the filter set $ZYJHK$ with wavelength
ranging from 0.83 $\mu$m to 2.37 $\mu$m and $H\_2$ (\citealp{2007MNRAS.379.1599L}). The area of LAS is about $4000$ $deg^2$ in the Northern Sky. 
For the four bands ($YJHK$) of LAS, the depths are 20.3, 19.5, 18.6 and 18.2 mag, respectively. 
The effective volume and the depth of the total UKIDSS is much larger
and deeper than the Two-Micron All-Sky Survey (2MASS; \citealp{2006AJ....131.1163S}).
The primary aim of UKIDSS is to provide a long-term astronomical legacy data base.

\subsection{Sample Selection}
\subsubsection{Sample Selection Method}
To obtain reliable and accurate $\mu_0$ of spiral galaxies, 
we select our sample by limiting distance, $fracDev_r$ and $M_r$ from SDSS DR7 main galaxy sample catalogue and UKIDSS LAS DR10. Our sample selection criteria are as follows:

(1) To obtain more accurate distance, 869,059 galaxies which have non-zero spectroscopic redshifts are selected from SDSS DR7 main galaxy sample.

(2) Then we further select 5,946 galaxies, which have the corrected distance within 57 Mpc to form a volume-limited sample. The method of correcting  distance will be described in detail in section 2.3.2. 

(3) To ensure that our sample galaxies are disk-dominated galaxies, whose surface brightness profiles
could be well described by an exponential profile (e.g. \citealp{2005AJ....129...61B}, \citealp{2006MNRAS.372..199C}, \citealp{2007ApJ...659.1159S}),
we select 4,725 galaxies which have the bulge-to-total ratios in $r$ band less than $50 \%$ ($fracDev_r \leq 0.5$). 

(4) To exclude dwarf galaxies, we select 2,363 galaxies which have $r$-band magnitudes brighter than 17.77 mag and absolute magnitudes in $r$-band brighter than -16.0 mag from the previous 4,725 galaxies. 

(5) By cross-matching with the UKIDSS LAS DR10 data, those 708 galaxies which have images in all $Y$, $J$, $H$ and $K$ bands are finally selected from the previous 2363 galaxies as our entire sample.

\subsubsection{Distance correction and $Mr$ calculation}
For galaxies with $z > 0.02$, 
the distance could be calculated using Doppler redshift formula and Hubble's law directly : 

\begin{equation}
    V_r $=$ z \times c, 
\end{equation}
and 
\begin{equation}
        V_r $=$ H_0 \times D,
        \label{eq:hubble's law}
\end{equation}
where $V_r$ is radial velocity, $D$ is the distance to the Earth in units of km, 
$z$ is redshift of the galaxy. 

For galaxies with $z \leq 0.02 $, we should correct the
effect of Virgo cluster on the observed velocity (\citealp{2000ApJ...529..786M}, \citealp{1996AJ....111..794K}).
We use Eq.~(\ref{eq:Vlg}) in \citealp{1996AJ....111..794K} to correct the observed heliocentric velocity to the Earth of our objects to that relative to the centroid of the Local Group, which is:
\begin{equation}
    V_{LG} $=$ V_h $+$ V_a(cosb \ cosb_a \ cos(l-l_a)$+$sinb \ sinb_a), 
    \label{eq:Vlg}
\end{equation}
where $V_{LG}$ is the velocity corrected to the centroid of the Local Group, 
$V_h$ is the observed heliocentric velocity estimated using $z \times c$, $l$ and $b$ are the galactic longitude
and galactic latitude of the object. 
For the solar apex with respect to the Local Group galaxies, the parameters we use here ($V_a = 316 km/s$, $l_a = 93 ^{\circ}$, $b_a = -4^{\circ}$) are provided 
by \citealp{2000ApJ...529..786M} and \citealp{1996AJ....111..794K}.
Then the Virgo infall is estimated using the infall model described in \citealp{1980AJ.....85..801S}. In the model, 
the estimated radial component (with respect to the Local Group) of peculiar velocity induced by an 
attractor is,  
\begin{equation}
    V_{infall} = V_{fid} cos \theta + V_{fid} \frac{V_o-V_v cos \theta}{r_{oa}} (\frac{r_{oa}}{V_v})^{1-\gamma},
    \label{eq:vinfall}
\end{equation}
where $V_{fid}$ is the amplitude of the infall pattern to Virgo cluster at the Local Group (here the value is 200 km/s), 
$V_o$ is the observed velocity of the object in the Local Group frame, $V_v$ is the observed distance of the Virgo cluster 
expressed as a velocity (here the value is $957 km/s$), $\gamma$ is the slope of the density profile of Virgo cluster (here the 
value is 2), $\theta$ is the projected angle between the object and Virgo cluster, and $r_{oa}$ is the estimated distance of 
the object from Virgo cluster expressed by velocity,
\begin{equation}
        r_{oa} = \sqrt{V_o^2 + V_v^2-2V_o V_v cos \theta}.
\end{equation}
There are two components in the Eq.~(\ref{eq:vinfall}). The first term is the vector contribution due to 
the Local Group's peculiar velocity into Virgo cluster, and the second one is the change in the velocity due to the 
infall of the object into Virgo cluster.
Finally, the corrected cosmic velocity with respect to the Local Group is 

\begin{subnumcases}
        {V_{cosmic}=}
        957,&$\theta \le 10^{\circ},V=600 \sim 2300$,\\
        V_{LG} + V_{infall},&$\theta > 10^{\circ}, V=others$.       
\end{subnumcases}

Then the luminosity distance of galaxies with $z \leq 0.02$ could be estimated using Eq.~(\ref{eq:hubble's law}). 

The $r-band$ absolute magnitude $M_r$ is calculated using
\begin{equation}
        M_r$=$m_r$+$5-5 \times log_{10}{D},
\end{equation}
where $m_r$ is the apparent magnitude in $r$ band, 
$D$ is the distance in units of $pc$ estimated by using the corrected velocity and spectroscopic redshift.  
\subsection{Sample}
\begin{figure*}
\centering
\begin{minipage}{\textwidth}
\includegraphics[width=.28\textwidth]{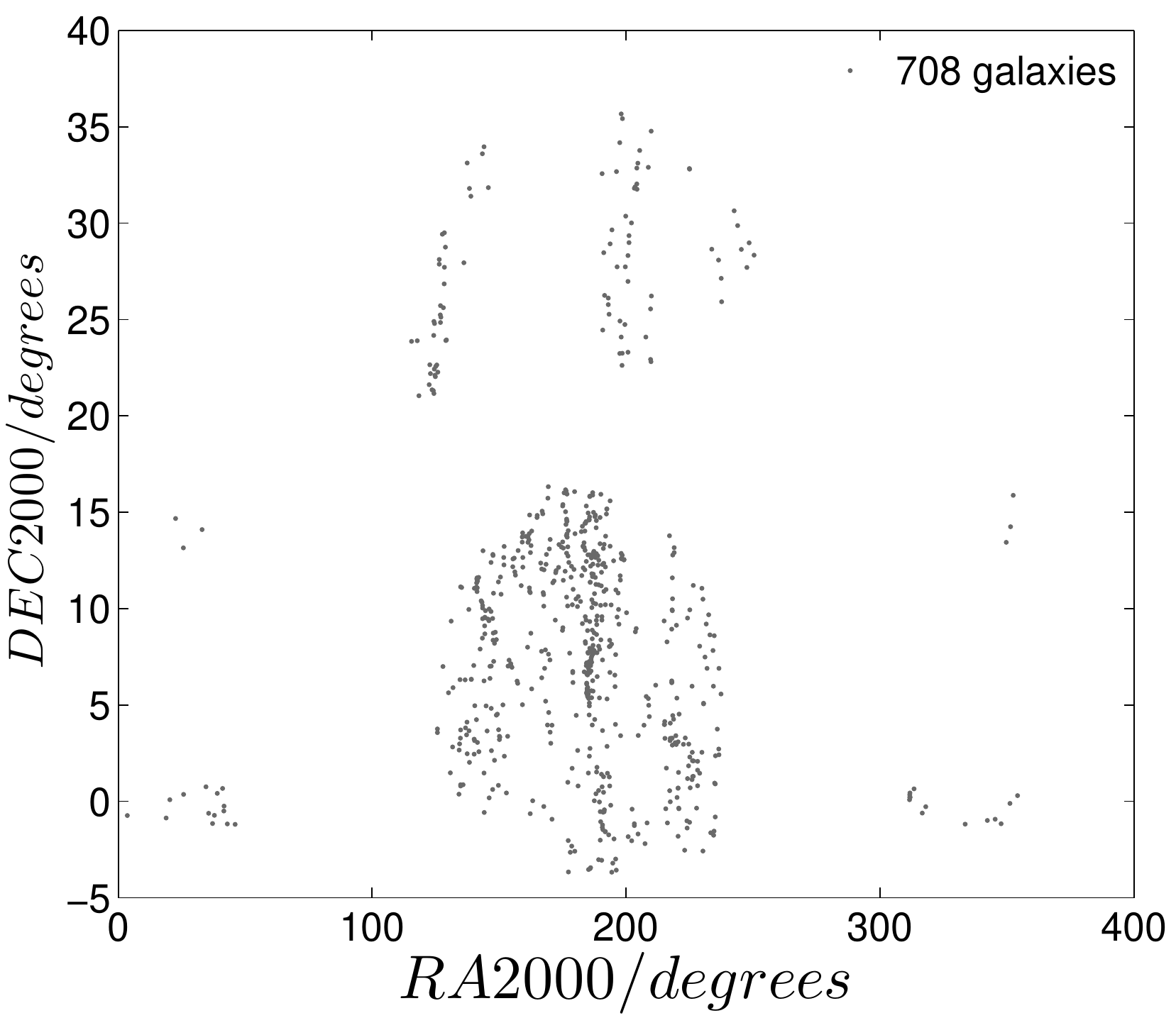}
\includegraphics[width=.32\textwidth]{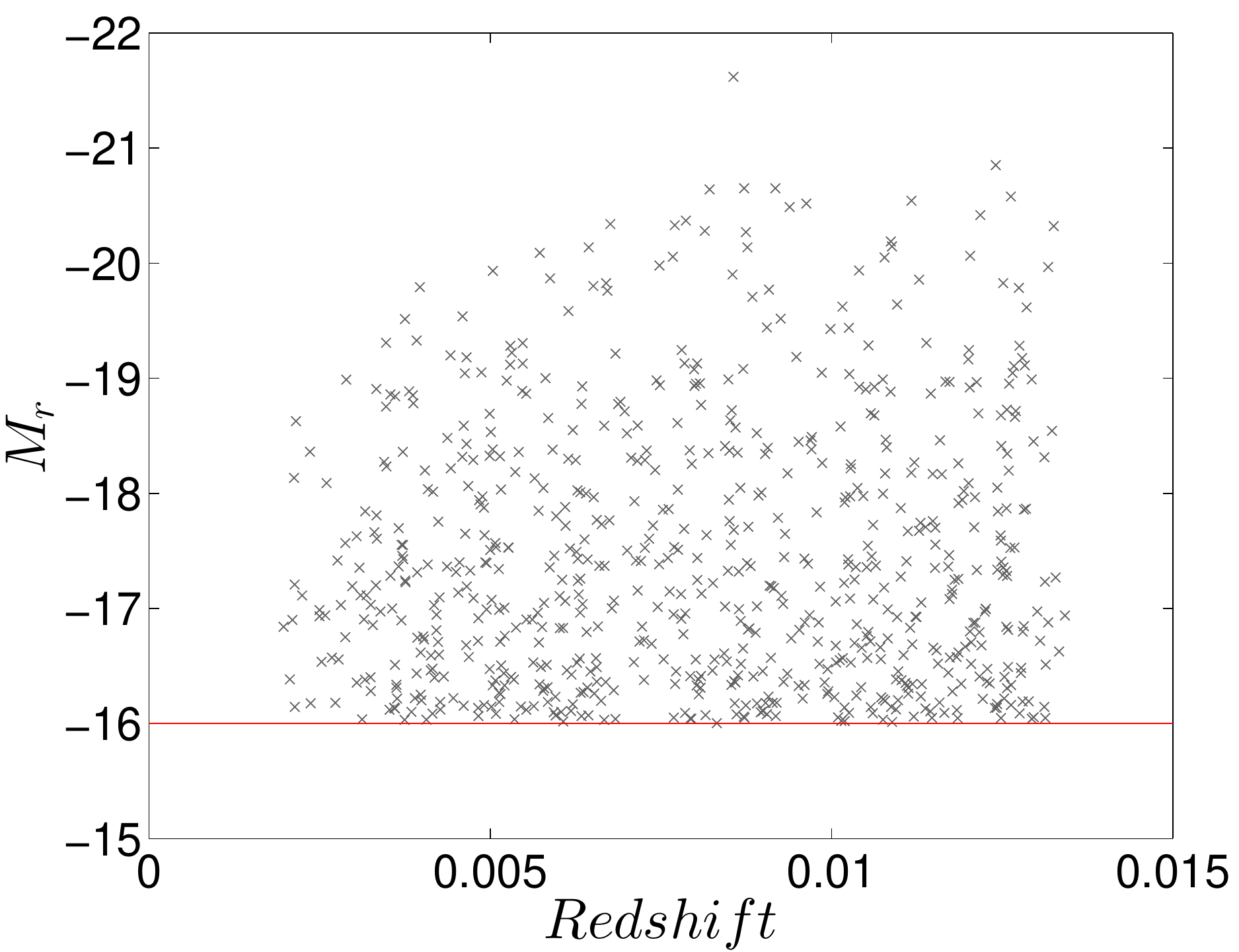}
\includegraphics[width=.32\textwidth]{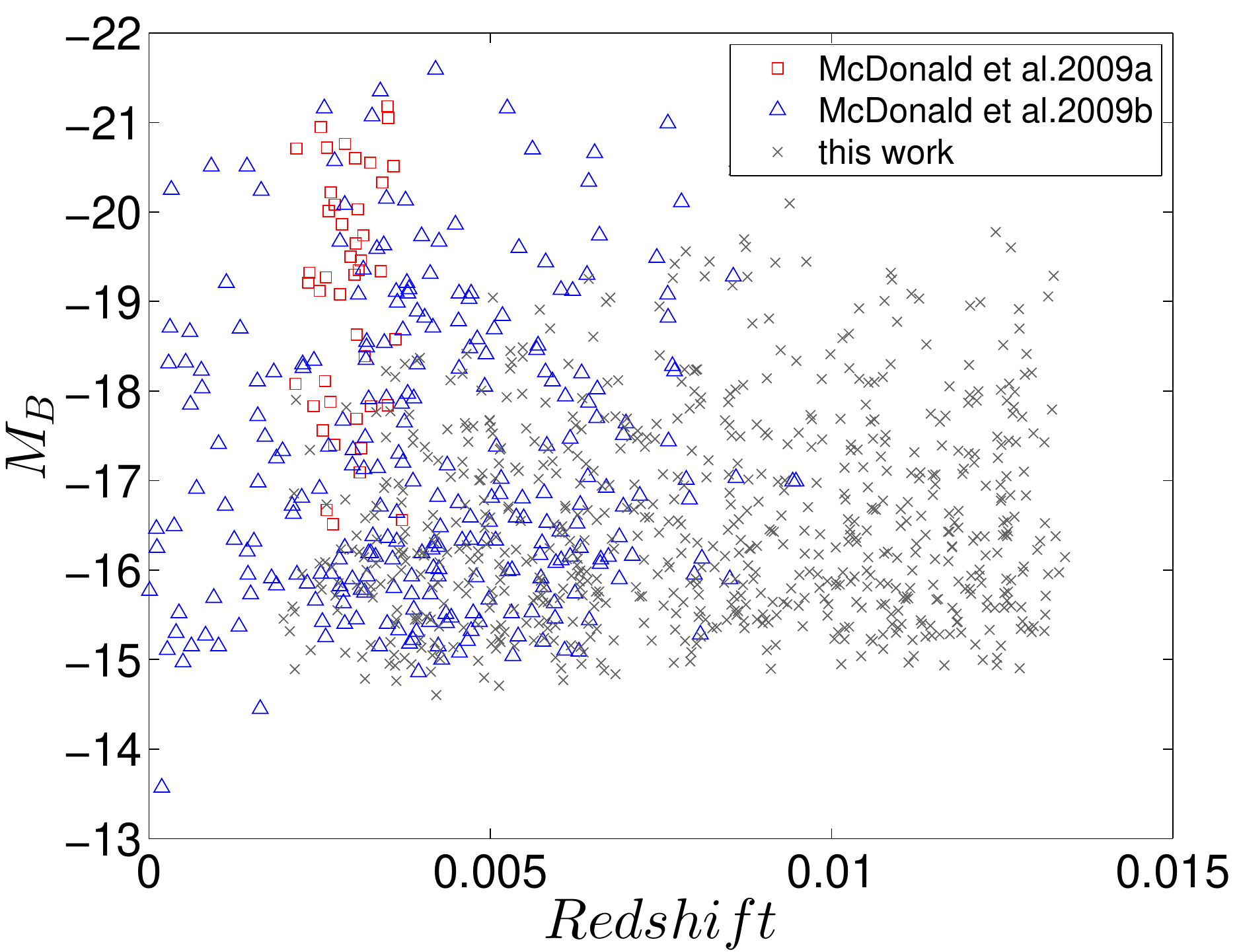}
\end{minipage}
\caption{The left panel shows the RA and DEC distributions of our 708 entire sample galaxies. The middle one shows the redshift and absolute magnitude in r band distribution of our entire sample. The red line represents $M_r = -16$ $mag$. The right panel presents the redshift and absolute magnitude in B band distributions of our entire sample and galaxies from previous studies.}
\label{fig.ra-dec-1}
\end{figure*}

Finally, by all the criteria, there are 708 galaxies left as our whole sample. 
The RA and DEC distribution of all the 708 sample galaxies are shown in the left panel of  Fig.~\ref{fig.ra-dec-1}.
The relation between redshift and $M_r$ is given in the middle panel of Fig.~\ref{fig.ra-dec-1}, which shows that redshifts of our sample galaxies range from 0 to 0.014, and the absolute magnitudes in $r$ band satisfy $M_r \leq -16$ $mag$.

Comparing with samples in previous papers such as \citealp{1997ApJ...484..145T, 2009MNRAS.393..628M, 2009MNRAS.394.2022M, 2013MNRAS.433..751S}, our sample is representative of a slightly different sample of galaxies. \citealp{1997ApJ...484..145T} has 62 galaxies in Ursa Major cluster with redshift lower than 0.005, \citealp{2009MNRAS.393..628M, 2009MNRAS.394.2022M} has 65 and 286 sample galaxies within Ursa Major cluster and Virgo cluster, respectively, and \citealp{2013MNRAS.433..751S} has 438 sample galaxies within the distance of 20 Mpc.
The redshift and $M_B$ distributions of our sample galaxies and those of \citealp{2009MNRAS.393..628M, 2009MNRAS.394.2022M} sample are presented in the right panel of Fig.~\ref{fig.ra-dec-1}.
Compared with \citealp{2009MNRAS.393..628M, 2009MNRAS.394.2022M}, 
our sample galaxies have a larger range of redshift, which range from 0 to 0.014.  

\begin{figure}
\centering
\begin{minipage}{\textwidth}
\includegraphics[width=45mm]{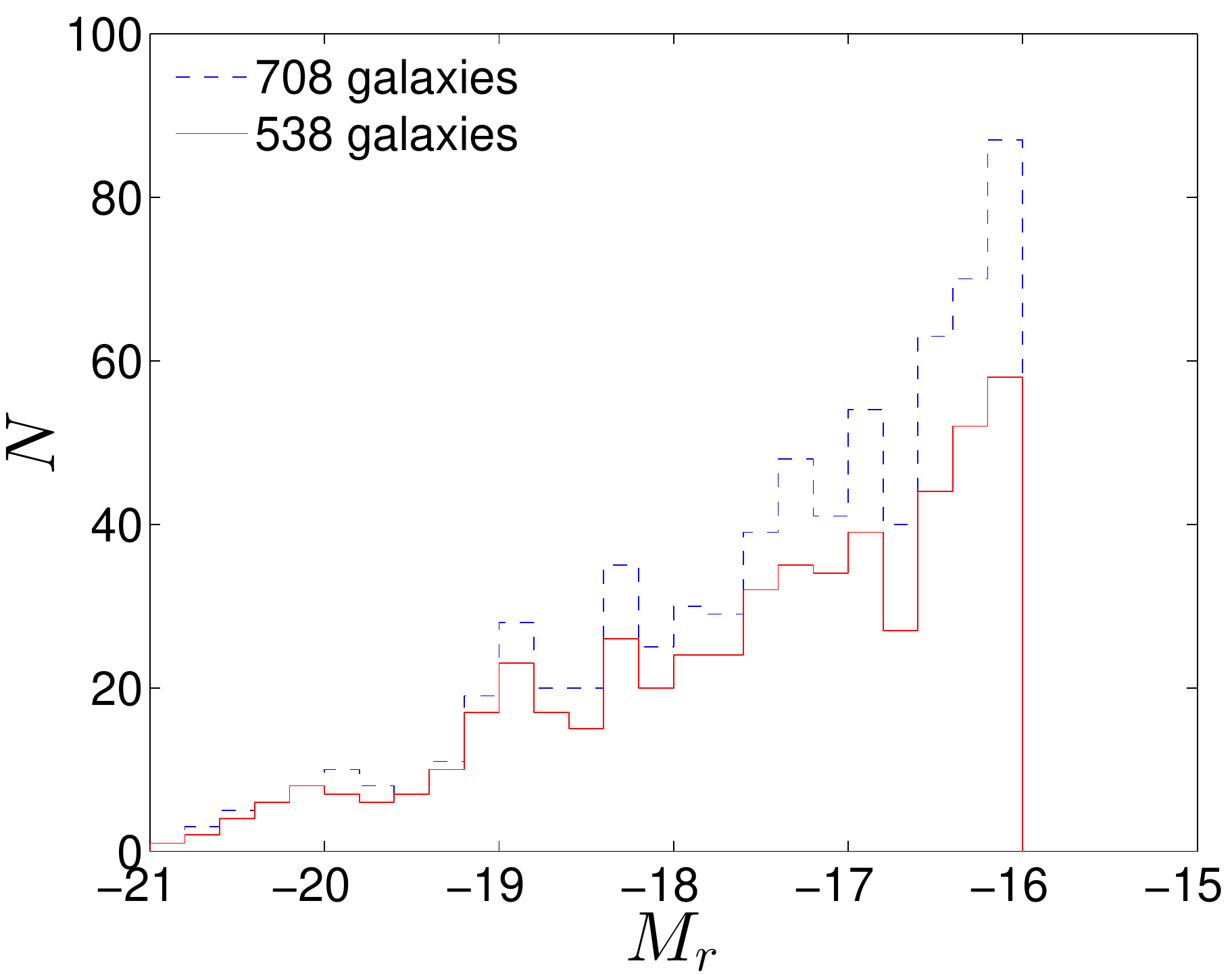}
\includegraphics[width=45mm]{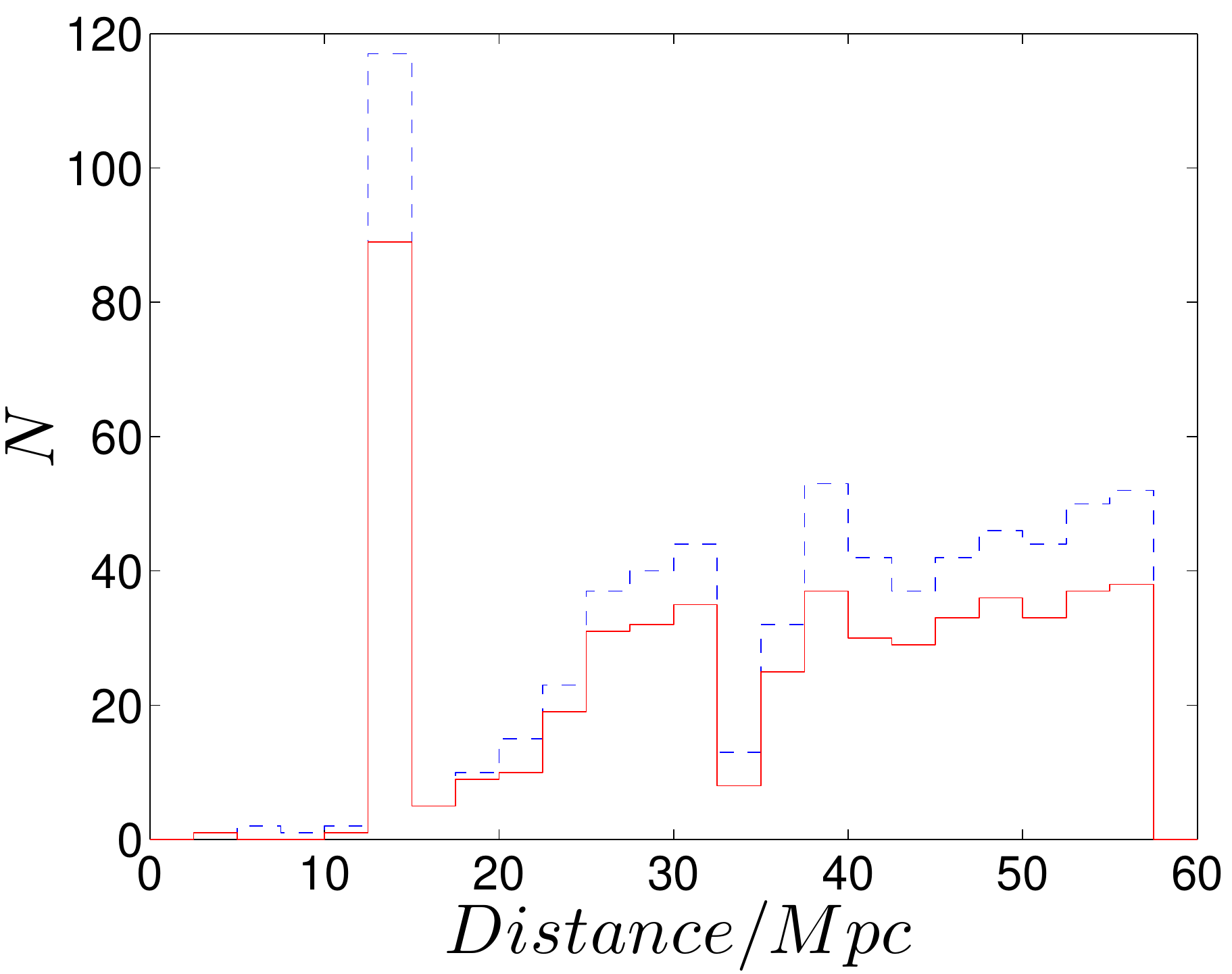}
\end{minipage}
\begin{minipage}{\textwidth}
\includegraphics[width=45mm]{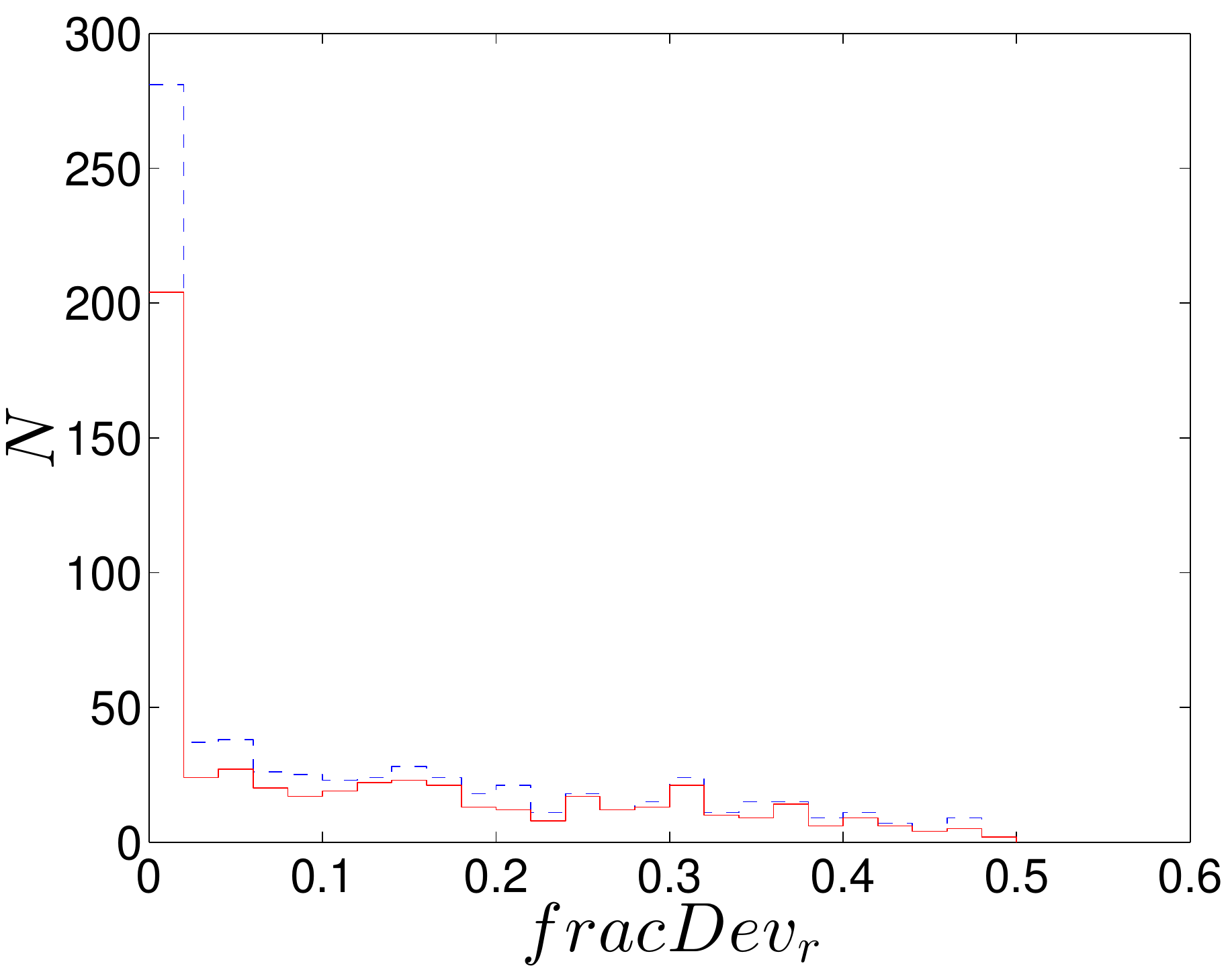}
\includegraphics[width=45mm]{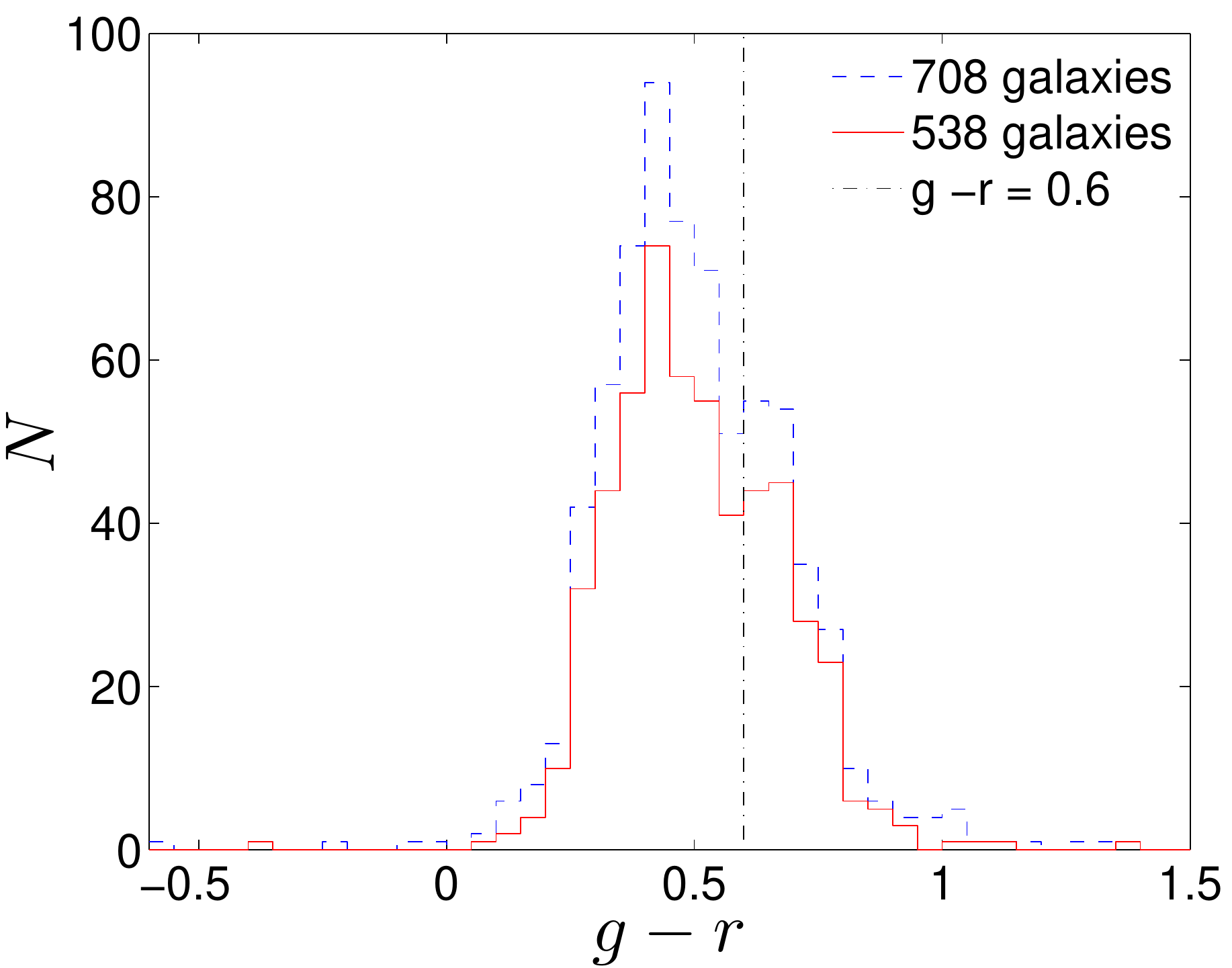}
\end{minipage}
\caption{Histograms of different parameters of sample galaxies. In each panel, blue dashed lines represent our entire sample of 708 galaxies, red solid lines represent our final sample of 538 galaxies to be analyzed.
Top-left: Absolute magnitude in $r$ band which is limited to -16 mag. 
Top-right: Corrected distance calculated using spectral redshift, which is limited to 57 Mpc.
Bottom-left: $fracDev_r$ value, which is limited to 0.5 to ensure that the bulge contribution to the total luminosity is less than $50 \%$. Bottom-right: color (g-r) distribution. $g - r = 0.6$ is presented using the black dot-dashed line. }
\label{fig.samples}
\end{figure}

In the Data Analysis section, we will finally obtain 538 galaxies through checking the Galfit fitting models and residual images. The $M_r$, distance, $fracDev_r$ and color ($g - r$) distributions of 708 entire sample galaxies and 538 analyzed galaxies are presented in Fig.~\ref{fig.samples}. It can be shown from Fig.~\ref{fig.samples} that the final 538 analyzed galaxies can well represent our entire sample.   
From the histograms, $M_r$ is limited to -16 mag, distance is limited to 57 Mpc and $fracDev_r$ is limited to 0.5.

\section{Data analysis} 
\subsection{Photometry}
SDSS has already subtracted sky background of images using Photometric Pipeline (PHOTO).
This process of sky subtraction is not accurate because it considers 
the large extended outskirts of bright objects as sky background and lead 
to sky background overestimation(\citealp{2007ApJ...662..808L}, \citealp{2008MNRAS.385...23L}, \citealp{2009MNRAS.394.1978H}, \citealp{2013ApJ...773...37H}). In view of this problem, we adopt a more precise method
(\citealp{1999AJ....117.2757Z}, \citealp{2002AJ....123.1364W}, 
 \citealp{2015AJ....149..199D} etc.) to subtract sky background of images in optical 
and near-infrared bands of our sample galaxies.

For preparation, we derive the corrected frames in $g, r, i, z$ bands from SDSS DR7 and images in $Y, J, H, K$ bands from UKIDSS DR10 LAS for all the 708 sample galaxies.
We filter the initial images with a Gaussian function which has a FWHM of 8 pixels, 
so that we can extend the area of every single object in the initial images to some extent. 
According to \citealp{2015AJ....149..199D}, the FWHM value of 8 pixels is much better than any other values for generating a 
good smoothed image. To avoid missing inner regions and wings of bright objects and faint stellar halos of galaxies, we apply smoothed images with $FWHM$ $=$ 8 for SExtractor to detect objects. 

\begin{figure}
\centering
\begin{minipage}{\textwidth}
\includegraphics[angle=0,width=3.9cm]{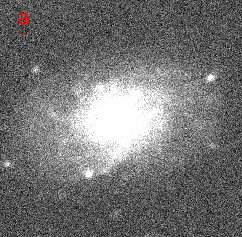}
\includegraphics[angle=0,width=3.9cm]{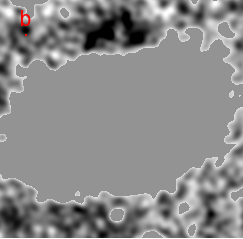}
\end{minipage}
\begin{minipage}{\textwidth}
\includegraphics[angle=0,width=3.9cm]{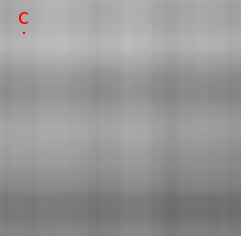}
\includegraphics[angle=0,width=3.9cm]{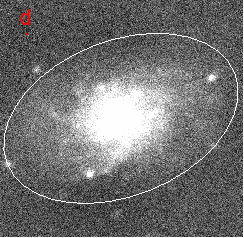}
\end{minipage}
\begin{minipage}{\textwidth}
\includegraphics[angle=0,width=3.9cm]{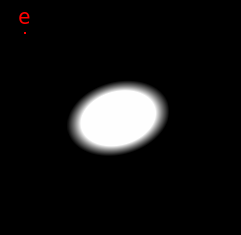}
\includegraphics[angle=0,width=3.9cm]{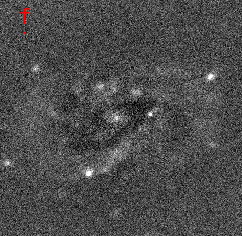}
\end{minipage}
\caption{Sky subtraction, photometry and exponential model fitting results for a sample galaxy (objID=587745544271102105) in $g$ band. This galaxy locates at the center of this image. The panel $a$ is the initial image of this sample. The panel $b$ is the complete object masking image. The panel $c$ indicates the sky background image from fitting. Kron elliptical aperture (SEx AUTO) for this sample galaxy is presented in panel $d$. Exponential model fitting results for this sample galaxy by Galfit are shown in panel $e$ and $f$. The panel $e$ presents the exponential profile model and the panel $f$ is the residual image, respectively.}
\label{fig.image}
\end{figure}

Then objects in the smoothed images with peak flux more than $1.5\sigma$ above the global sky background value 
are detected and masked using the software SExtractor (\citealp{1996AAS..117..393B}). 
As \citealp{2015AJ....149..199D} pointed, if we replace the smoothed images with 
initial images to do object-masking process using SExtractor directly, 
the wings of bright objects and the faint stellar halos of galaxies may be mistaken as the sky background. 
The panel $b$ in Fig.~\ref{fig.image} is the complete masking image for the initial image, which is the panel $a$ in Fig.~\ref{fig.image}.
According to the good object-masking images, all objects in the masked area of the initial images are subtracted.
In the object-subtracted image, only sky pixels are left. 
We could derive accurate and reliable sky background from these sky pixels. 

\begin{figure}
\centering
\includegraphics[angle=0,width=7.8cm]{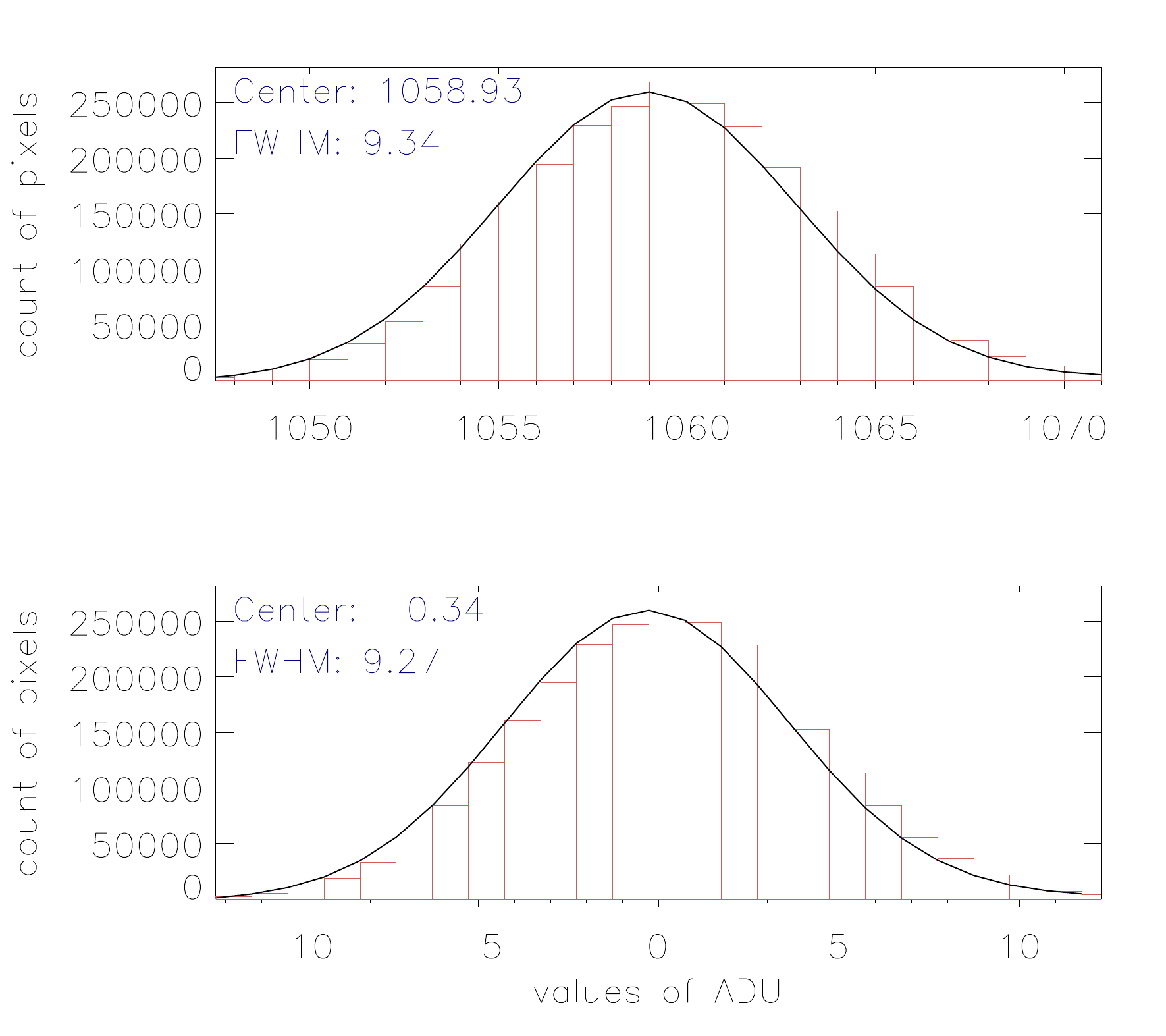}
\caption {The ADU histogram of the image before (top panel) and after (bottom panel) sky subtraction.}
\label{fig.image_histogram}
\end{figure}

At last, the final sky background of the initial image should be modeled using 
a low order least-squares polynomial fitting process, which is performed row-by-row and column-by-column,
to the sky pixels in the object-subtracted image. 
Also, we average the row-fitted and column-fitted sky background maps, 
which is smoothed with the box size of $31\times31$ pixels  
to remove any mistakes and make the sky background map more accurate in the modeling process. 
As \citealp{1999AJ....117.2757Z} pointed, the fitting method of straight 2D background fitting may be under-fitted or over-fitted
to some areas of images. However, this row-by-row and column-by-column bidirectional fitting method 
(\citealp{1999AJ....117.2757Z}, \citealp{2002AJ....123.1364W}, \citealp{2003AJ....126.1286L}, \citealp{2005AJ....129.2628L}, \citealp{2006AJ....132.1581D}, \citealp{2008MNRAS.385...23L}, \citealp{2011AJ....142..166C}, \citealp{2014ApJ...789...76M},  \citealp{2015AJ....149..199D}) is able to 
forecast the sky background, which is underneath the object-masked areas, in a mutually orthogonal way.    
Then we should replace the masked object pixels with the fitted values. 
To avoid introducing spurious fluctuations to the masked areas in the fitted sky background map by interpolations, 
we restrict the polynomial fits to a low order. 
The panel $c$ of Fig.~\ref{fig.image} shows a gradient across the frame. 
It indicates the smoothed sky background image, which is considered as sky background.
The panel $d$ of Fig.~\ref{fig.image} is the sky-subtracted image of the original image. 
We have precisely derived the sky background images and subtracted the sky background from the original images of 
our sample galaxies in multi-bands ($g, r, i, z, Y, J, H, K$). Fig.~\ref{fig.image_histogram} shows the count distributions 
of the image before (top panel) and after (bottom panel) sky subtraction. After sky subtraction, the mean value of ADU is very close to 0.

Here we do surface photometry for the sky-subtracted images of our sample galaxies 
with the elliptical apertures using the SExtractor software (\citealp{1996AAS..117..393B}).
Compared with SDSS Petrosian circular aperture, which is not optimal choice for galaxies 
with large angular extent, irregular morphology or edge-on shape, the automatic aperture 
magnitude (AUTO) from SExtractor package could estimate the "total magnitudes" much more precisely.
SDSS petrosian aperture could be so large that it includes the light from adjacent objects, or too small 
to include all the intrinsic light from our target objects.

Automatic aperture magnitudes (AUTO) is inspired by Kron's "first moment" algorithm(\citealp{1980ApJS...43..305K}). 
The AUTO aperture R is an elliptical aperture with elongation, $\epsilon$, and position angle, $\theta$, 
 which are defined by the second-order moments of the object's light distribution. Within this aperture, the first moment of an image is defined with 

\begin{equation}\label{R1}
    R_1 $=$ \frac{\Sigma R I(R)}{\Sigma I(R)}
\end{equation}
According to \citealp{1980ApJS...43..305K} and \citealp{1987AA...183..177I}, more than 90\% of the flux is expected 
to lie within an aperture of radius $kR_1$ for stars and galactic profiles convolved with Gaussian seeing.
Here we adopt $k $=$ 2.5$ during the automatic elliptical aperture photometry, which is the default setting of Sextractor (\citealp{1996AAS..117..393B}). 
The panel $d$ in Fig.~\ref{fig.image} shows the Kron elliptical apertures (SEx AUTO) in the aperture photometry process 
for the sample galaxy from SDSS DR7 (objID $=$ 587745544271102105) in $g$ band. 
The elliptical aperture is Kron radii.
 In this way,  we could estimate the AUTO magnitude 
in multi-bands for all our sample galaxies.

\subsection{Geometry by Galfit}
We use Galfit software (\citealp{2002AJ....124..266P}), which is good at galactic geometric fitting, to estimate some useful
geometric parameters.

There are several different radial profile functions, e.g., Sersic, exponential, Nuker and other models. 
Through the sample selection, we remove the bulge-dominated galaxies, and 
only disk-dominated galaxies with fracDev $\leq$ 0.5 are left.  
Therefore, we fit all of our sample galaxies only with exponential profile. 
The parameters derived from SExtractor photometry process, 
like galactic magnitude, disk scalelength, axis ratio and position angle, are 
set as initial input values for the set of Galfit.
Then we could derive the best-fit values of parameters, like axis ratio (q), 
disk scalelength in pixel ($\alpha$), 
and inclination angle (i), and also generate three images, 
including the initial galaxy image, the exponential model(the panel $e$ in Fig.~\ref{fig.image}), 
and the residual image (the panel $f$ in Fig.~\ref{fig.image}) for each galaxy in our sample.  

\subsection{Central surface brightness}
Normally, $\mu_0$ is used to classify galaxies into low or high 
surface brightness regime (\citealp{1970ApJ...160..811F}, \citealp{1996ApJS..105..209I}, \citealp{1997AJ....113.1212O}, \citealp{2008MNRAS.391..986Z}, \citealp{2015AJ....149..199D}). 
In this subsection, we calculate $\mu_0$ in optical and near-infrared bands 
for each galaxy in our sample. 
The surface brightness profiles of disk galaxies could be estimated using an exponential profile: 
\begin{equation}
        \Sigma(r) $=$ \Sigma_0 exp(-r/a),
        a $=$ x \times y,
\end{equation}
where $\Sigma(r)$ and $\Sigma_0$ are the surface brightness in units of $M_\odot pc^{-2}$ 
at the radius of r and at the center of the disk, respectively. 
The parameter of $a$ is the disk scalelength in units of arcsecond.
In the description of $a$, the parameter $x$ is the disk scalelength in units of pixel derived from Galfit, 
and $y$ is the pixel size in units of arcsec/pixel, which is 0.396 for the SDSS images and 0.4 for UKIDSS LAS DR10 images. 
Considering the disk of a galaxy is infinite thin, then 
\begin{equation}
        F_{tot} $=$ 2{\pi}{{a}^2}\Sigma_0.
\end{equation}
Combining with 
\begin{equation}
        F $=$ 10^{-0.4m},
\end{equation}
the disc central surface brightness is derived in the form of 
\begin{equation}
        \mu_0 $=$ m + 2.5log(2\pi{a^2}),
\end{equation}
where $\mu_0$ is the disc central surface brightness in units of
$mag$ $arcsec^{-2}$, and $m$ means the total apparent magnitude, which is the AUTO aperture magnitude 
estimated using SExtractor.   
As \citealp{1997AJ....113.1212O}, \citealp{2006AA...458..341T}, \citealp{2008MNRAS.391..986Z} and \citealp{2015AJ....149..199D} pointed out, the central surface brightness should 
be corrected by inclination and cosmological dimming effects in the form of 
\begin{equation}
        \mu_0 $=$ m + 2.5log_{10}(2\pi{a^2}q) - 10log_{10}(1+z).
\end{equation}    
In this equation, $q$ is the axis ratio estimated from the exponential profile fitting by Galfit, 
and $z$ is the spectral redshift from SDSS DR7. For the multi-bands, we use the same $q$ in r-band to calculate $\mu_0$.
Using this equation, $\mu_0$ in multi-bands ($g, r, i, z, Y, J, H, K$) 
are calculated.

We check the Galfit fitting models and residual images of every galaxy in $g, r, i, z, Y, J, H$ and $K$ band. There are 170 galaxies that cannot be fitted well due to the pollution of bright stars or the irregular shape of the galaxy. We remove these 170 badly-fitted galaxies and 538 galaxies are left as our final sample to study the $\mu_0$ distributions. 
In the final 538 galaxies, there are cluster galaxies and field galaxies. 
\section{Results}
\begin{figure*}
\begin{minipage}{\textwidth}
\includegraphics[width=.24\textwidth]{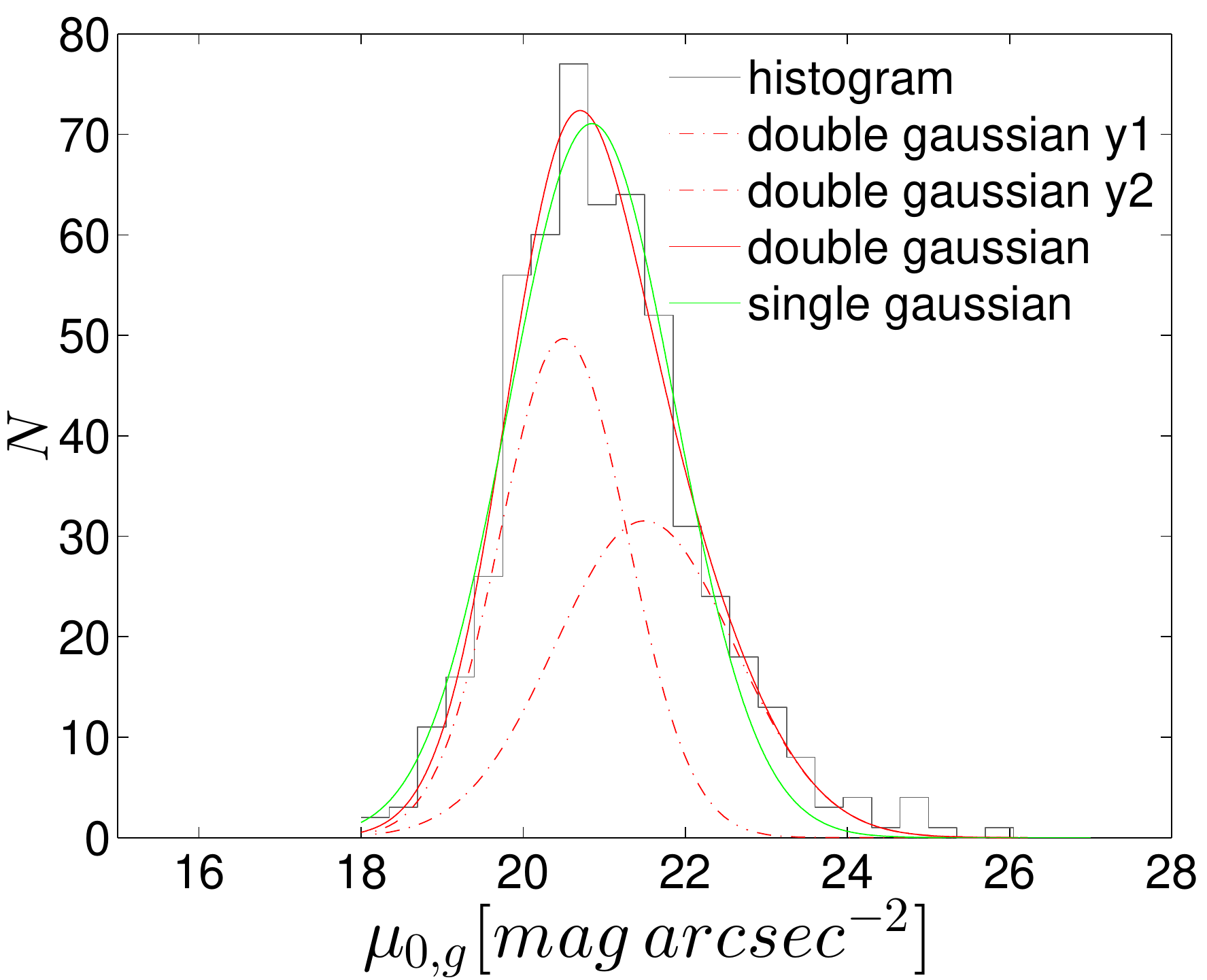}
\includegraphics[width=.24\textwidth]{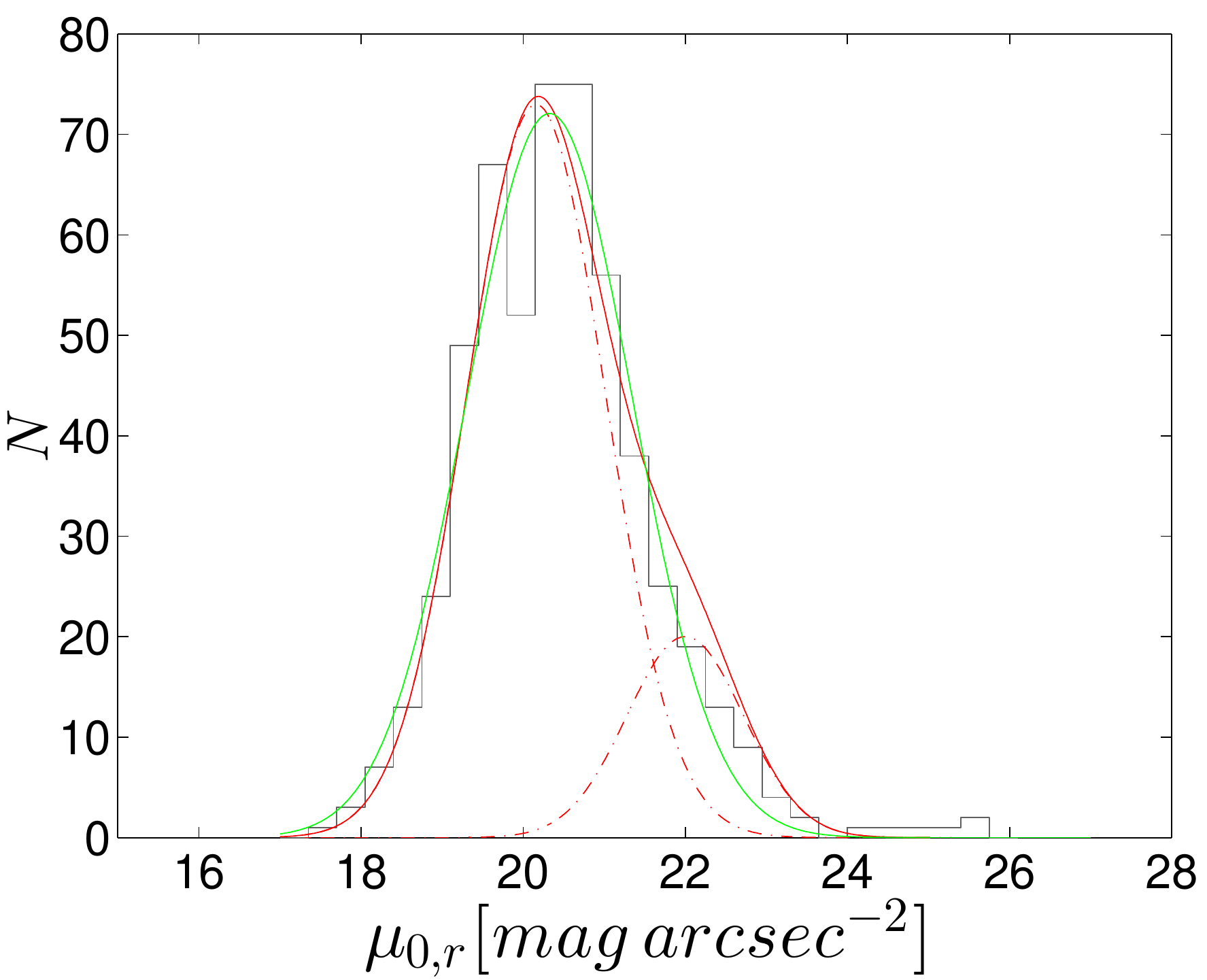}
\includegraphics[width=.24\textwidth]{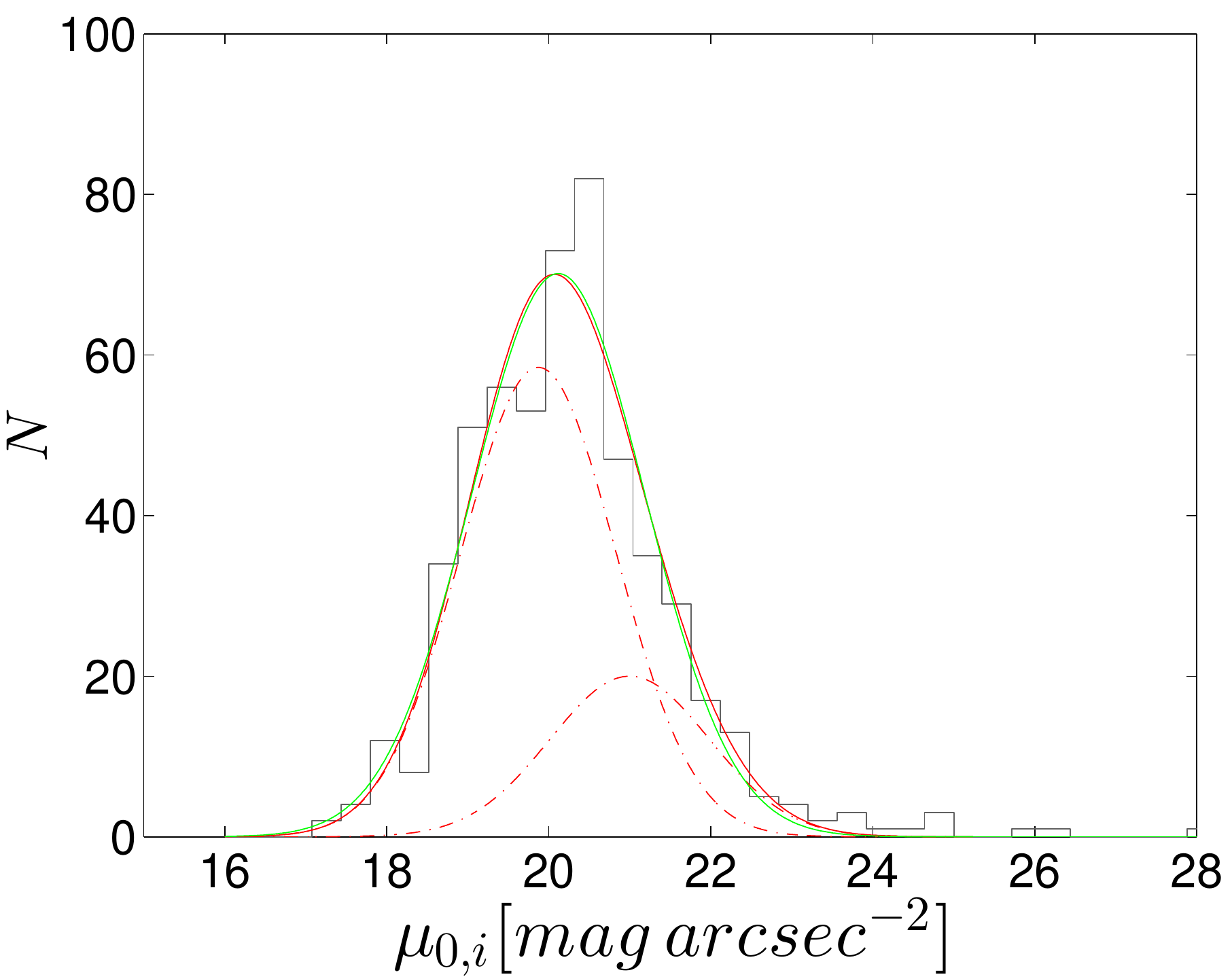}
\includegraphics[width=.24\textwidth]{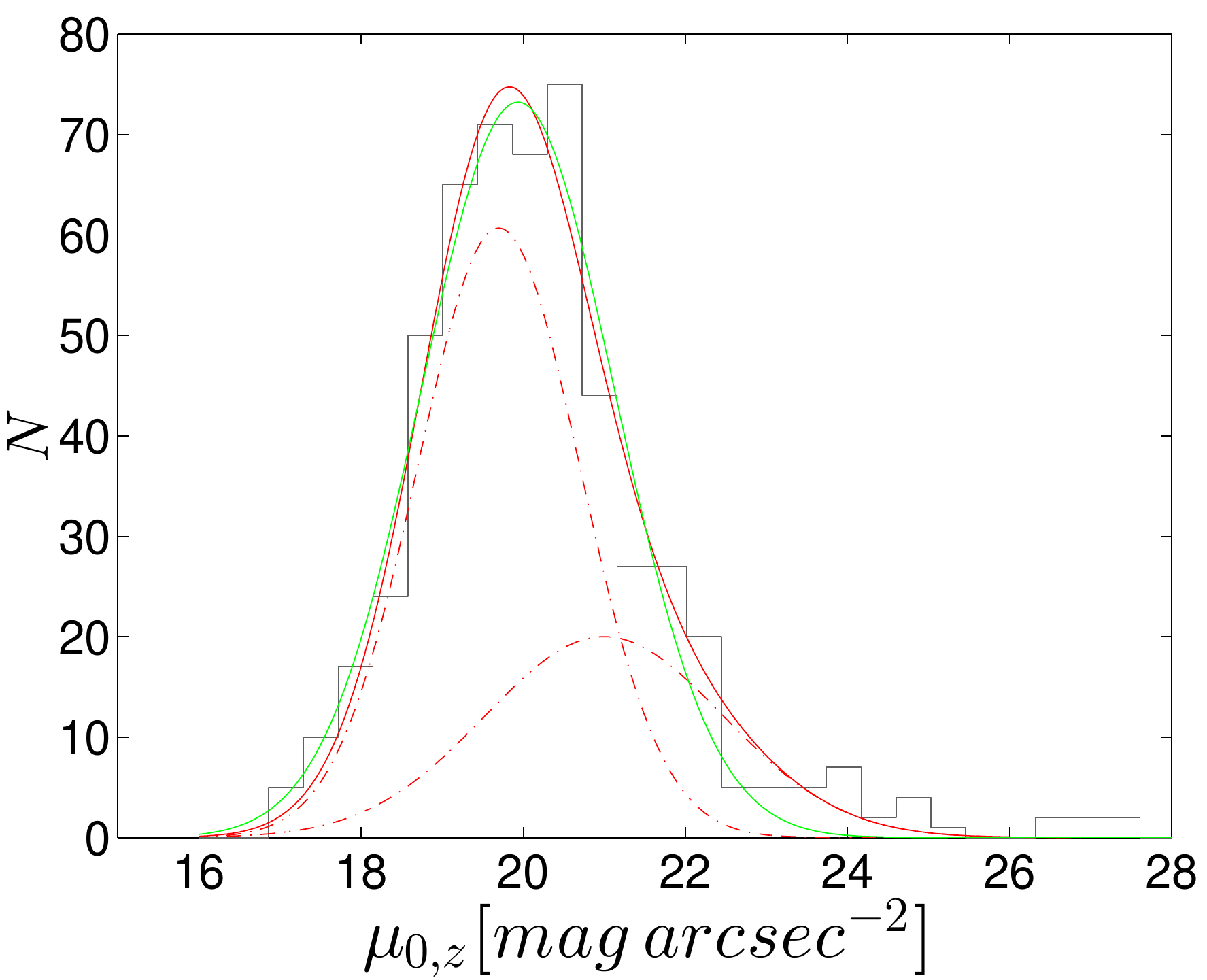}
\end{minipage}
\begin{minipage}{\textwidth}
\includegraphics[width=.24\textwidth]{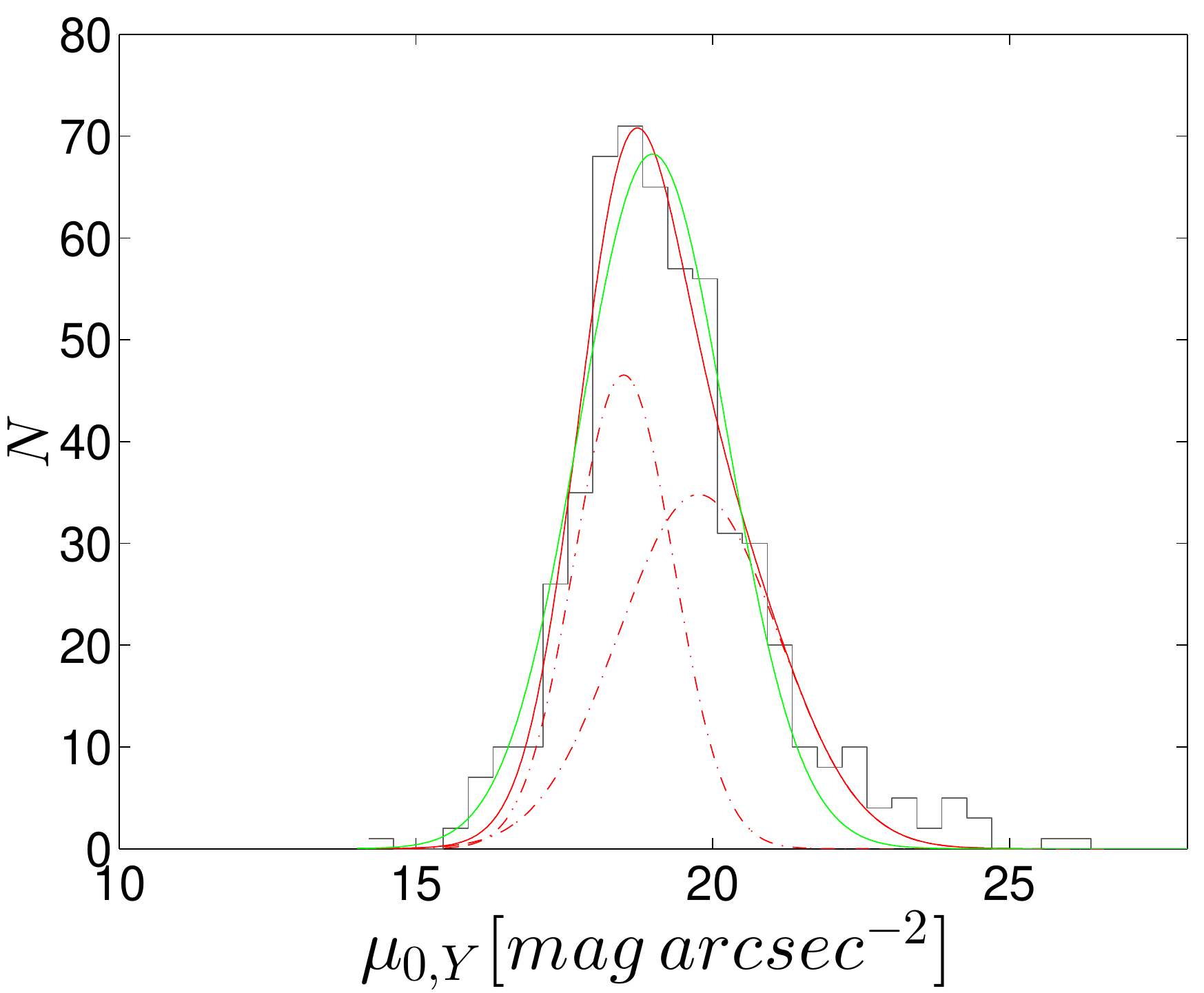}
\includegraphics[width=.24\textwidth]{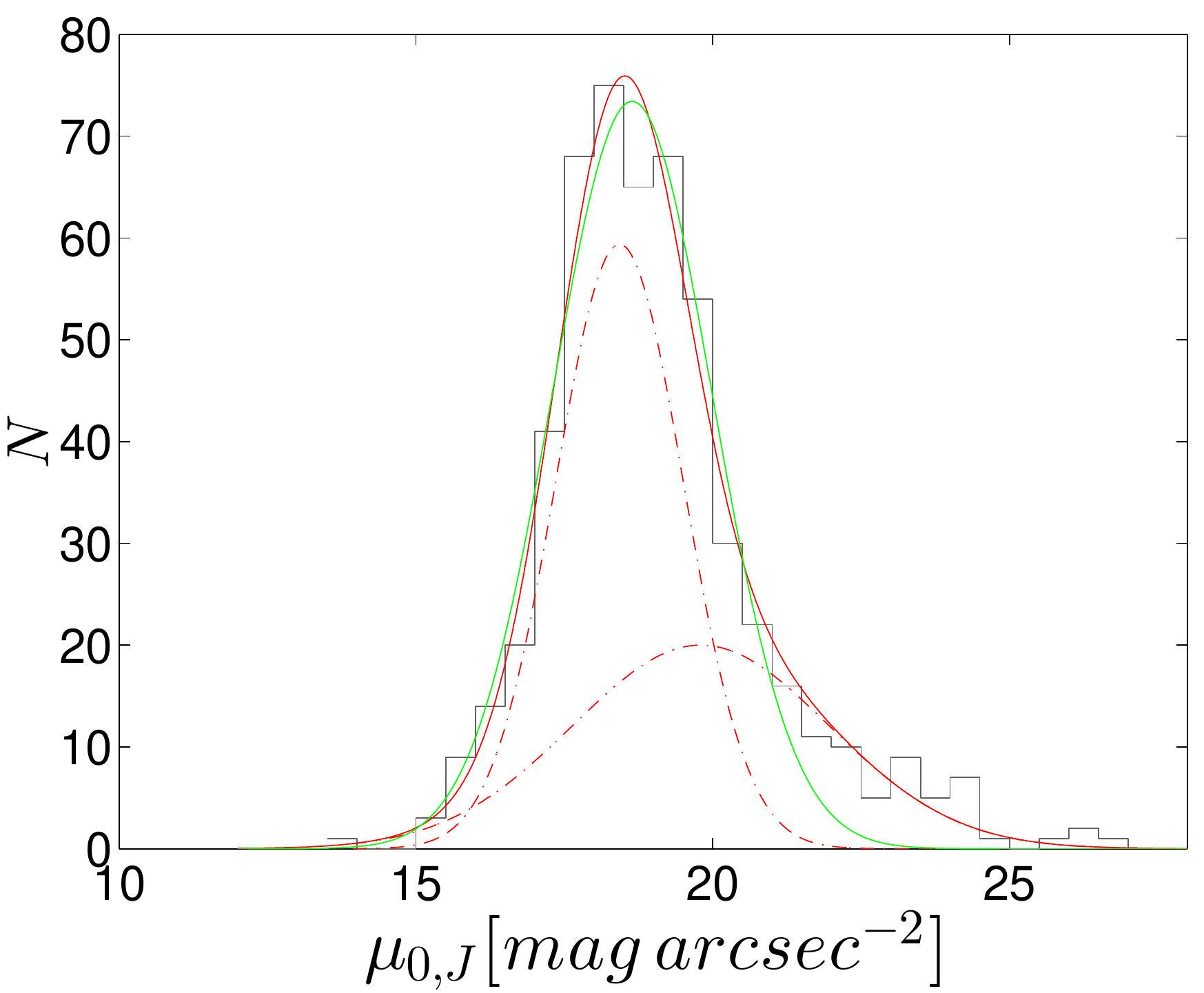}
\includegraphics[width=.24\textwidth]{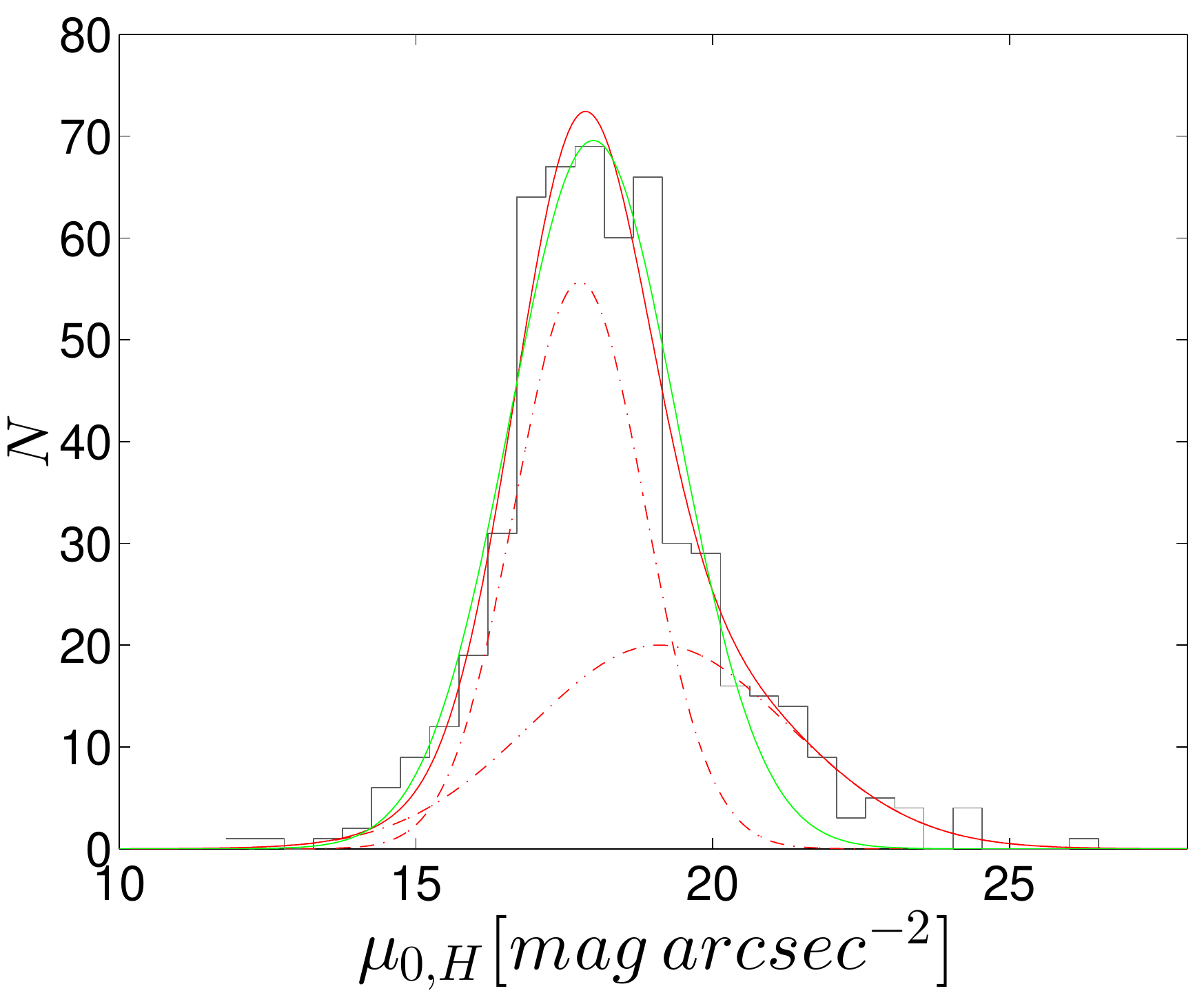}
\includegraphics[width=.24\textwidth]{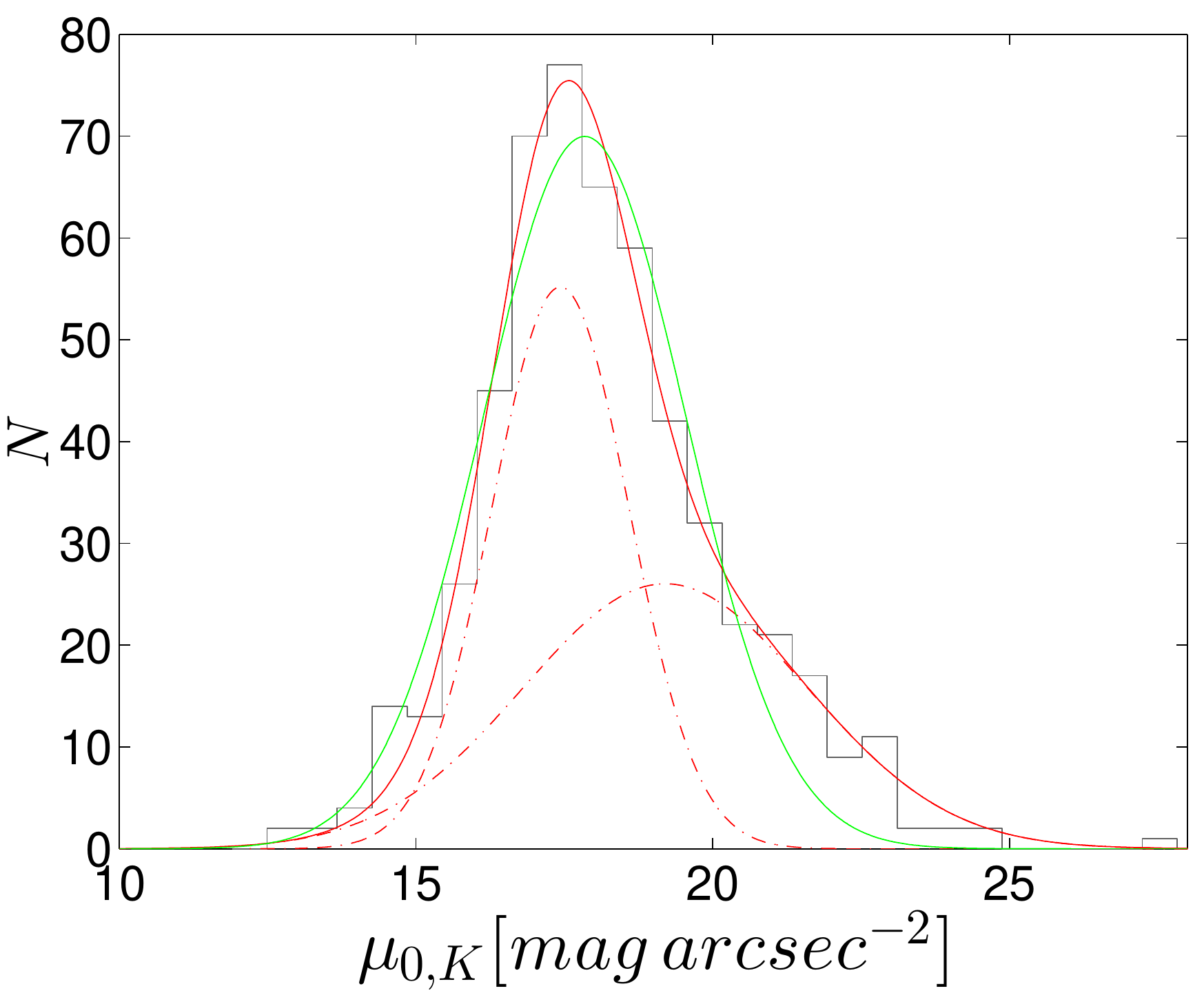}
\end{minipage}
\caption{Histograms of $\mu_0$ in multi-bands ($g, r, i, z, Y, J, H, K$ bands) with the optimal bin size.
Single Gaussian profile fitting to the distribution is described by green lines, while a sum of double Gaussian profiles is described
using red solid lines and two separate Gaussian profiles are described using red dotted lines.}
\label{fig.surface brightness05}
\end{figure*}

In this section, we examine the $\mu_0$ distributions of our 
final sample in multi-bands ($g, r, i, z, Y, J, H, K$) and show them in 
Fig.~\ref{fig.surface brightness05} with optimal bin size.

We adopt the optimal bin size in each band. According to \citealp{fr81}, the optimal bin size is $h = 2IQR(X)n^{-1/3}$, where n is the number of observations on X, which is 538 for our sample. IQR(X) is the interquartile range, which is the difference between the upper (top $75 \%$) and lower (bottom $25 \%$) quartiles.
A single Gaussian fitting to the distribution is drawn by green lines, while a sum of double Gaussian profiles is described
using red solid lines and two separate Gaussian components are described using red dashed-dotted lines. 

To estimate the quality of a model and select the best model for a given data, 
we apply the Akaike information criterion (AIC/AICc) and Bayesian information criterion (BIC) 
in this work. When the sample size is small, AIC should be corrected into AICc. The likelihood will increase with more parameters added 
in the process of fitting models, but it could result in over fitting at the same time. 
Both BIC and AICc attempt to resolve this problem. 
According to \citealp{ra95}, the model with lower BIC represents better fitting than models with higher BIC. 
We set $\bigtriangleup BIC = BIC_{single} - BIC_{double}$ and $\bigtriangleup AICc = AICc_{single} - AICc_{double}$. 
The evidence that double Gaussian being better than single Gaussian fitting is positive when $ 2 \le \bigtriangleup BIC (or \bigtriangleup AICc) < 6 $, 
strong when $6 \le \bigtriangleup BIC (or \bigtriangleup AICc) < 10$, and very strong when $\bigtriangleup BIC (or \bigtriangleup AICc)\ge 10$.

\begin{table*}
\centering
\caption{The AIC and BIC values of all the sample galaxies with single and double Gaussian fitting with the optimal bin size in multi-bands.}
\label{table.AIC_BIC_all}
\begin{tabular}{|ccccccccc|}
\hline
Band & $g$ & $r$ & $i$ & $z$ & $Y$ & $J$ & $H$ & $K$ \\
\hline
$AICc_{s}$ &69.9 &83.7 &109.8 &85.2 &92.6 &85.3 &100.1 & 90.8 \\
$AICc_{d}$ &68.3 &96.7 &111.5 &86.3 &87.4 &75.5 &90.9  &61.2\\
$BIC_{s}$ &71.8 &86.2 &113.5 &88.0 &95.6 &88.0 &103.2 &93.5 \\
$BIC_{d}$ &69.8 &99.7 &116.8 &90.1 &91.5 &78.5 &94.7 &64.2 \\
$\bigtriangleup AICc$ & 1.6 & -13.0 & -1.7 & -1.1 & 5.2 & 9.8 & 9.2 & 29.6 \\ 
$\bigtriangleup BIC$ & 2.0 & -13.5 & -3.3 & -2.1 & 4.1 & 9.5 & 9.5 & 29.3 \\
bin size & 0.35 & 0.35 & 0.36 & 0.43 & 0.42& 0.50& 0.49& 0.59 \\
\hline
\end{tabular}
\end{table*}

From all these histograms in Fig.~\ref{fig.surface brightness05} and the values of AICc and BIC in Table~\ref{table.AIC_BIC_all}, it is much better for the $\mu_{0}$ distribution fitting with a double Gaussian profile than a single Gaussian profile in all the 
NIR bands ($Y, J, H$ and $K$ band), but it is not in optical bands ($g, r, i$ and $z$ band). The difference of $\mu_{0}$ distribution between optical and near-infrared bands could be caused by the effect of dust extinction in optical bands. Especially in $K$ band, the evidence of a double Gaussian profile being better than a single Gaussian profile is very strong, which has $\bigtriangleup BIC \ge 10$ and $\bigtriangleup AICc \ge 10$. 
For $K$ band, the $\mu_0$ distribution has double Gaussian peaks with a separation of $\delta \mu_0 \approx 1.8$ $mag$ $arcsec^{-2}$. The double Gaussian peaks are at $\sim$ 17.4 $mag$ $arcsec^{-2}$ and $\sim$ 19.2 $mag$ $arcsec^{-2}$, respectively. 
The gap position between two Gaussian peaks in this study locates at 18.3 $mag$ $arcsec^{-2}$. In \citealp{1997ApJ...484..145T}, the double Gaussian distribution of $\mu_0$ shows double peaks at 17.3 and 19.7 $mag$ $arcsec^{-2}$ and the gap position locates at 18.5 $mag$ $arcsec^{-2}$ in the $K'$ band in the Vega system. In \citealp{2009MNRAS.393..628M} and  \citealp{2009MNRAS.394.2022M}, the double peaks locate at about 17.8 and 20 $mag$ $arcsec^{-2}$ and the gap position locates at 19 $mag$ $arcsec^{-2}$ in the $K'$ and $H$ band in the Vega system. In AB system, \citealp{2013MNRAS.433..751S} has found double peaks at 20.5 and 22.5 $mag$ $arcsec^{-2}$ and a dearth at 21.5 $mag$ $arcsec^{-2}$ in the 3.6 $\mu m$ band. The double peak locations and gap positions in this study are similar with those in the previous studies. 

\section{Discussion}
The evidence of a double Gaussian profile being better than a single Gaussian profile for $\mu_0$ distribution is very strong in $K$ band. Therefore, we will discuss possible reasons for the result that the double Gaussian being better than a single Gaussian by taking $K$ band for example.
In this section, we analyze the effect of sample incompleteness, axis ratio ($b/a$), disk scalelength, $fracDev_r$, bin size, color and statistical fluctuations on the $\mu_0$ distribution in $K$ band. 

\subsection{Effect of sample incompleteness}
Our sample is not absolutely complete due to the sample selection criteria. 
Generally, the criteria in absolute magnitude (brighter than -16 mag in r
band) tends to preclude more galaxies with low surface brightness, and our criteria in distance
(less than 57 Mpc) and bulge-to-total ratio (less than 0.5) tends to preclude more of the large and
bright galaxies, which are generally have high surface brightness. That is to say, it is likely that
our selection criteria have only reduced the number of galaxies with low surface brightness and
high surface brightness more severely than the number of the galaxies with intermediate surface brightness.

Even by fitting the $\mu_0$ of our sample, which has been flattened
at the high and low surface brightness part more largely than the intermediate surface brightness
part due to our selection criteria, we could get a double Gaussian fitting in $K$ band, which means the high surface
brightness part and low surface brightness part could be separated. 
So, if the lost galaxies at
the high surface brightness and low surface brightness parts could be recovered back into our
sample, the peak of the number of galaxies with high or low surface brightness would become more
distinct and thus would still show a double Gaussian distribution in central surface brightness in $K$ band. 

Therefore, the sample 
incompleteness does not change our conclusion about double Gaussian 
distribution of $\mu_0$ in $K$ band.

\subsection{Inclination}
Dust and projection geometry may affect the estimation of 
$\mu_0$ (\citealp{bell2000}). As \citealp{1994essg.book.....H} pointed out, when 
averaging ellipse surfaces at high inclinations, the estimation for the value of $\mu_0$  is systematically smaller than the real value of $\mu_0$.
The assuming of a thin, uniform, slab disc at high inclinations 
could result in incorrect conclusions due to the effect of three dimensionality of stellar structure on the inclination correction (\citealp{2013MNRAS.433..751S}).
The integration along the line of sight may hide the effects of sub-structures, like bars and spiral arms (\citealp{2010MNRAS.401..559M}), 
and there is no precise method to correct the inclination up to now, so it is not easy to 
estimate accurate $\mu_0$, especially for edge-on galaxies. 

To avoid the effect of incorrect inclination correction to $\mu_0$, 
we separate our samples into two parts in terms of inclination, one is galaxies with high inclination ($b/a \leq 0.35$) and the other is galaxies with low inclination ($b/a > 0.35$). 
The value of 0.35 is adopted by \citealp{2013MNRAS.433..751S}. 
According to Figure.2 in \citealp{bell2000}, the surface brightness distribution of high-inclination galaxies with $b/a \leq 0.4$ 
has larger error bar and is more inaccurate due to internal extinction. The uncertainties on inclinations are about $4^{\circ}$ - $5^{\circ}$, 
therefore, choosing 0.35 ($73^{\circ}$) instead of 0.4 ($69^{\circ}$) will not change the conclusions (\citealp{2013MNRAS.433..751S}).    

\begin{figure*}
\begin{minipage}{\textwidth}
\includegraphics[width=.24\textwidth]{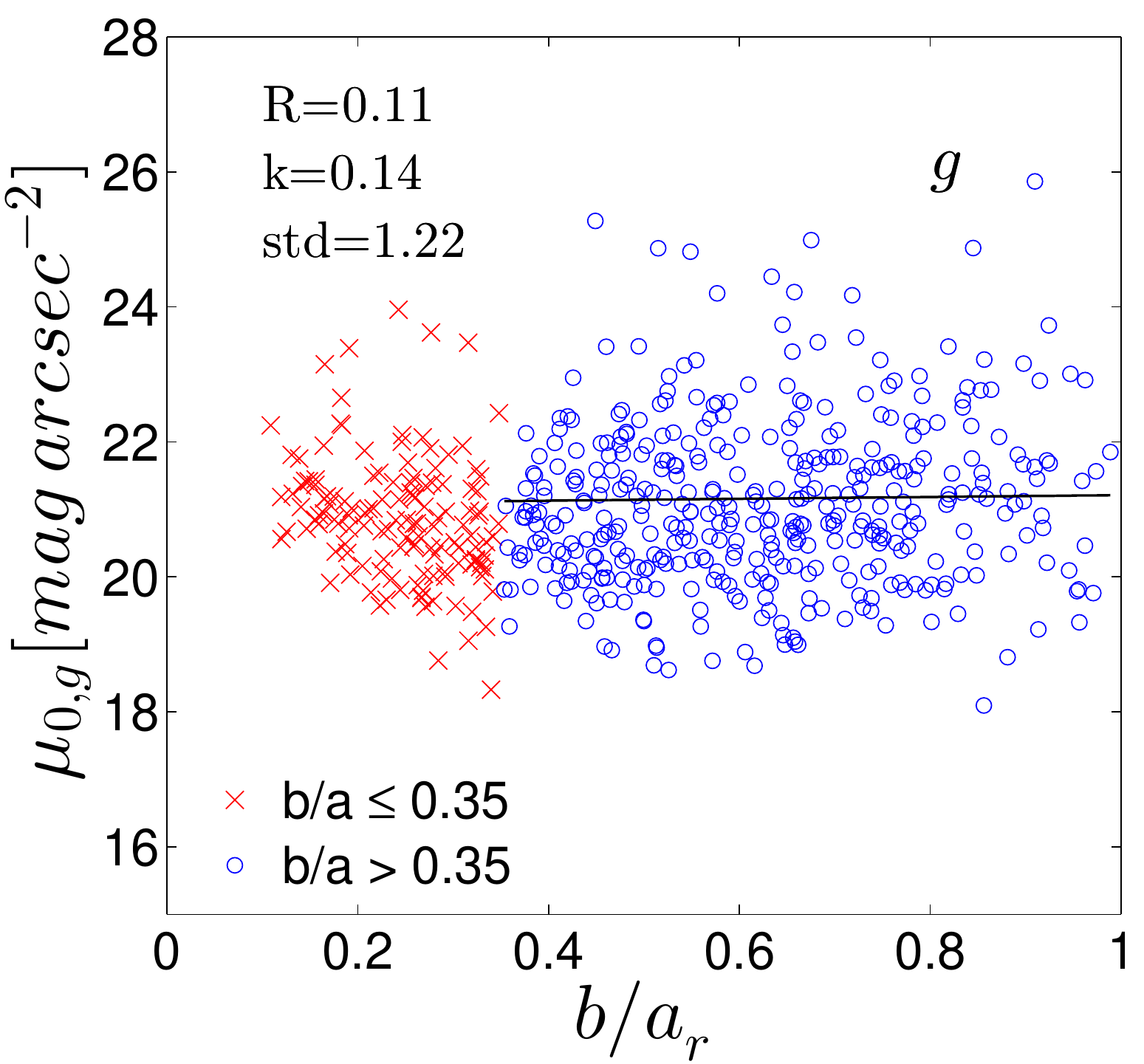}
\includegraphics[width=.24\textwidth]{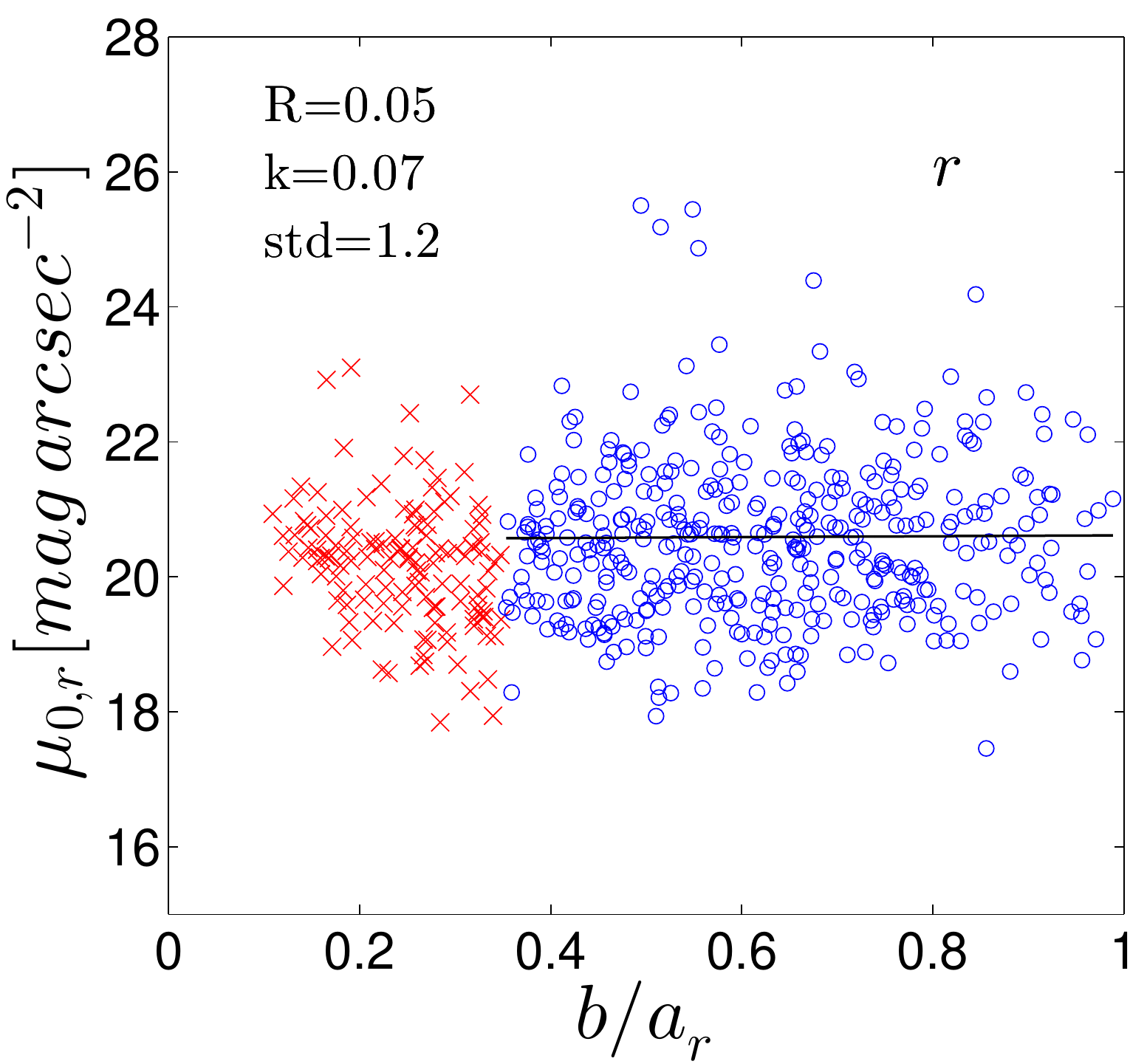}
\includegraphics[width=.24\textwidth]{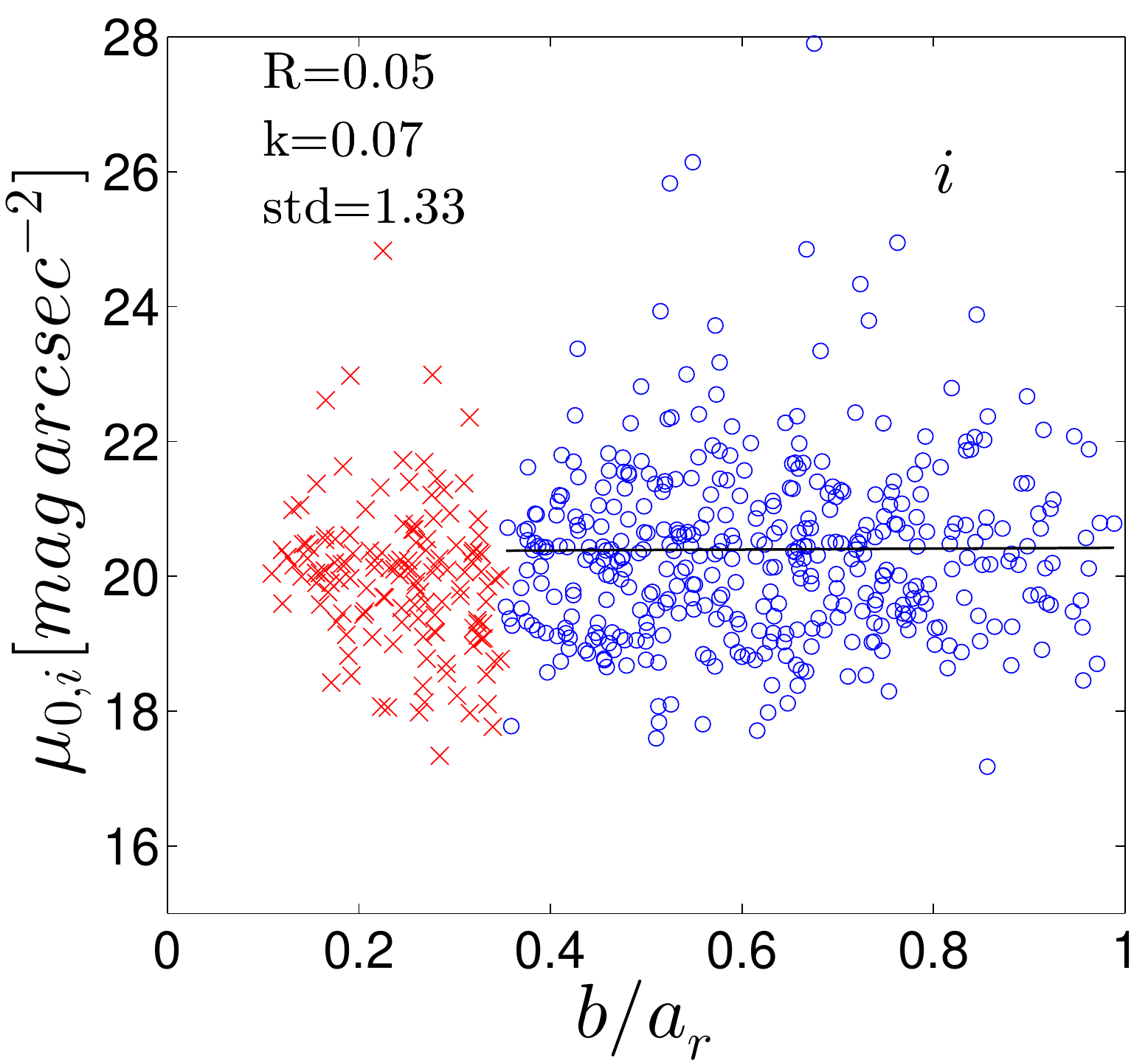}
\includegraphics[width=.24\textwidth]{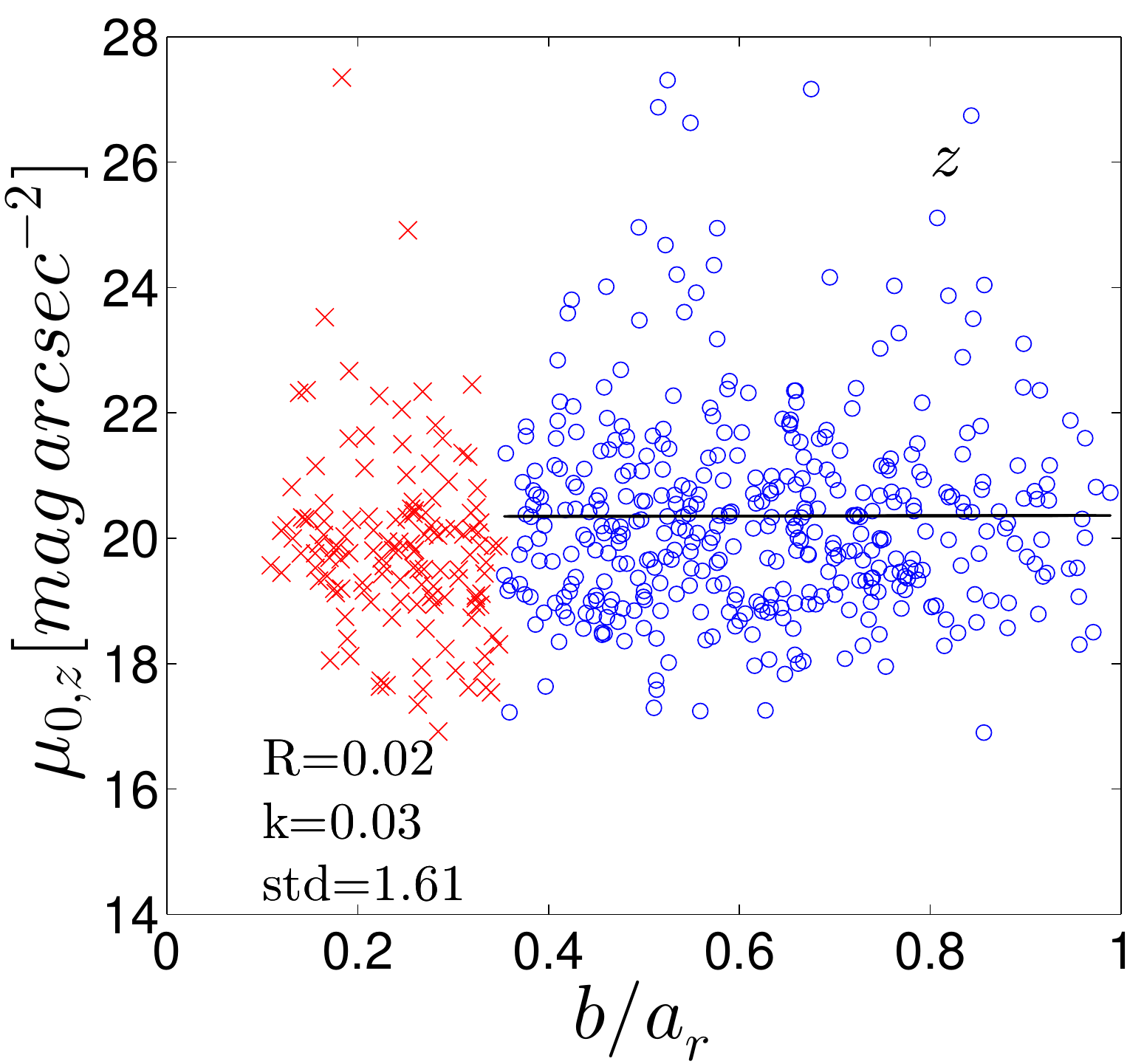}
\end{minipage}
\begin{minipage}{\textwidth}
\includegraphics[width=.24\textwidth]{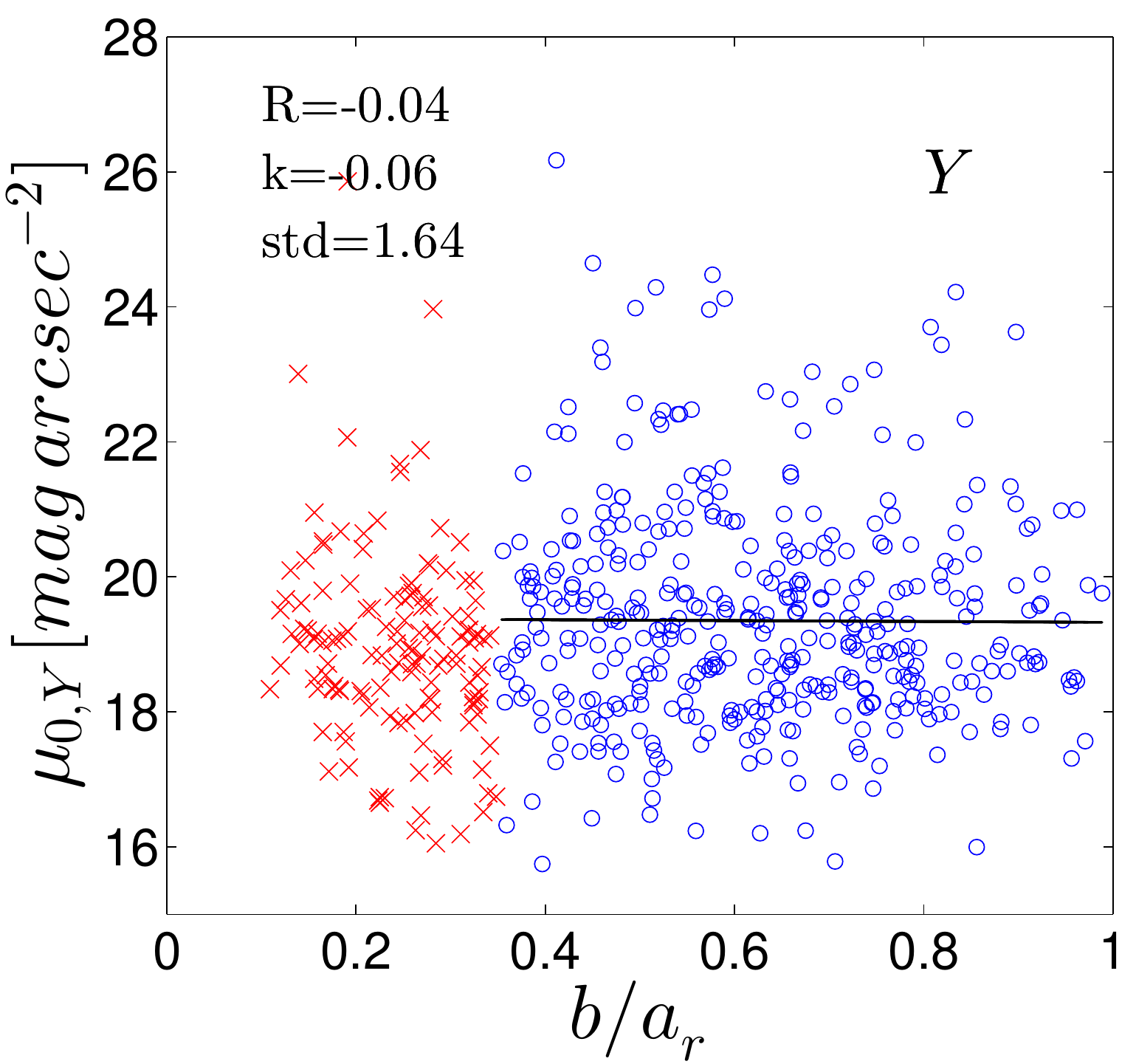}
\includegraphics[width=.24\textwidth]{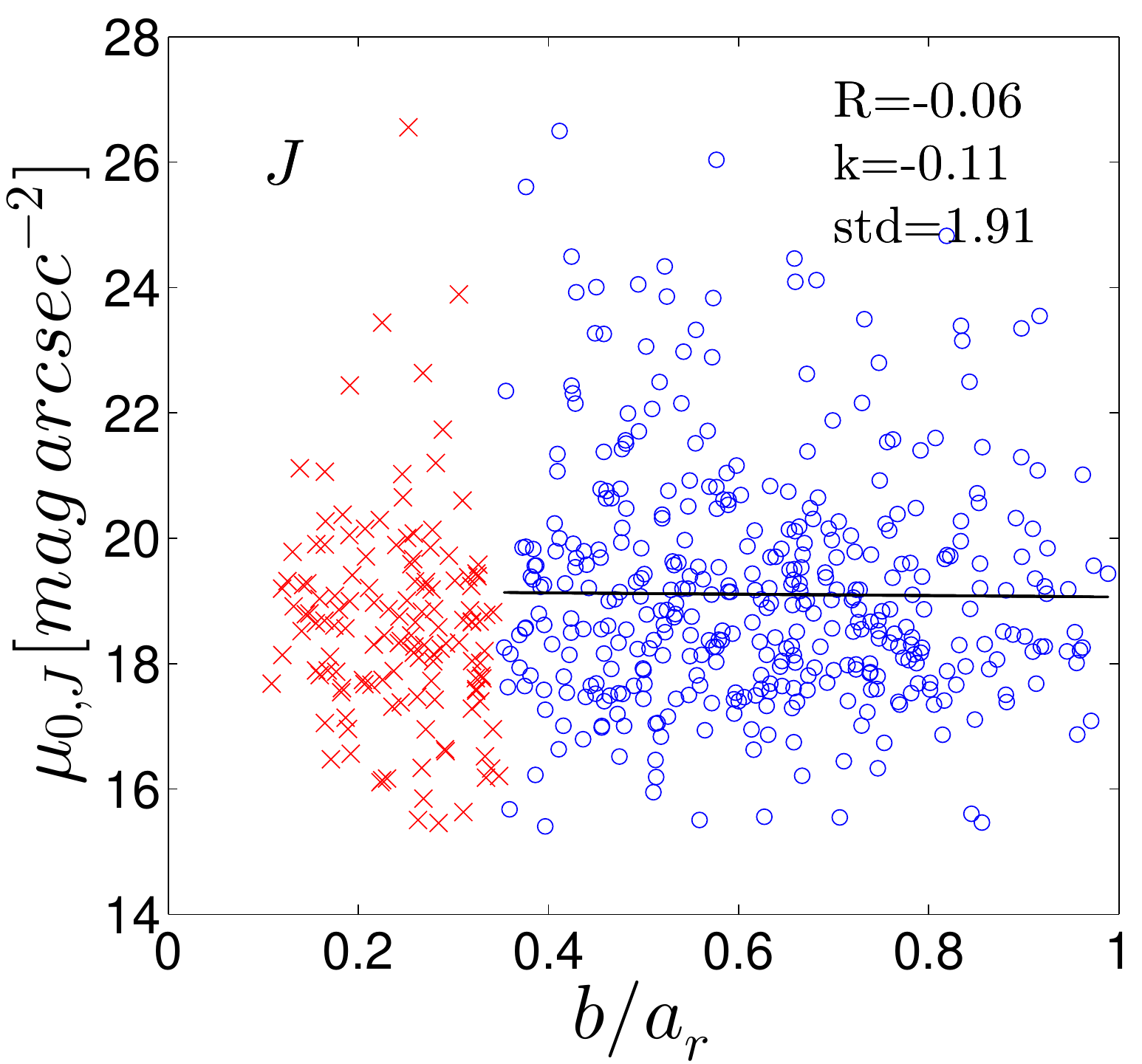}
\includegraphics[width=.24\textwidth]{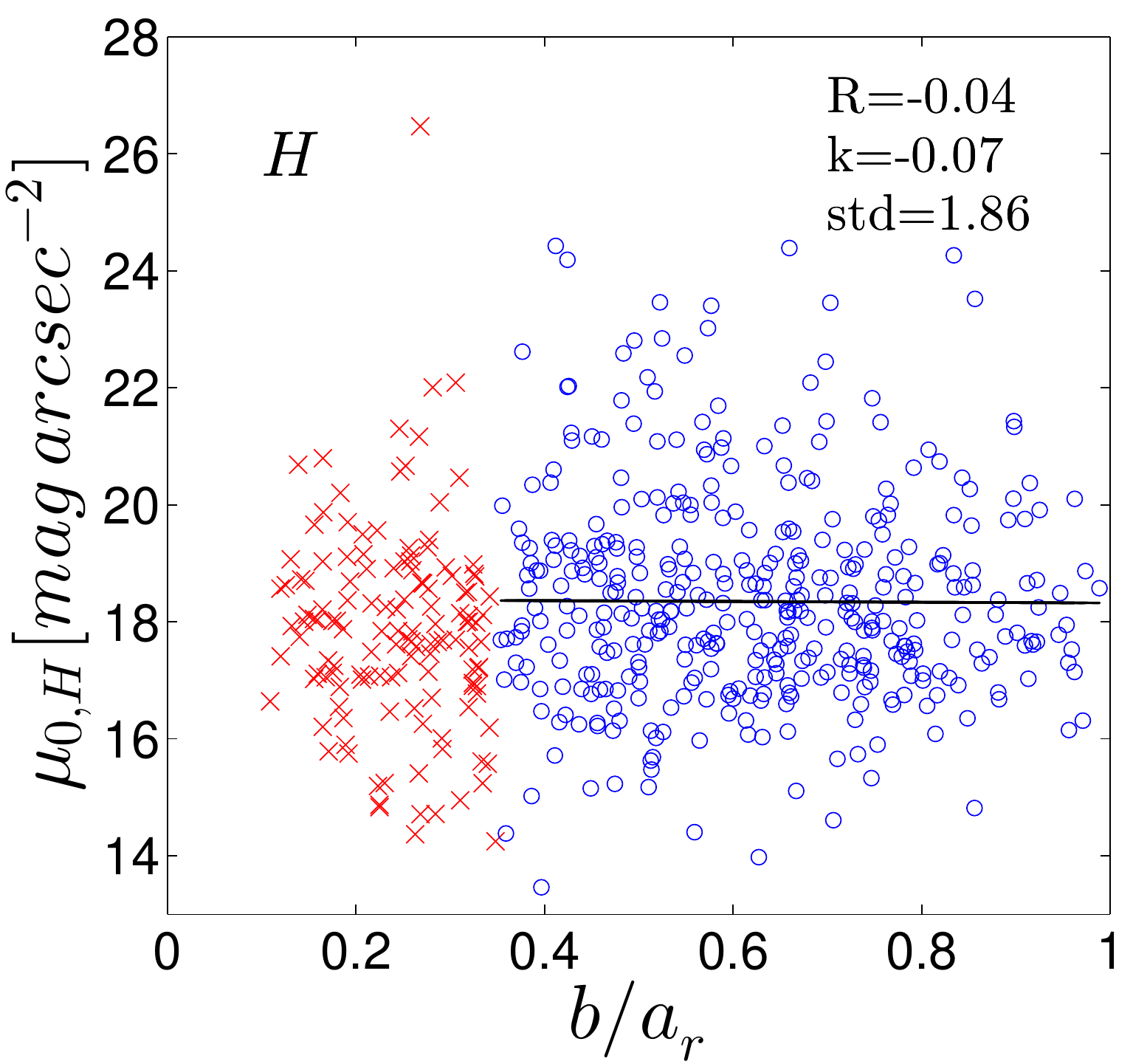}
\includegraphics[width=.24\textwidth]{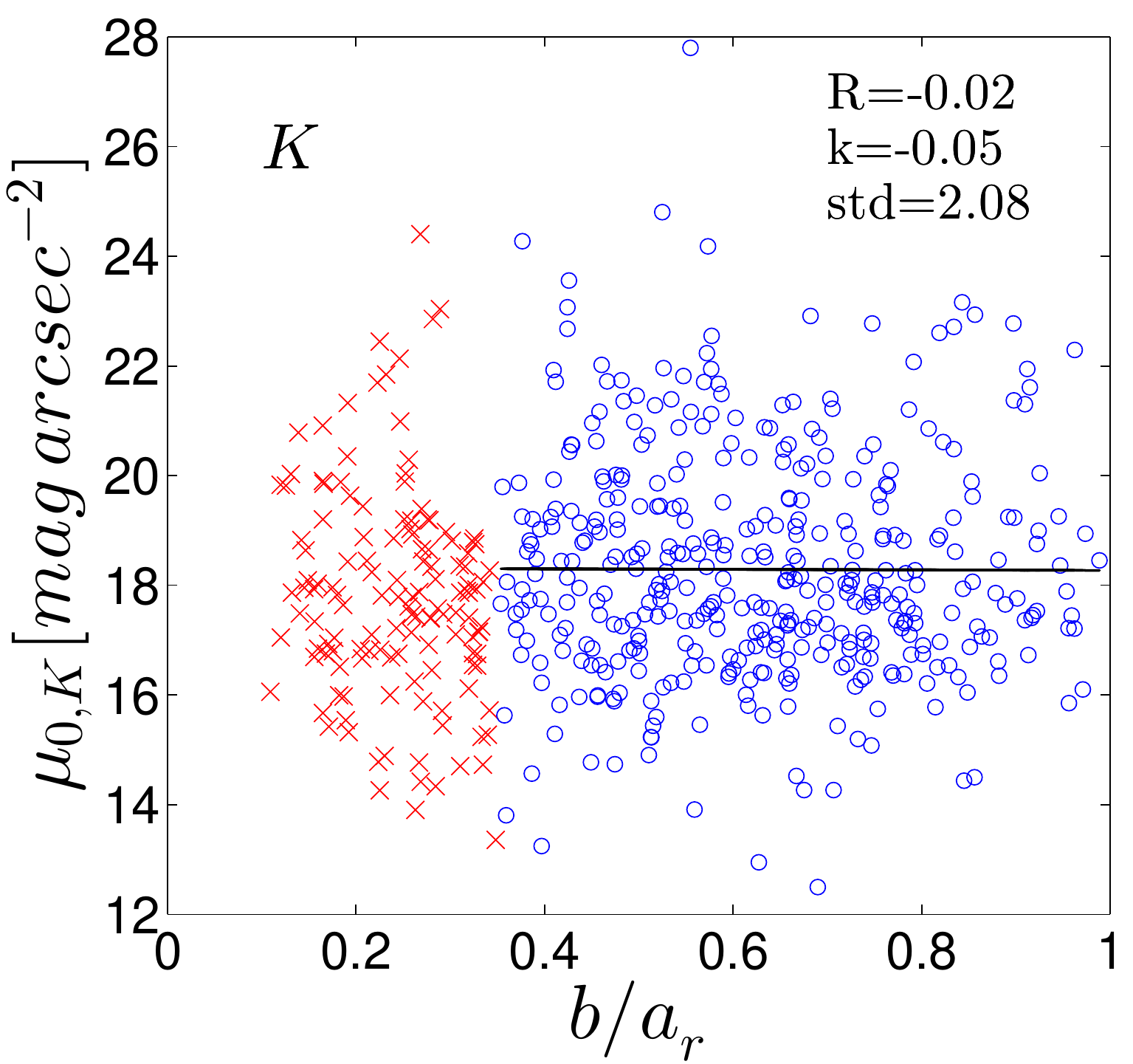}
\end{minipage}
\caption{The relations between axial ratio in r-band ($b/a_r$) and $\mu_0$ in multi-bands ($g, r, i, z, Y, J, H, K$ bands). 
Red crosses represent highly inclined galaxies with $b/a \leq 0.35$ and blue circles represent less inclined galaxies with $b/a > 0.35$. Black lines represent the fitted lines of less inclined galaxies ($b/a_r > 0.35$). In each panel, R means the correlation coefficient, k means the slope of fitted lines, and std means standard deviation, which are the same as Fig.~\ref{fig.u0-scalelength_kpc} and Fig.~\ref{fig.plot_u0_fracdev_lt_ht}.}
\label{fig.plot_u0_ba_lt_ht}
\end{figure*}

\begin{figure}
\centering
\includegraphics[angle=0,width=7.8cm]{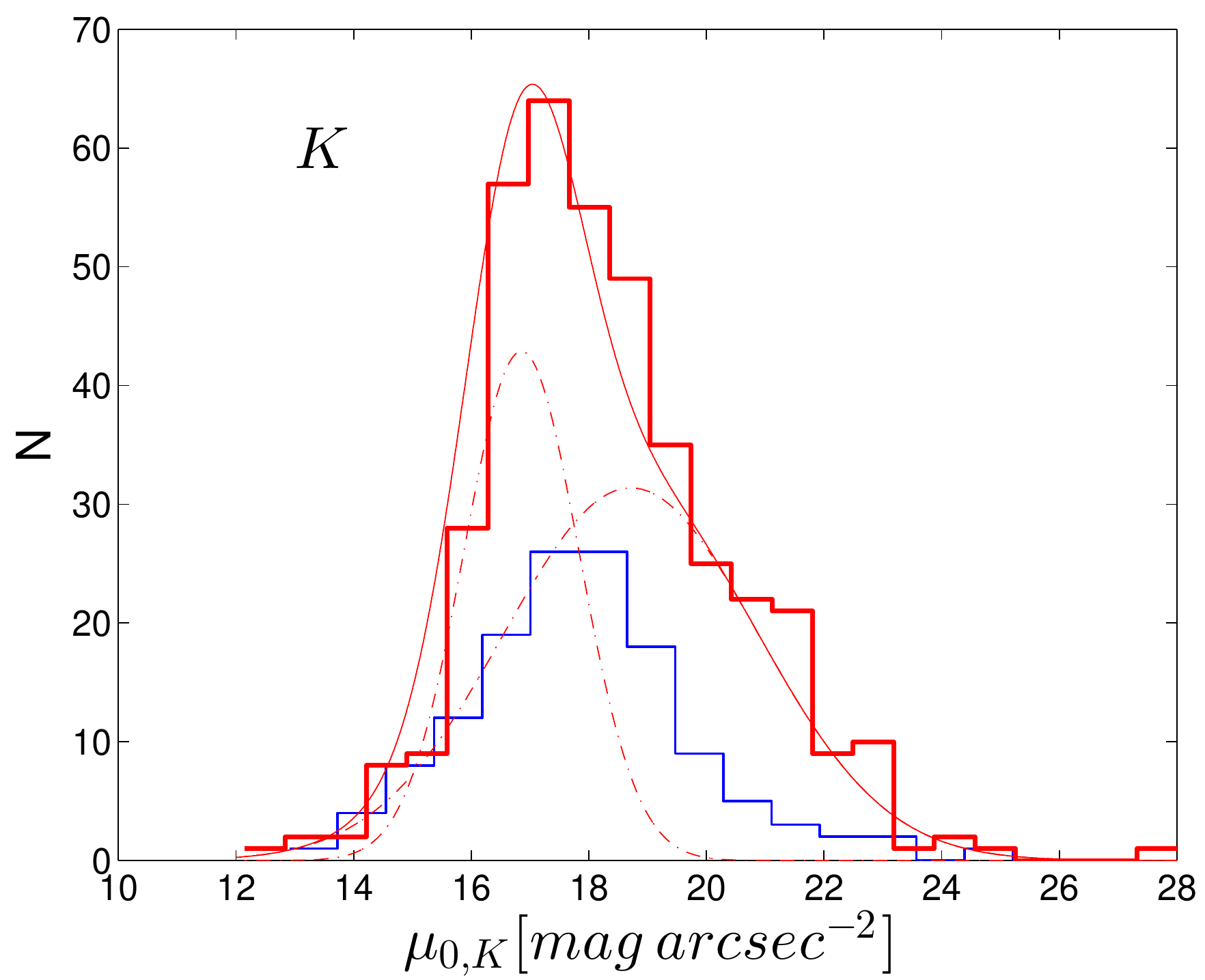}
\caption{Histograms of the $\mu_0$ distribution with the optimal bin size in $K$ band for galaxies with $b/a \leq 0.35$ (shown with blue thin lines) and $b/a > 0.35$ (shown with red thick lines), respectively. The red dot-dashed lines represent two separate Gaussian components for galaxies with $b/a > 0.35$.}
\label{fig.u0-ba}
\end{figure}

The relations between ${b/a}_r$ and $\mu_0$ in multi-bands are shown in Fig.~\ref{fig.plot_u0_ba_lt_ht} and distribution of $\mu_0$ for both parts in $K$ band is shown in Fig.~\ref{fig.u0-ba}. In every panel of Fig.~\ref{fig.plot_u0_ba_lt_ht}, $R$ means the correlation coefficient, $k$ means the slope of fitted lines, and $std$ means standard deviation, which are the same as Fig.~\ref{fig.u0-scalelength_kpc} and Fig.~\ref{fig.plot_u0_fracdev_lt_ht}. From Fig.~\ref{fig.plot_u0_ba_lt_ht}, the correlation coefficients are lower than 0.2, so there is no apparent correlation between $\mu_0$ and axis ratio. 

Fig.~\ref{fig.u0-ba} is a histogram of the optimal fitting for $\mu_0$ with the optimal bin size in $K$ band 
for $b/a \leq 0.35$ (blue lines) and $b/a > 0.35$ (red lines), respectively. 
Given to the incorrect inclination correction for edge-on galaxies, we just ignore the galaxies with $b/a \leq 0.35$ (edge-on galaxies).
According to Table~\ref{table.ba}, the evidence of double Gaussian being better than a single Gaussian for fitting subsample galaxies with $b/a > 0.35$ (low inclinations) is still very strong for the $\mu_0$ distribution in $K$ band.

Therefore, it is not the inclination that lead to the result that double Gaussian being better than a single Gaussian fitting of $\mu_0$ distribution in $K$ band.
  
\begin{table}
\centering
\caption{The AIC and BIC values of our subsample with $b/a > 0.35$ fitting with single and double Gaussian profiles in $K$ band.}
\label{table.ba}
\begin{tabular}{|cc|}
\hline
Band & $K$ \\
\hline
$AICc_{single}$ &84.3\\
$AICc_{double}$ &54.1\\
$BIC_{single}$ &86.4\\
$BIC_{double}$ &55.1\\
$\bigtriangleup AICc$ &30.2\\
$\bigtriangleup BIC$ &31.3\\
\hline
\hline
\end{tabular}
\end{table}

\subsection{Disk scalelength}

\begin{figure*}
\begin{minipage}{\textwidth}
\includegraphics[width=.24\textwidth]{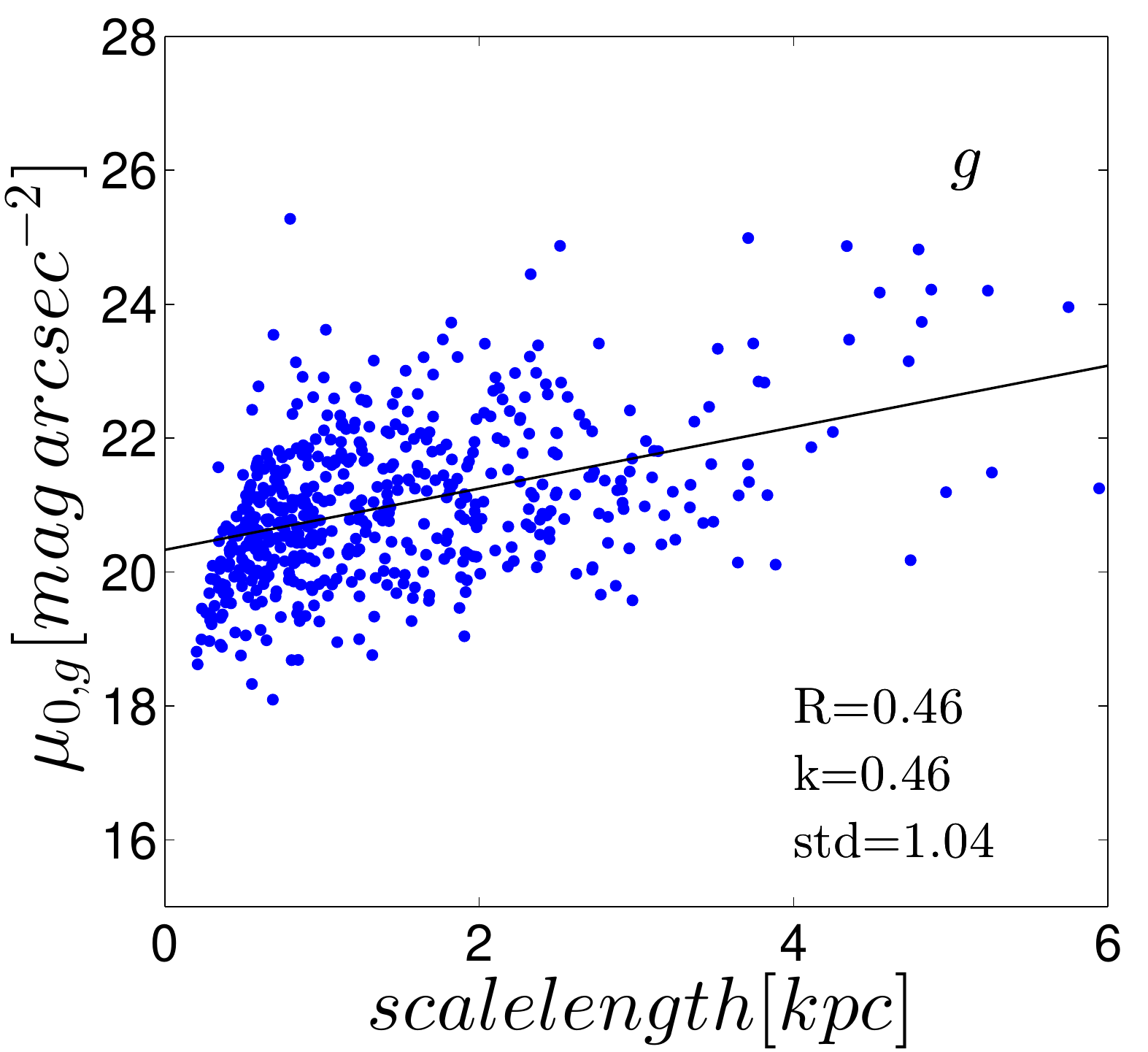}
\includegraphics[width=.24\textwidth]{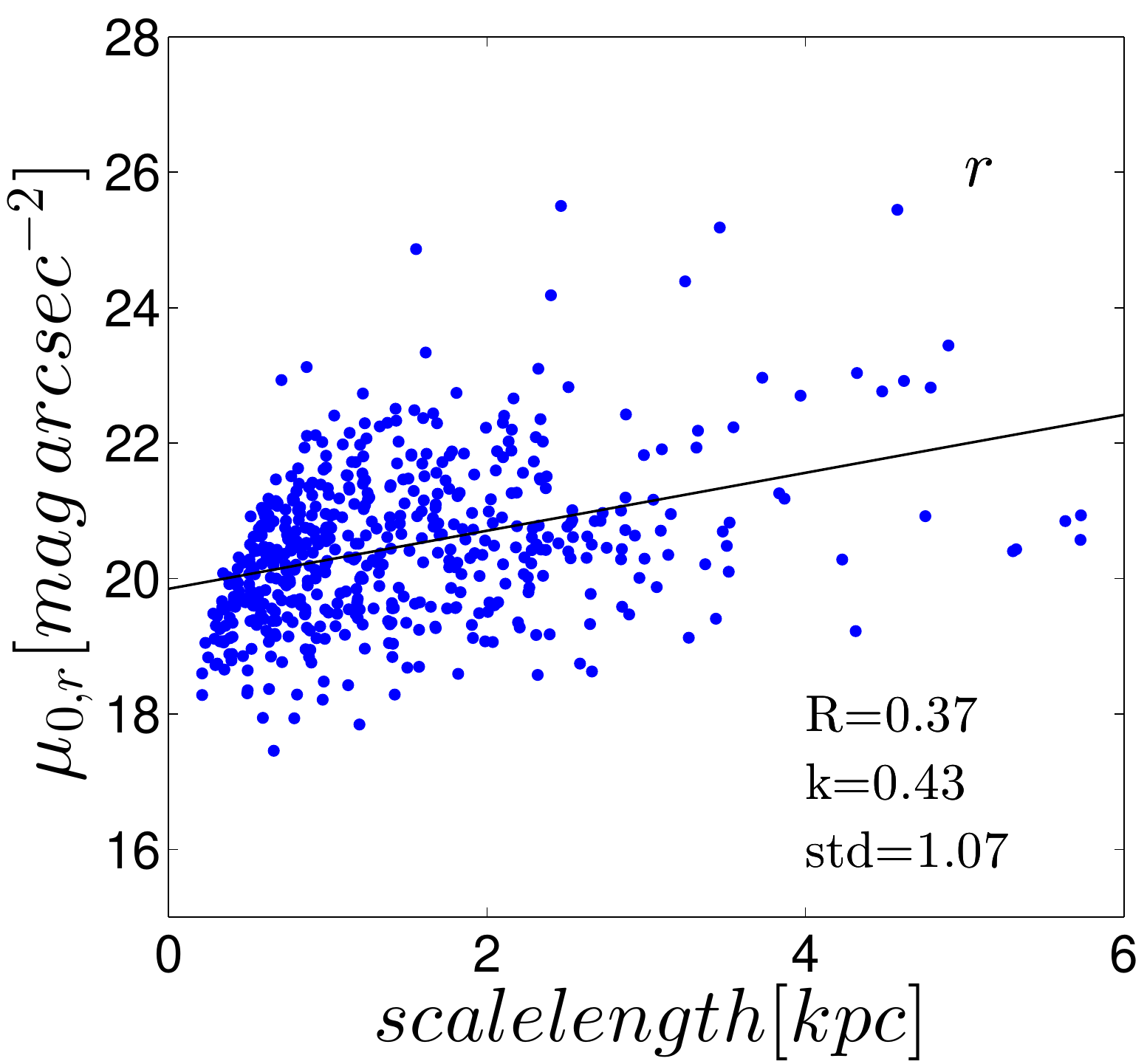}
\includegraphics[width=.24\textwidth]{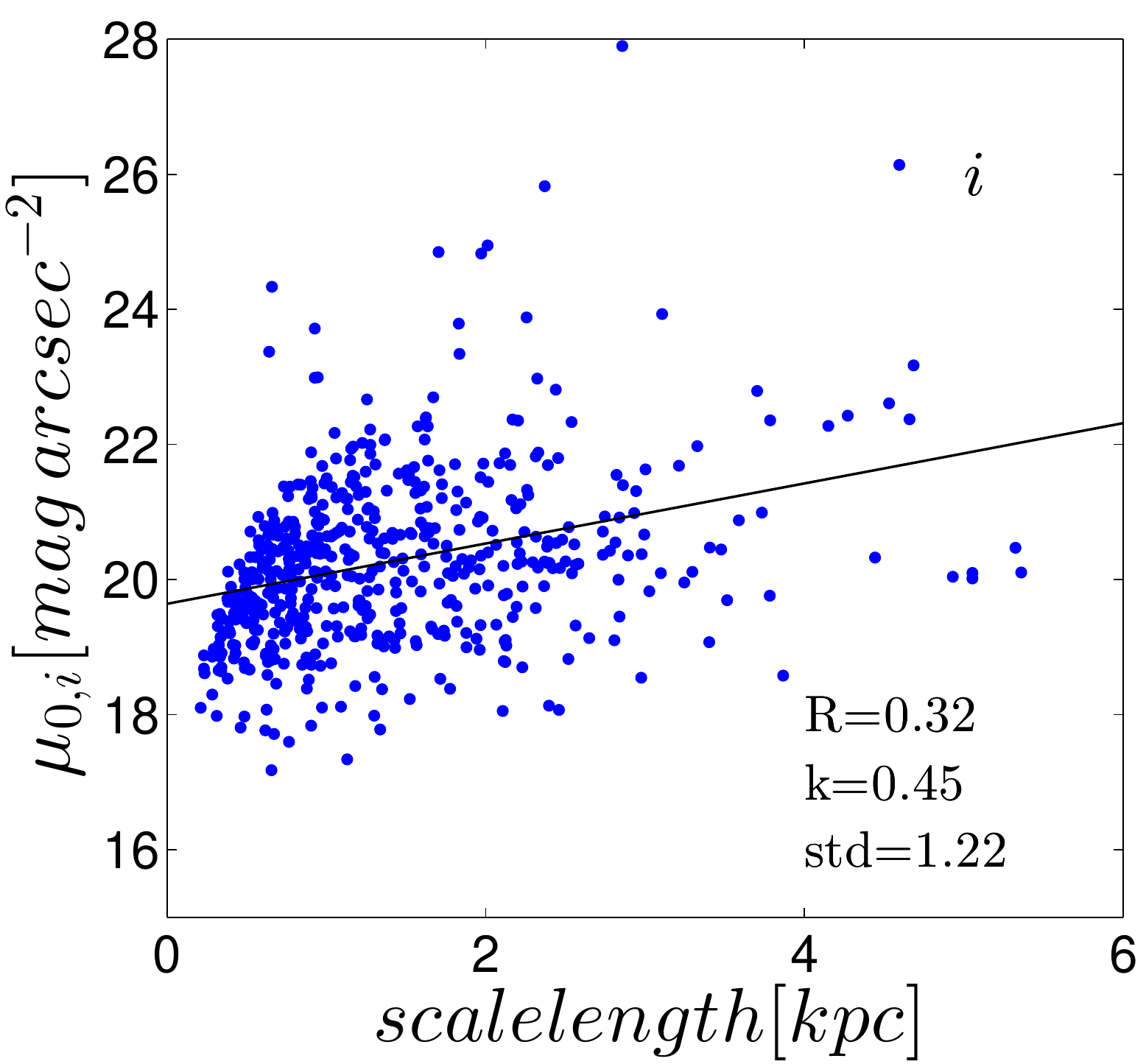}
\includegraphics[width=.24\textwidth]{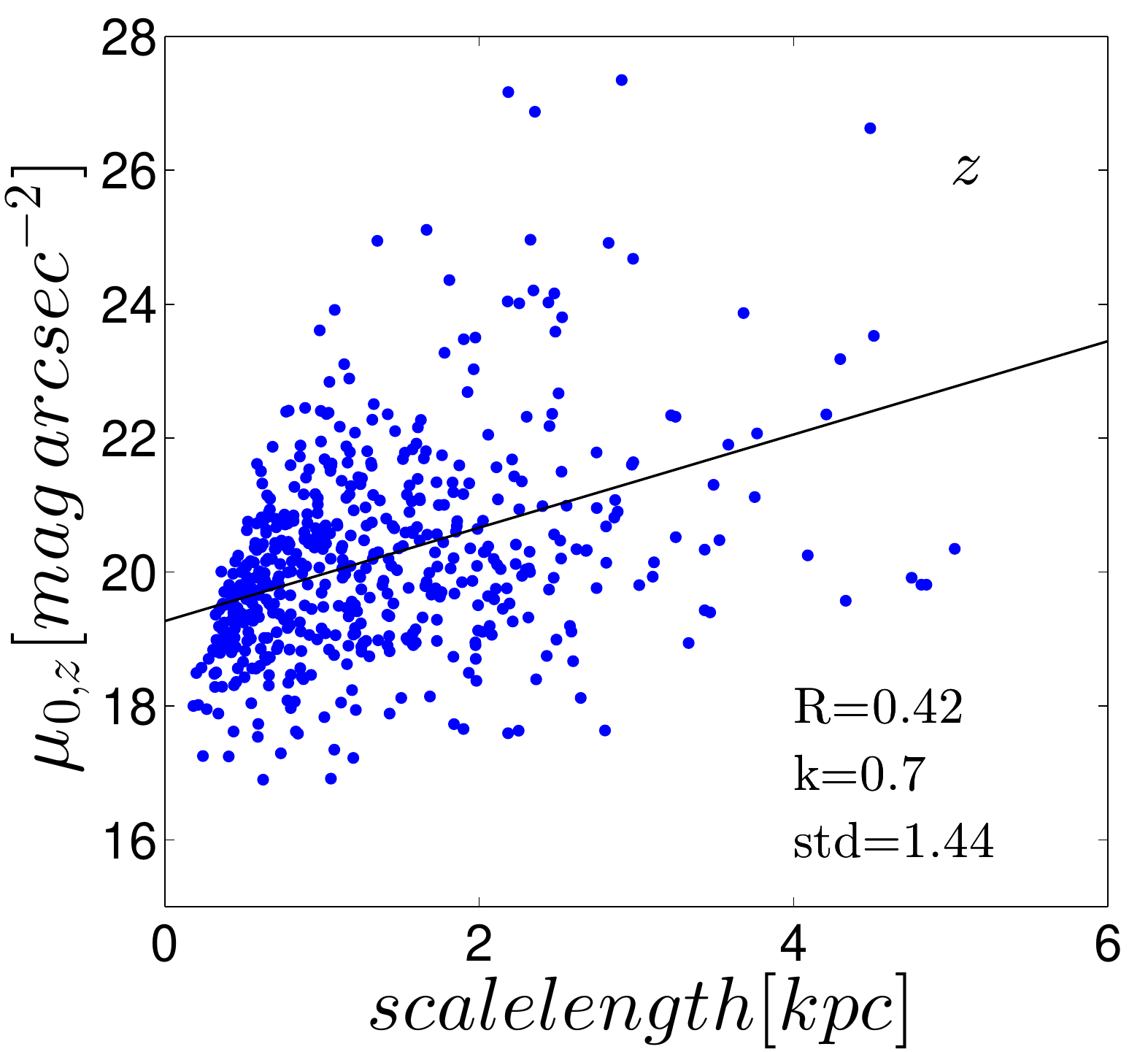}
\end{minipage}
\begin{minipage}{\textwidth}
\includegraphics[width=.24\textwidth]{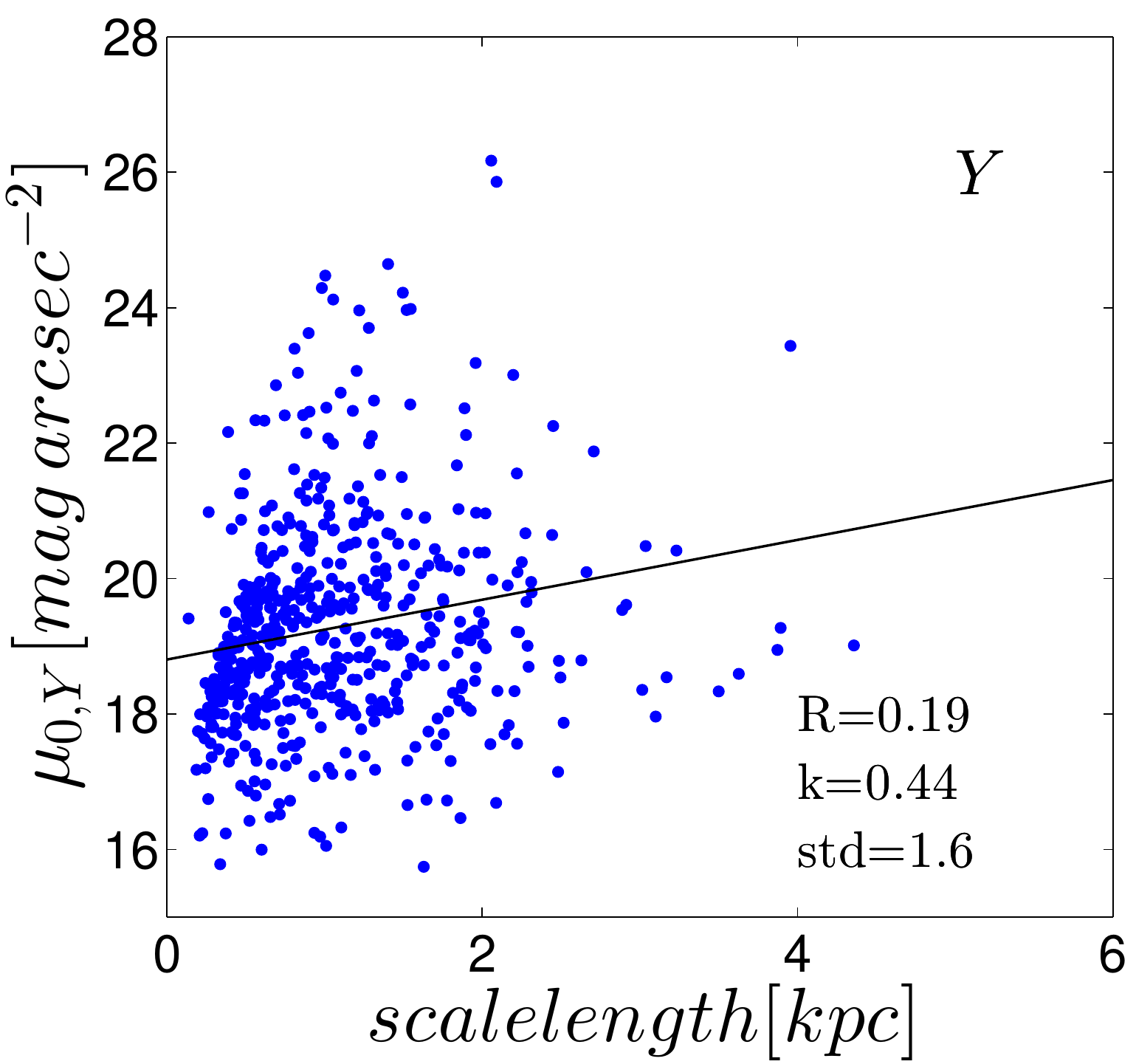}
\includegraphics[width=.24\textwidth]{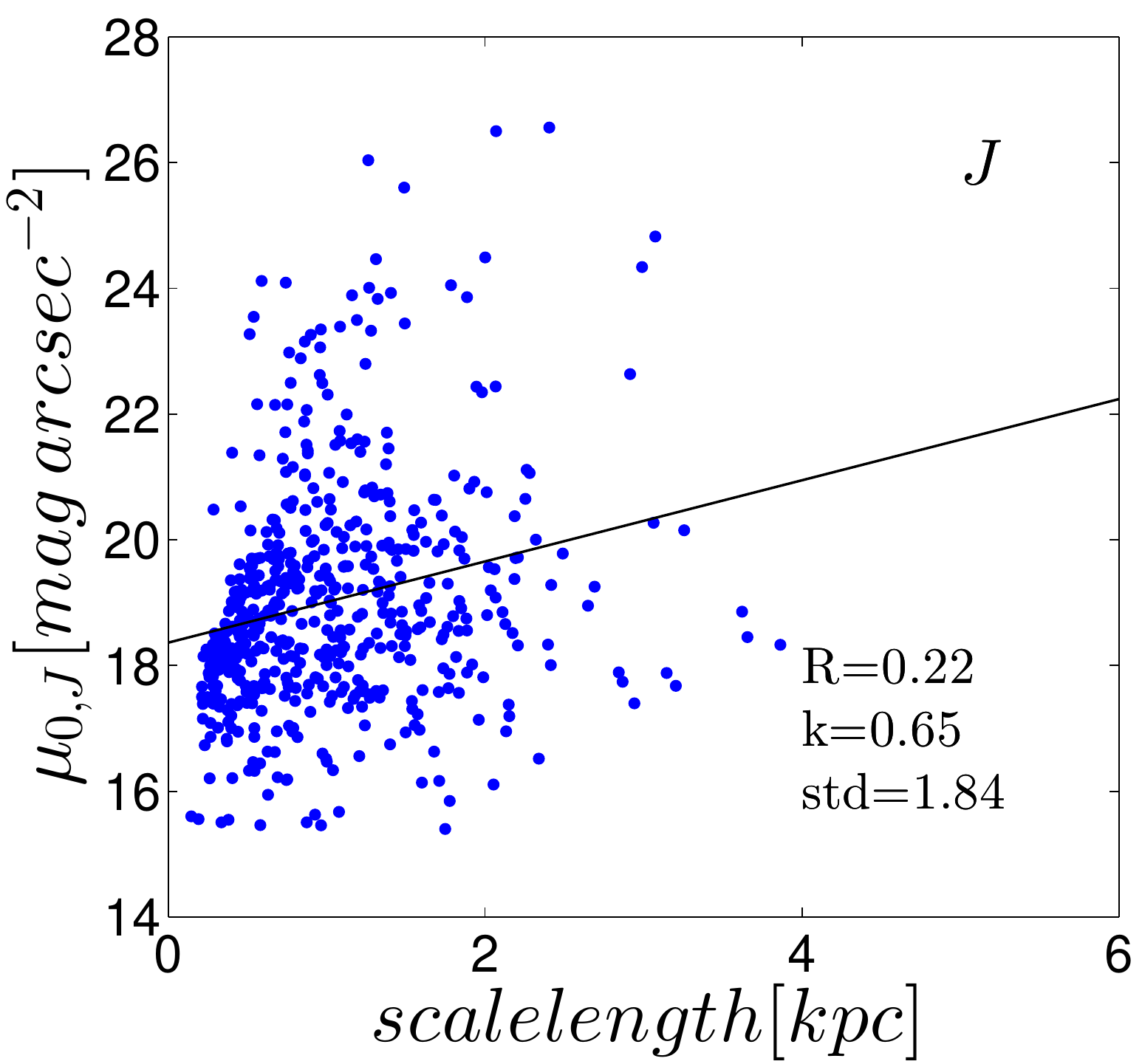}
\includegraphics[width=.24\textwidth]{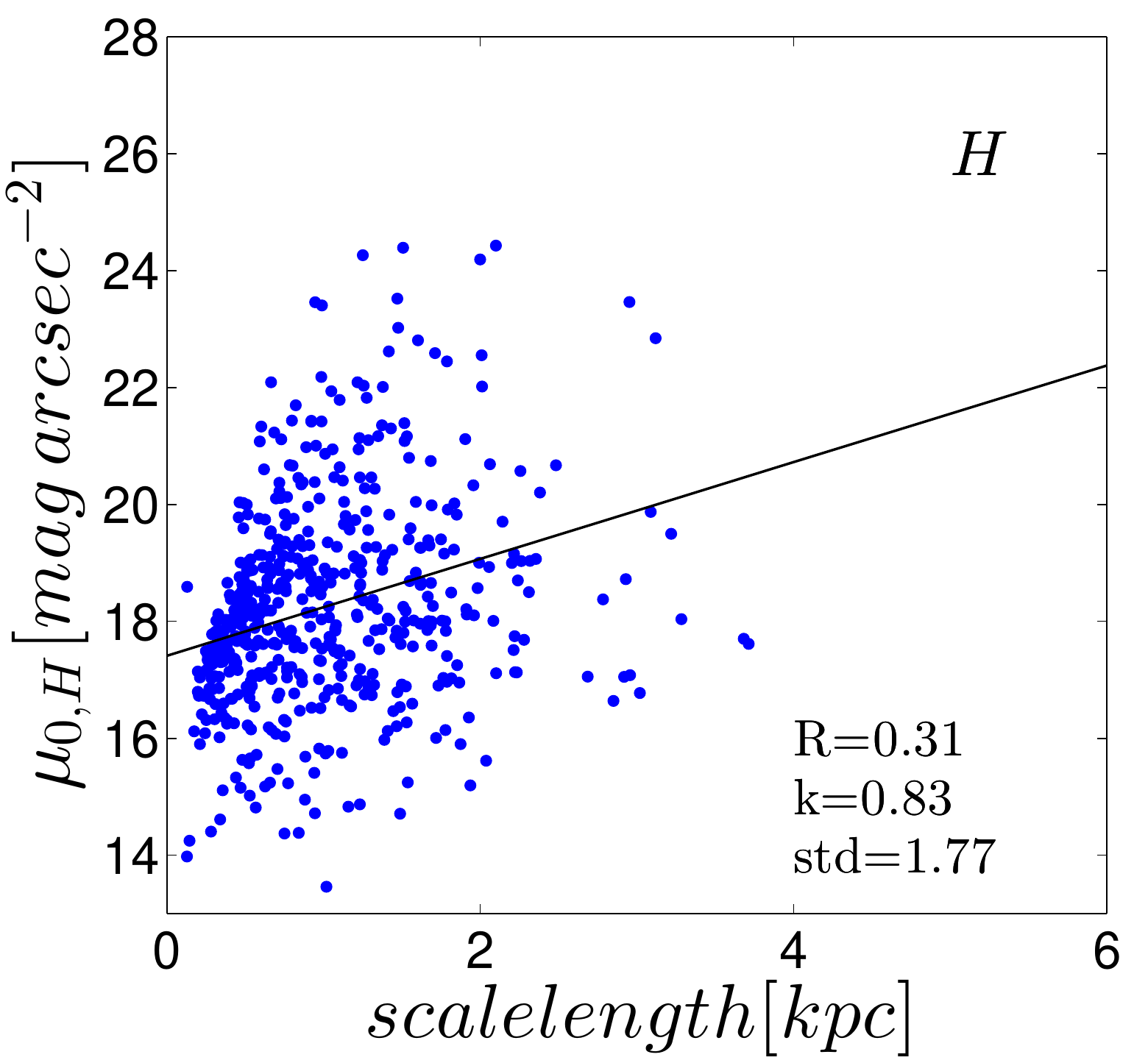}
\includegraphics[width=.24\textwidth]{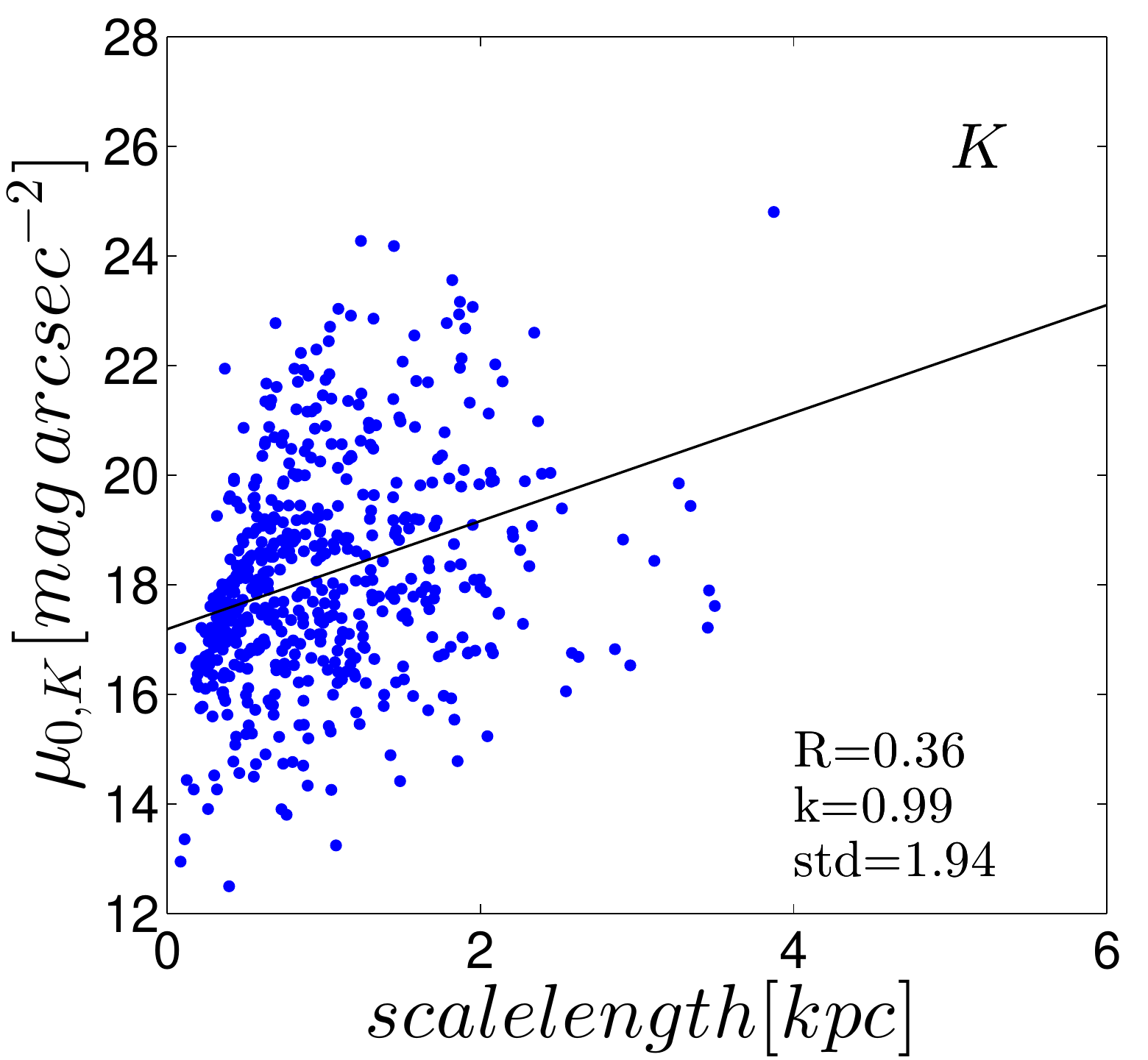}
\end{minipage}
\caption{The relations between $\mu_0$ and disk scalelength (in units of kpc)
in multi-bands ($g, r, i, z, Y, J, H, K$ bands).}
\label{fig.u0-scalelength_kpc}
\end{figure*}

The disk scalelength is derived from Galfit fitting procedure result.
The correlation coefficients in each panel of Fig.~\ref{fig.u0-scalelength_kpc} range from 0.2 to 0.5, so there is a weak correlation between $\mu_0$ and disk scalelength in units of kpc. 
It is clear that LSB galaxies are more extend than HSB galaxies. 
The $\mu_0$ is larger for galaxies with larger disk scalelength and the galaxies are fainter, 
and the $\mu_0$ is smaller for galaxies with smaller disk scalelength and the galaxies are brighter.

\subsection{Morphology}
\begin{figure*}
\begin{minipage}{\textwidth}
\includegraphics[width=.24\textwidth]{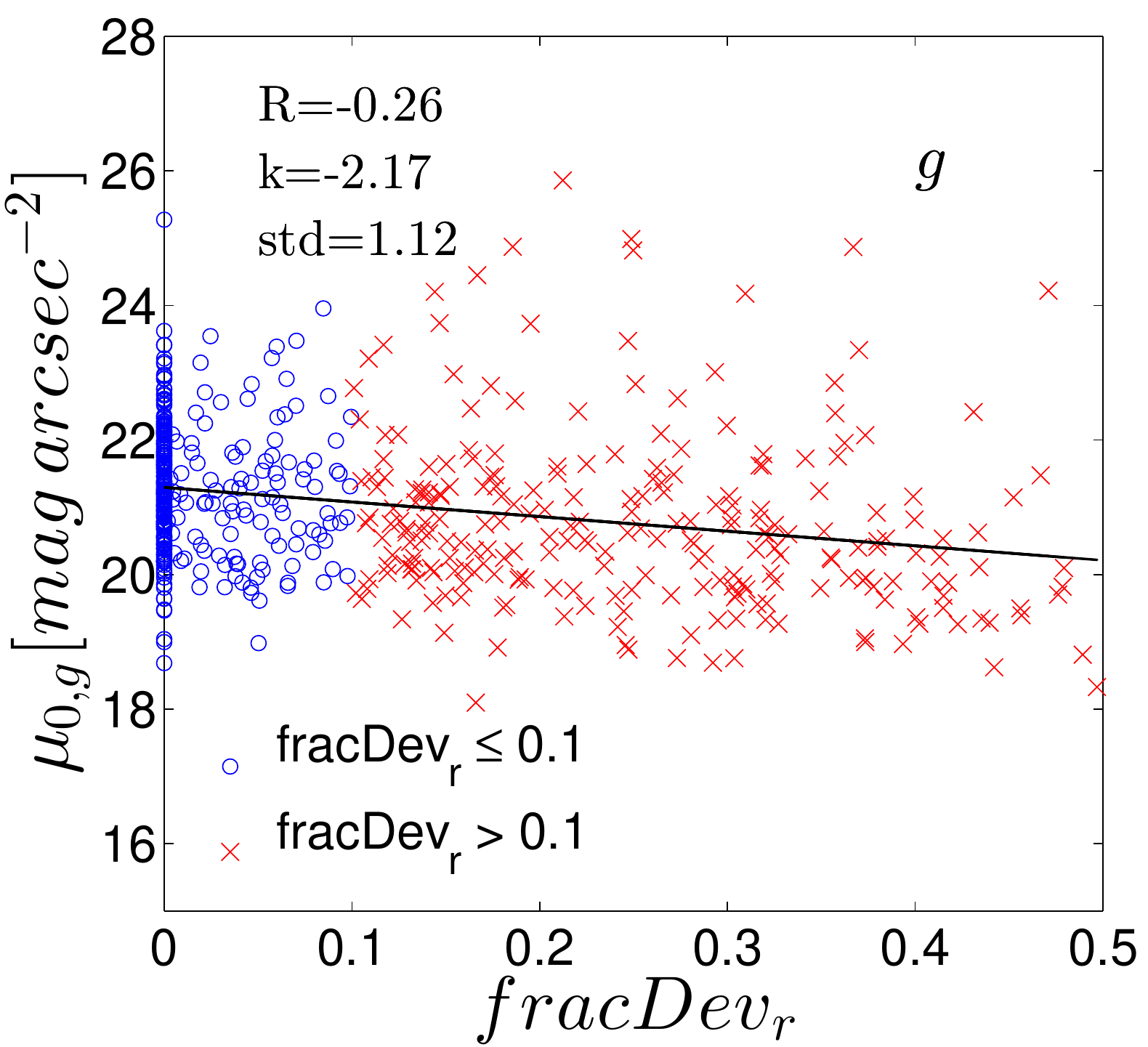}
\includegraphics[width=.24\textwidth]{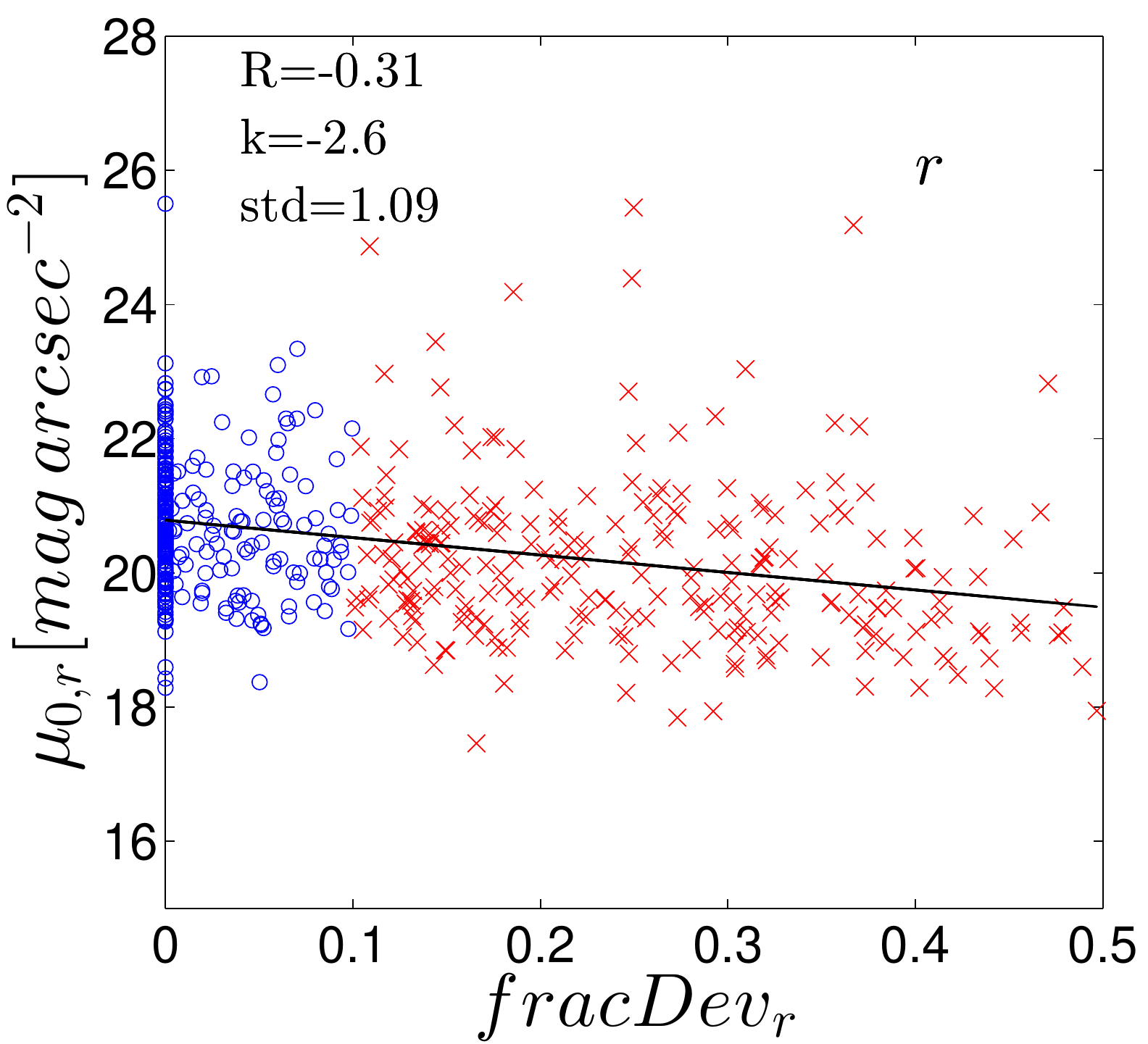}
\includegraphics[width=.24\textwidth]{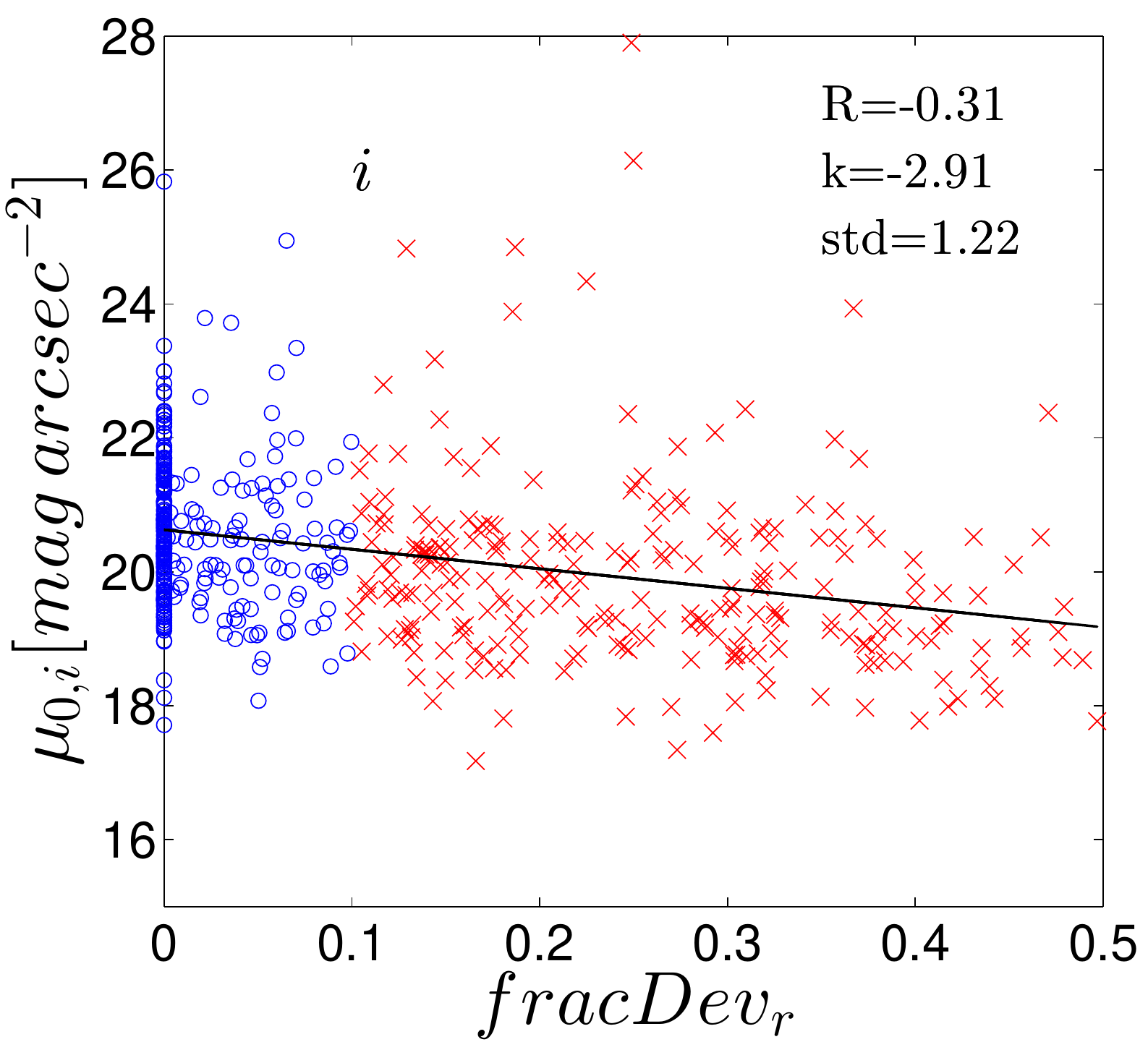}
\includegraphics[width=.24\textwidth]{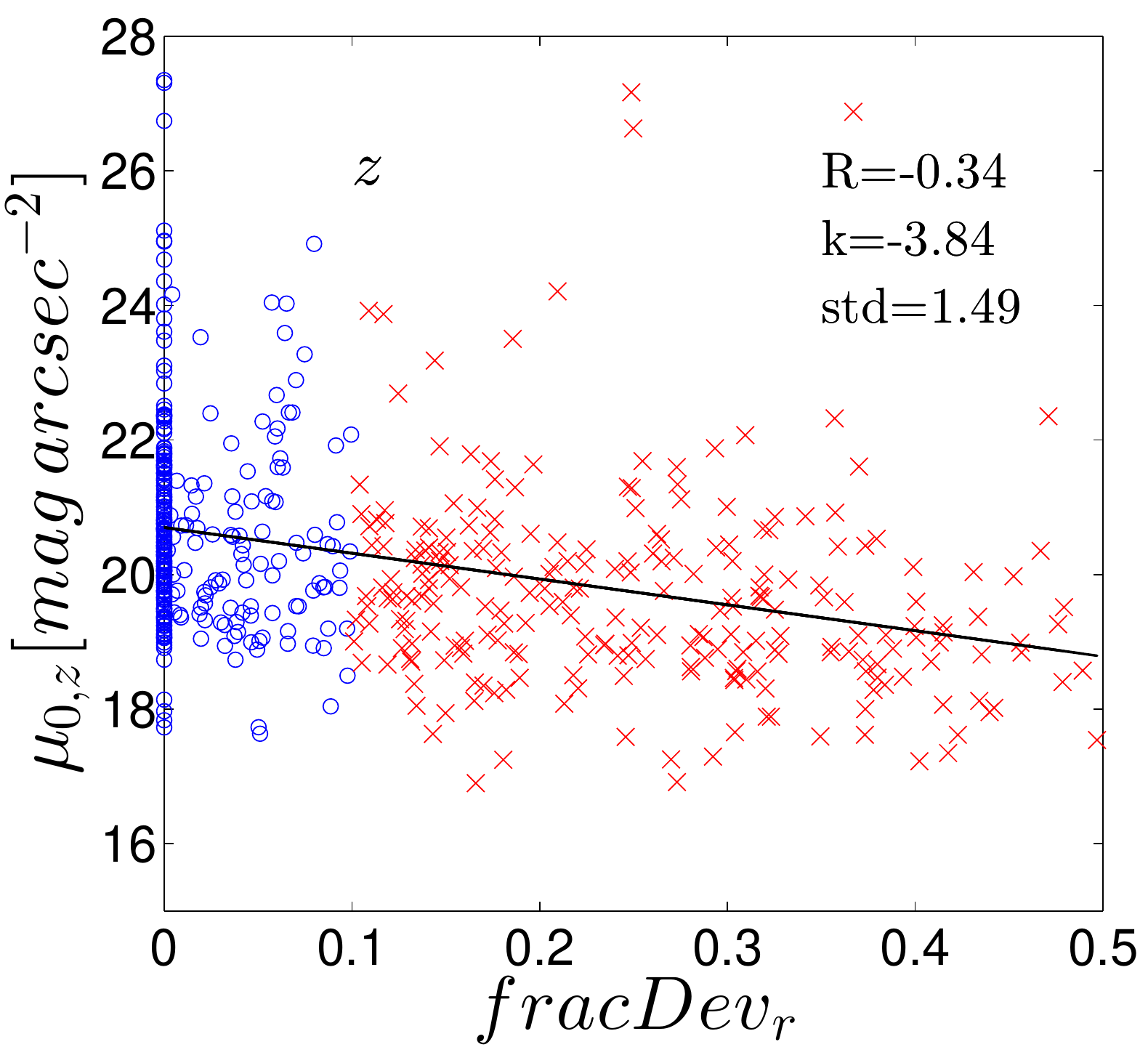}
\end{minipage}
\begin{minipage}{\textwidth}
\includegraphics[width=.24\textwidth]{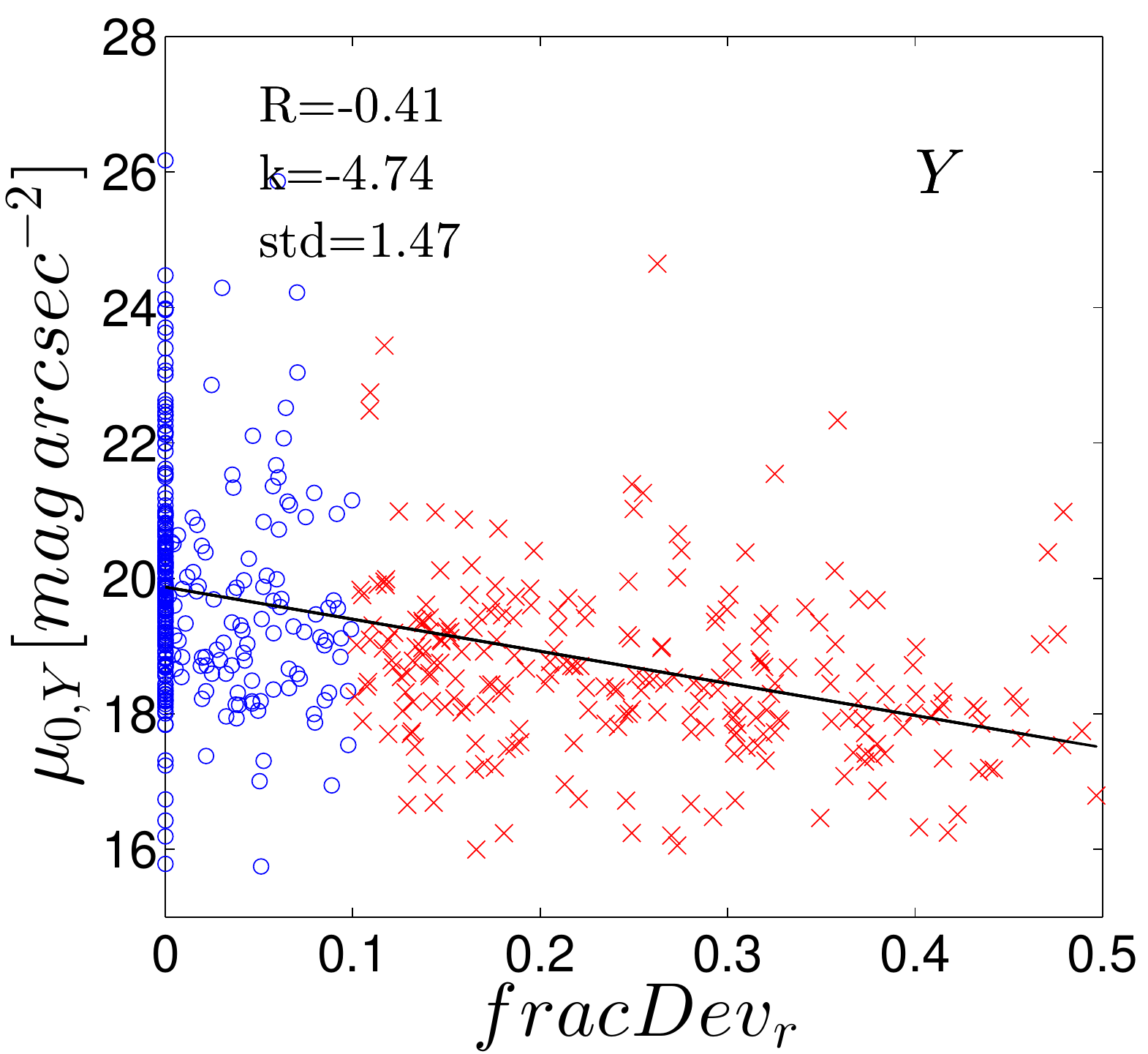}
\includegraphics[width=.24\textwidth]{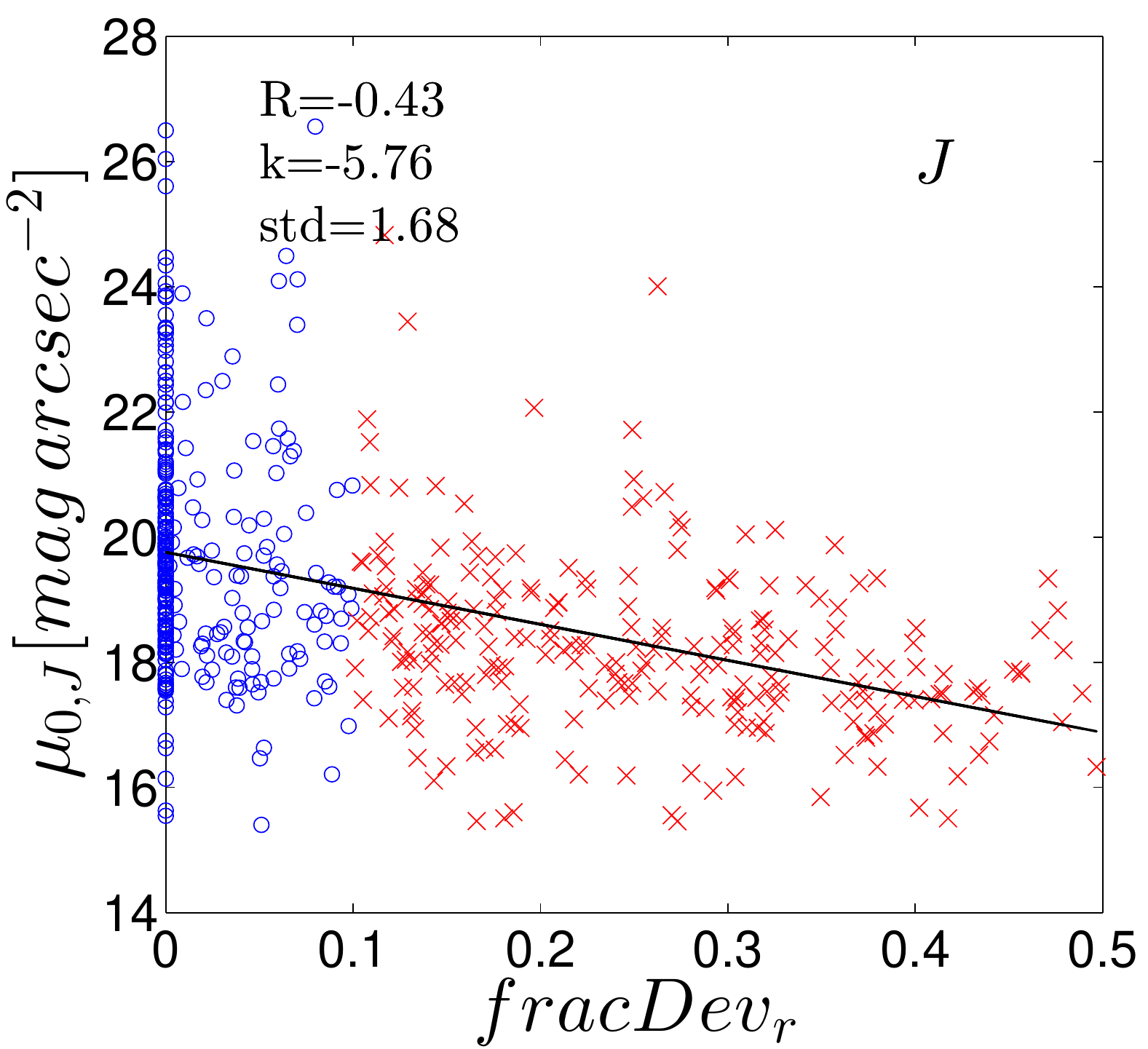}
\includegraphics[width=.24\textwidth]{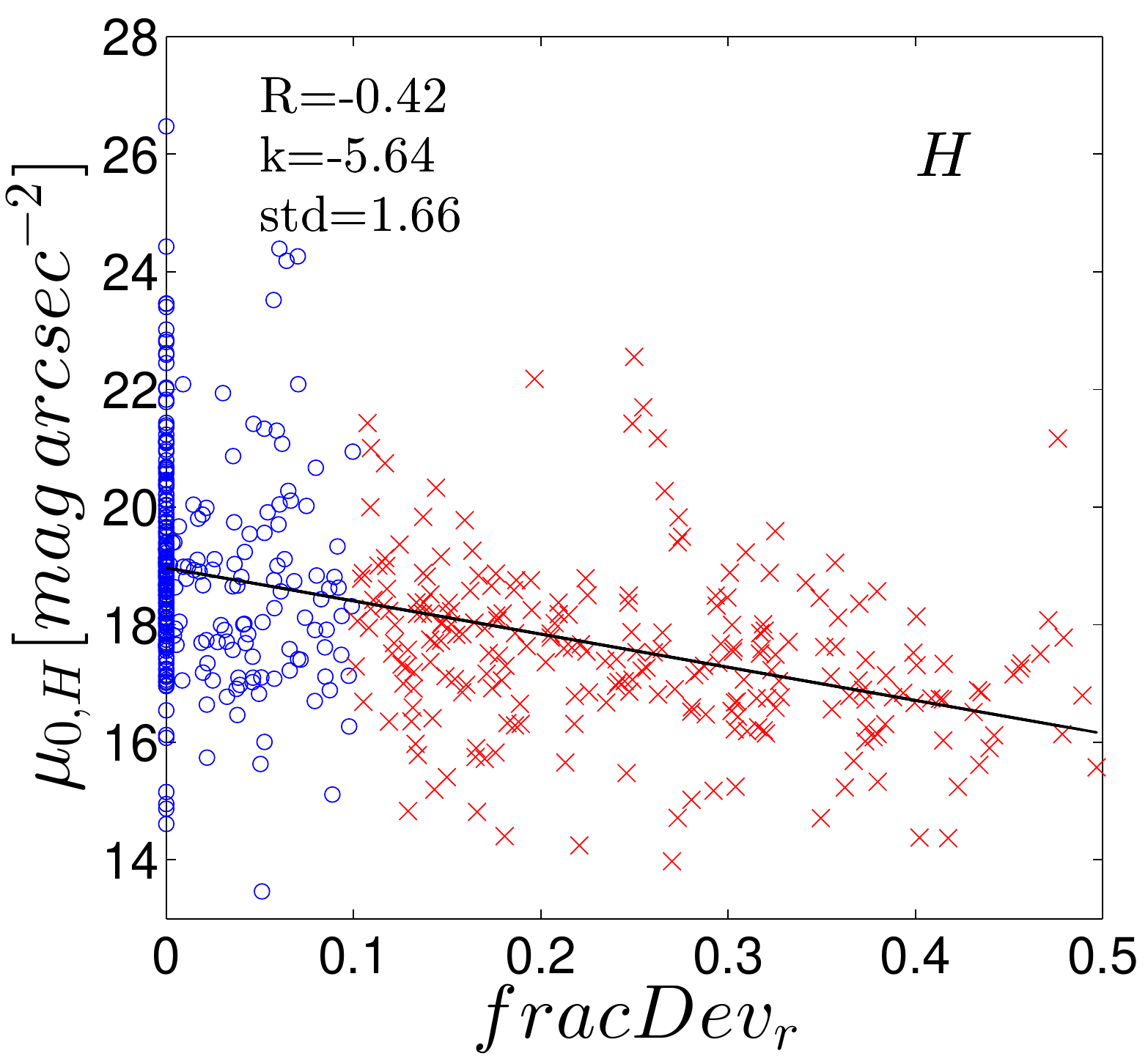}
\includegraphics[width=.24\textwidth]{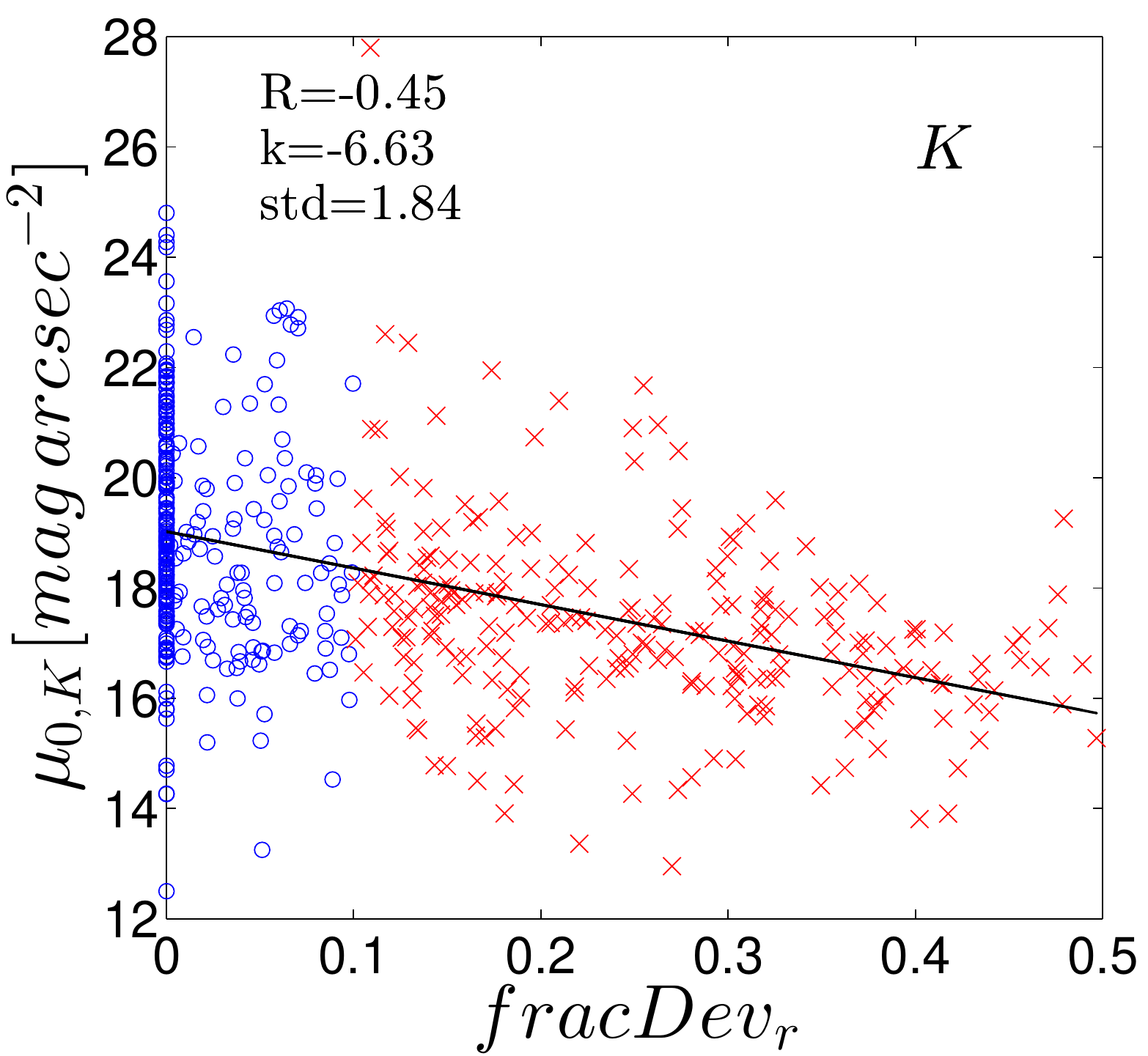}
\end{minipage}
\caption{Distributions of $\mu_0$ as a function of $fracDev_r$ for our sample galaxies.
Blue circles represent disc galaxies with $fracDev_r \leq 0.1$ and red crosses represent disc galaxies with $0.5 \geq fracDev_r > 0.1$.}
\label{fig.plot_u0_fracdev_lt_ht}
\end{figure*}

Decomposition is used to estimate the ratio of bulge light and disk light 
of galaxies. The definition of $fracDev$ \footnote{http://classic.sdss.org/dr7/algorithms/photometry.html} is as follows:  
\begin{equation}
      F_{composite} $=$ fracDev F_{dev} $+$ (1 - fracDev) F_{exp},
\end{equation} 
In this equation, $F_{composite}$, $F_{dev}$ and $F_{exp}$ are the composite, de Vaucouleurs and exponential flux of the object, 
$fracDev$ is the weight of de Vaucouleurs component in best composite model.
The value of $fracDev$ could be obtained from SDSS database. When a galaxy is in the case of $fracDev_r$ $=$ 0, 
it corresponds to pure disk galaxy without a bulge. Here we classify our sample into two subsamples. 
Subsample 1 is composed of galaxies with $fracDev_r \leq 0.1$ and subsample 2 is composed of galaxies with $ 0.5 \geq fracDev_r > 0.1$.
As \citealp{2009MNRAS.393..628M} and \citealp{2009MNRAS.394.2022M} expected, early-type disc galaxies dominate the HSB peak, while late-type disc galaxies
and irregulars are present in both HSB and LSB peaks.
The correlation coefficients in each panel of Fig.~\ref{fig.plot_u0_fracdev_lt_ht} range from -0.5 to -0.2, so there is a weak negative correlation between $\mu_0$ and $fracDev_r$.  
It shows that with the increasing of $fracDev_r$, the $\mu_0$ appear to decrease, 
which means with the increasing of the fraction of bulge light, the galaxies tend to be brighter, 
which is consistent with \citealp{2009MNRAS.393..628M} and \citealp{2009MNRAS.394.2022M}.

\begin{figure}
\centering
\includegraphics[angle=0,width=7.8cm]{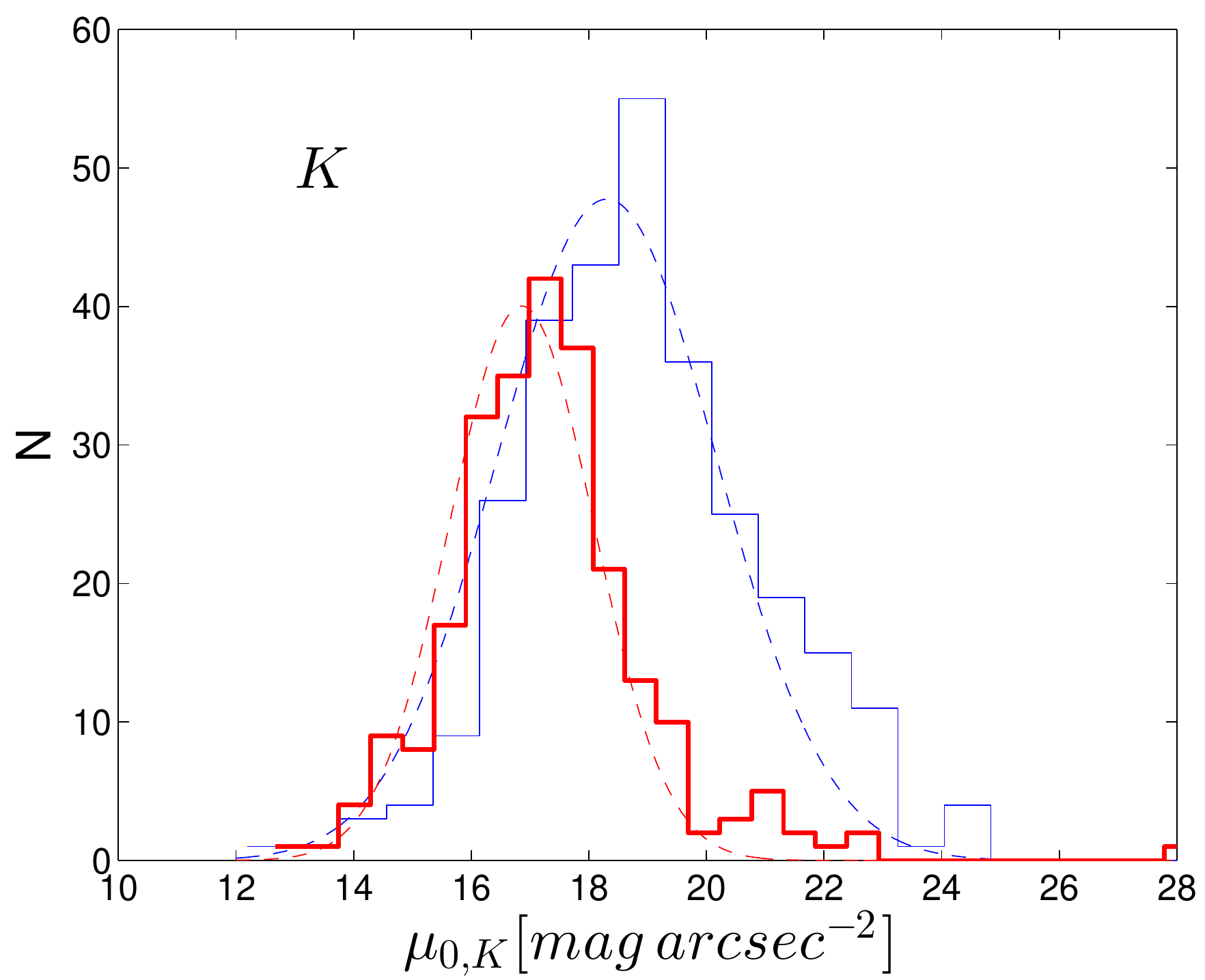}
\caption{Histograms of the $\mu_0$ distribution with the optimal bin size in $K$ band for galaxies with $fracDev \leq 0.1$ (shown with blue thin lines) and $0.5 \geq fracDev > 0.1$ (shown with red thick lines), respectively.}
\label{fig.u0-fracdev05}
\end{figure}

The optimal fitting of $\mu_0$ distribution for subsample 1 and subsample 2 in $K$ band  
are presented in Fig.~\ref{fig.u0-fracdev05}. 
The location of peaks of $\mu_0$ distribution for subsample 2 (red dashed lines) are smaller than that for subsample 1 (blue dashed lines),
which means that disc galaxies with larger fraction of bulge are dominated by galaxies with higher surface brightness and disc galaxies with smaller fraction of bulge are dominated by galaxies with lower surface brightness.

Therefore, there may be some effect for the morphology (disc galaxies with larger or smaller fraction of bulge) on the fact of double Gaussian being better than a single Gaussian fitting for $\mu_0$ distribution in $K$ band.

\begin{table}
\centering
\caption{The AIC and BIC values of our subsamples with $fracDev_r \leq 0.1$ \& $0.5 \geq fracDev_r > 0.1$ fitting with single and double Gaussian profiles in $K$ band.}
\label{table.fracdev}
\begin{tabular}{|ccc|}
\hline
Value & $fracDev_r \leq$ 0.1& $0.5 \geq fracDev_r >$ 0.1 \\
\hline
\hline
$AICc_{single}$ &51.6 & 47.8\\
$AICc_{double}$ &51.6 & 48.2\\
$BIC_{single}$ &52.3 &  50.8\\
$BIC_{double}$ &48.0 & 51.6 \\
$\bigtriangleup AICc$ & 0.0 & -0.4 \\
$\bigtriangleup BIC$ & 4.3 & -0.8 \\
\hline
\hline

\end{tabular}
\end{table}

\subsection{Bin size}

\begin{figure*}
\begin{minipage}{\textwidth}
\includegraphics[width=.24\textwidth]{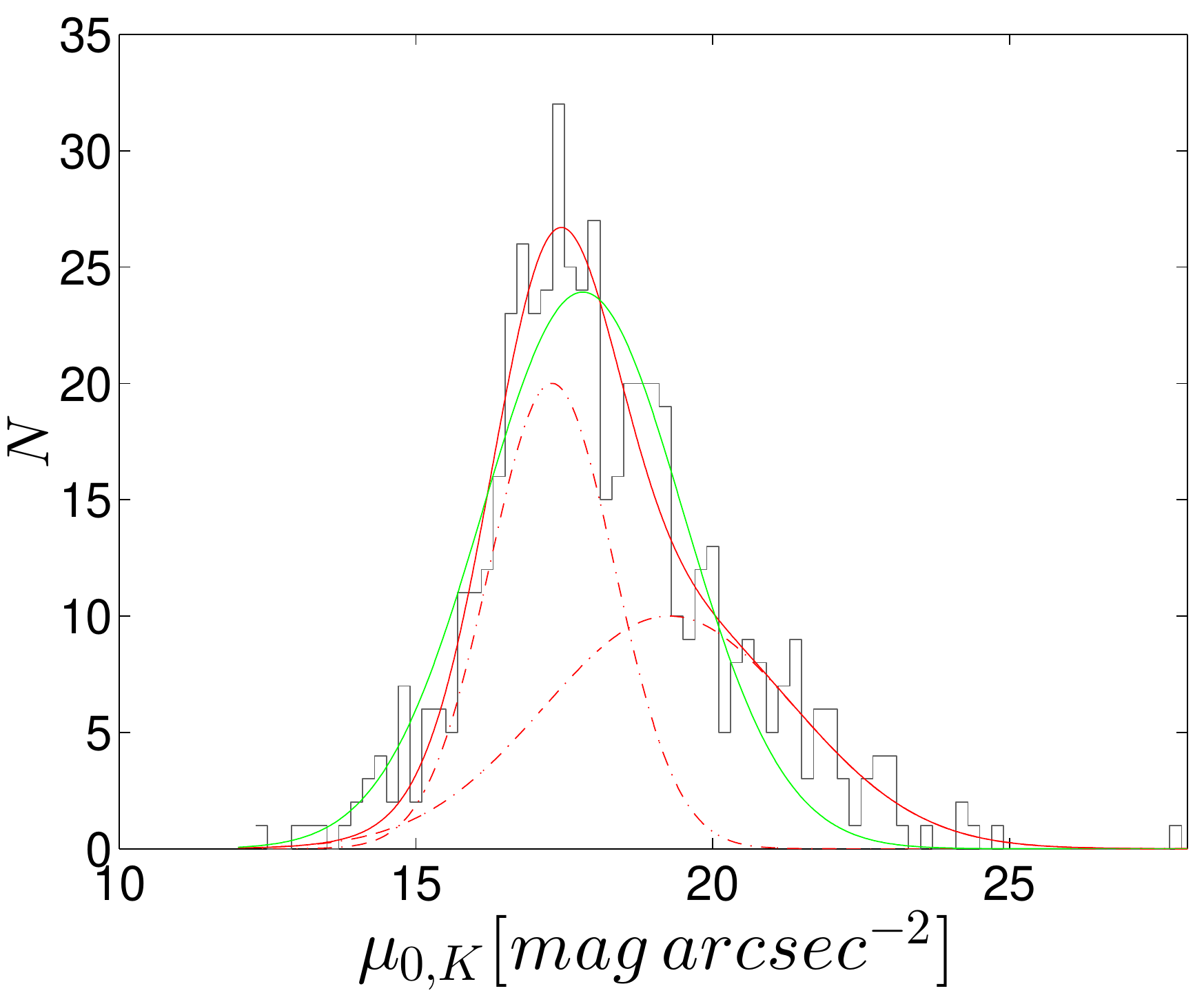}
\includegraphics[width=.24\textwidth]{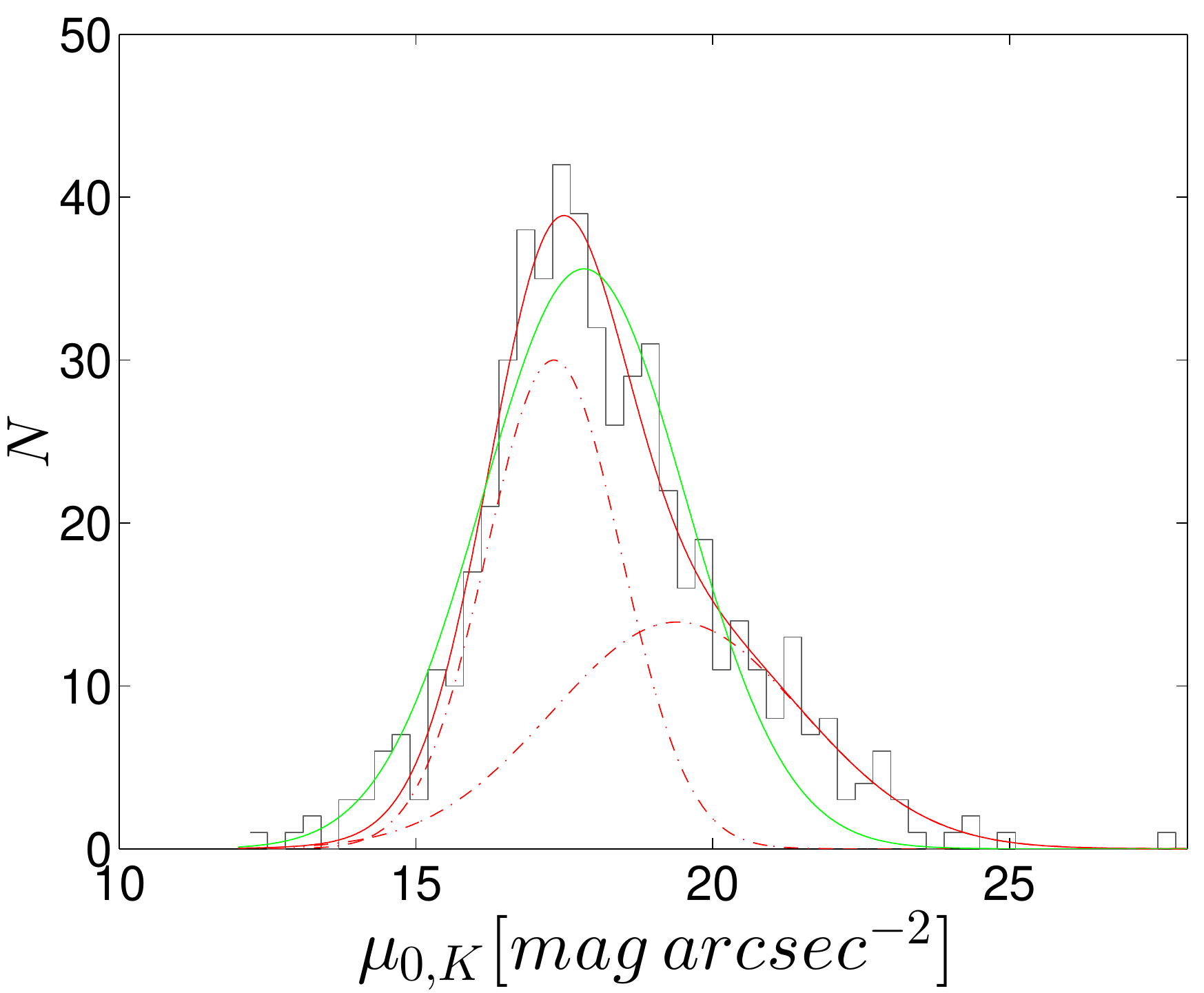}
\includegraphics[width=.24\textwidth]{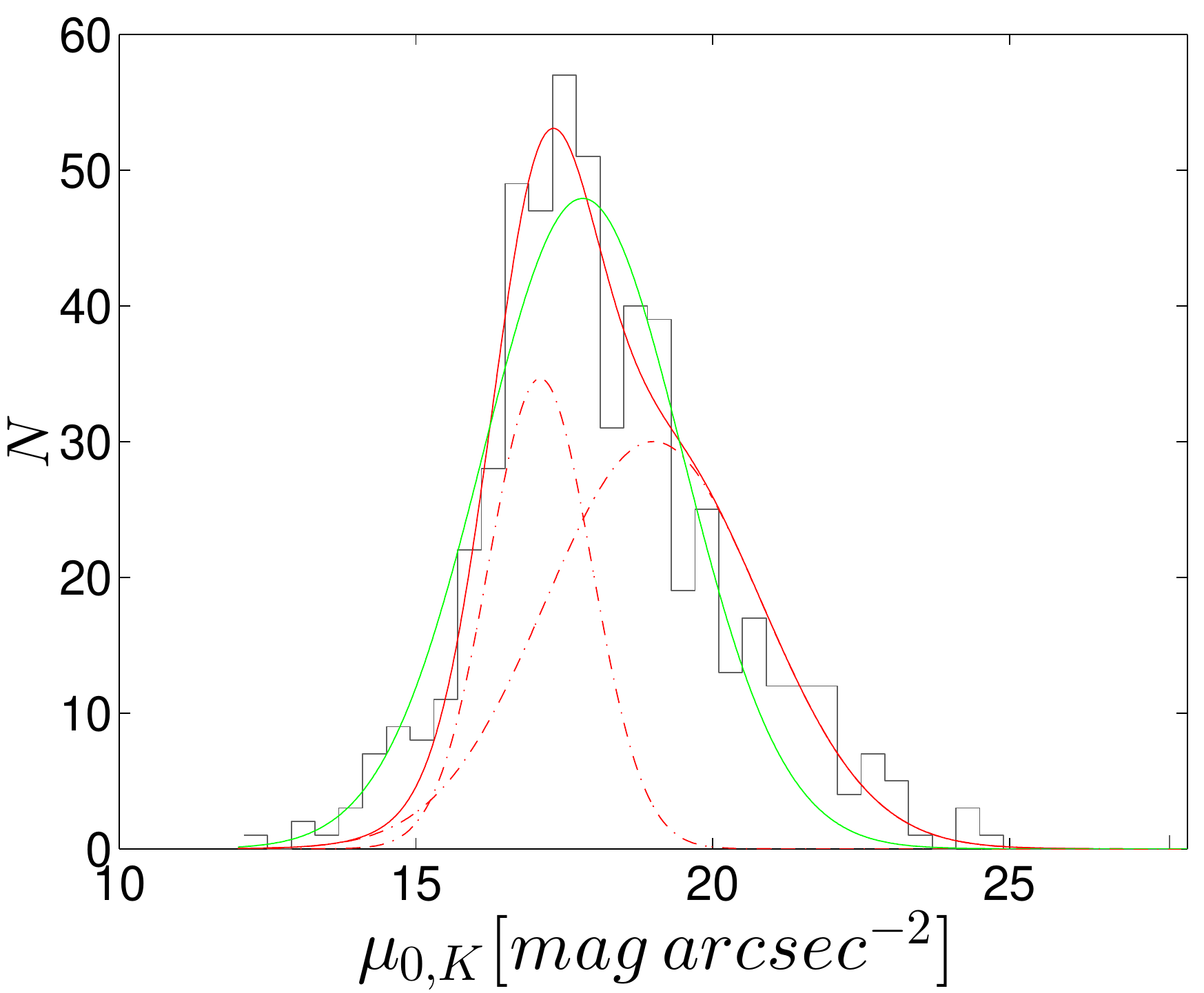}
\includegraphics[width=.24\textwidth]{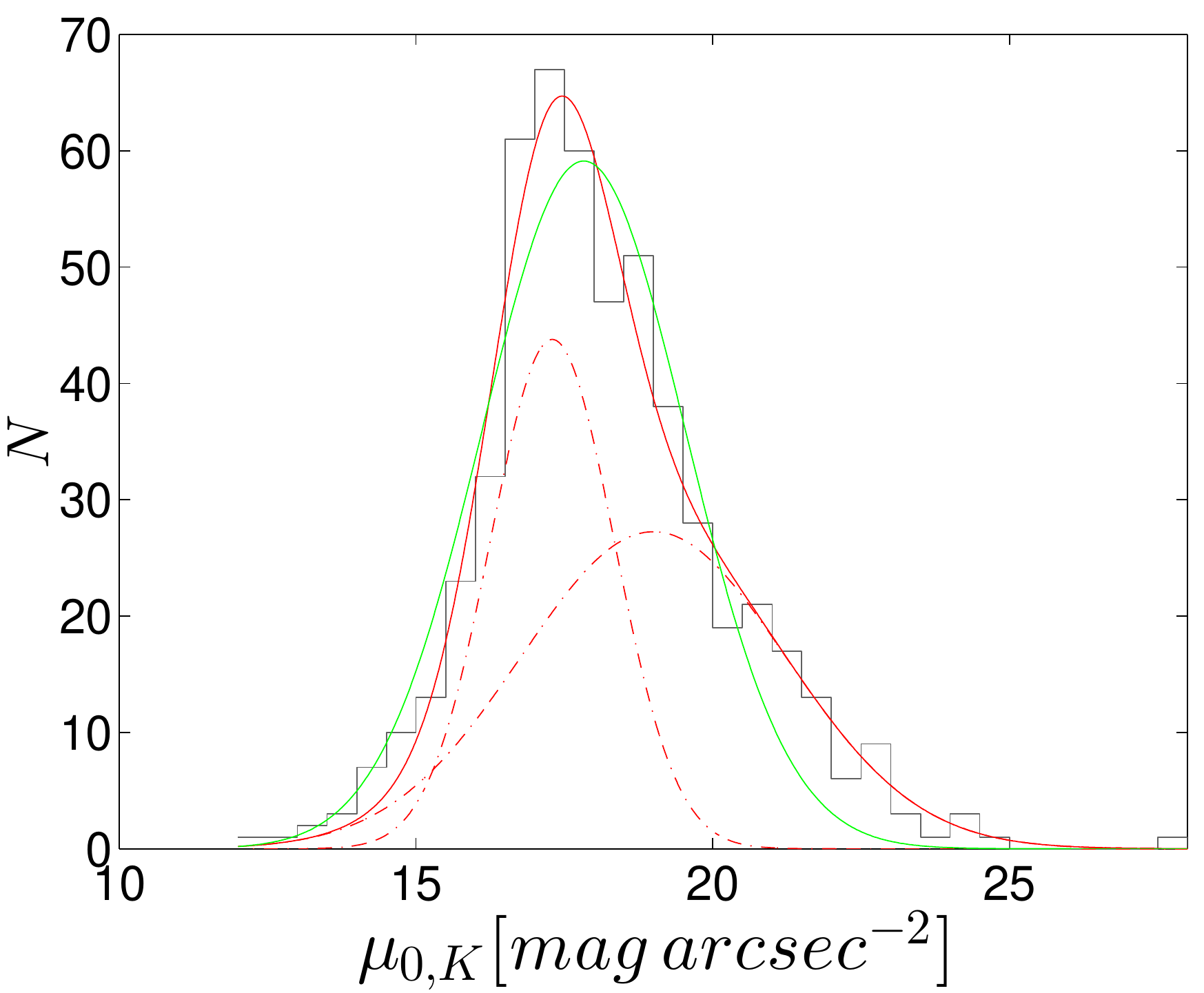}
\end{minipage}
\begin{minipage}{\textwidth}
\includegraphics[width=.24\textwidth]{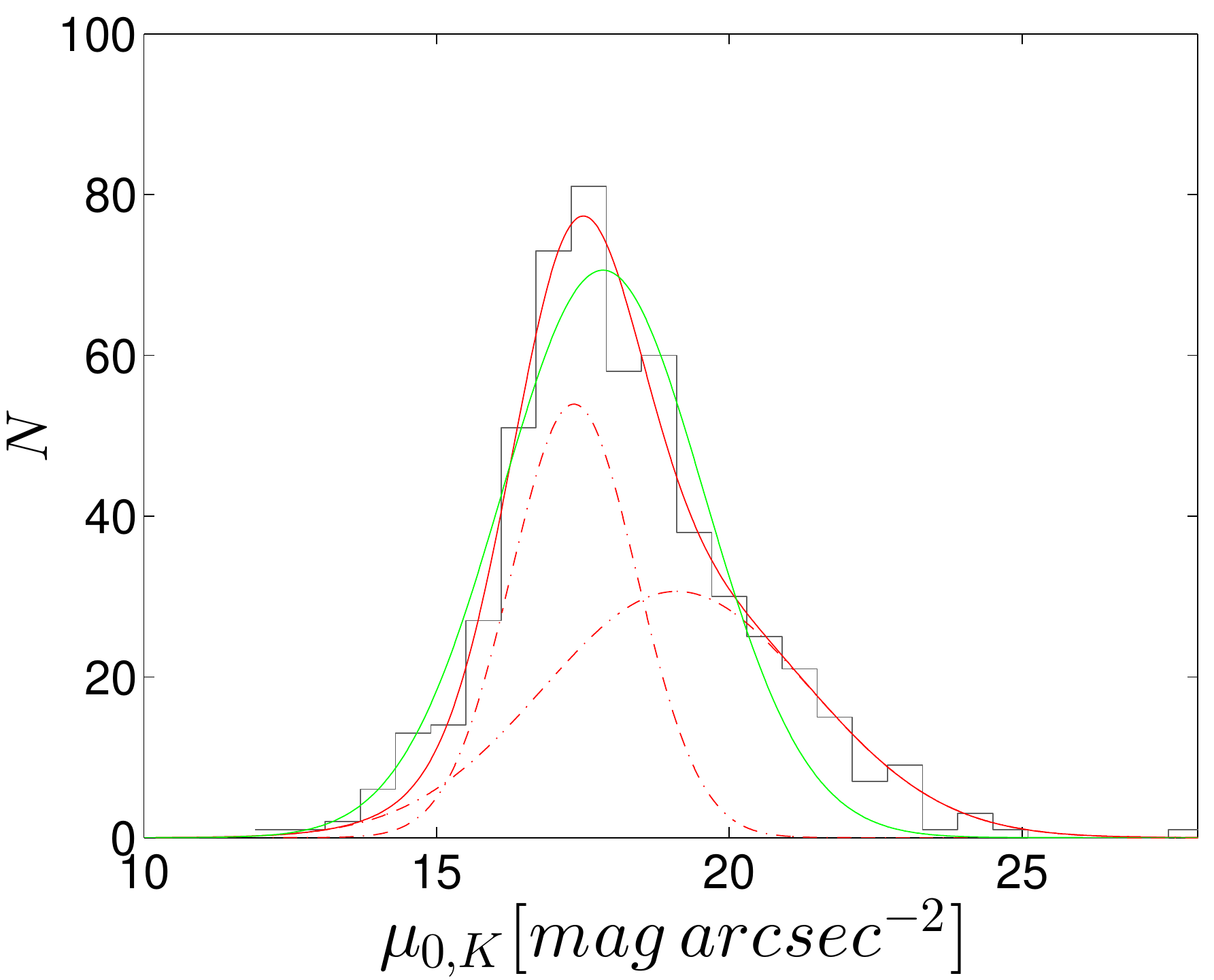}
\includegraphics[width=.24\textwidth]{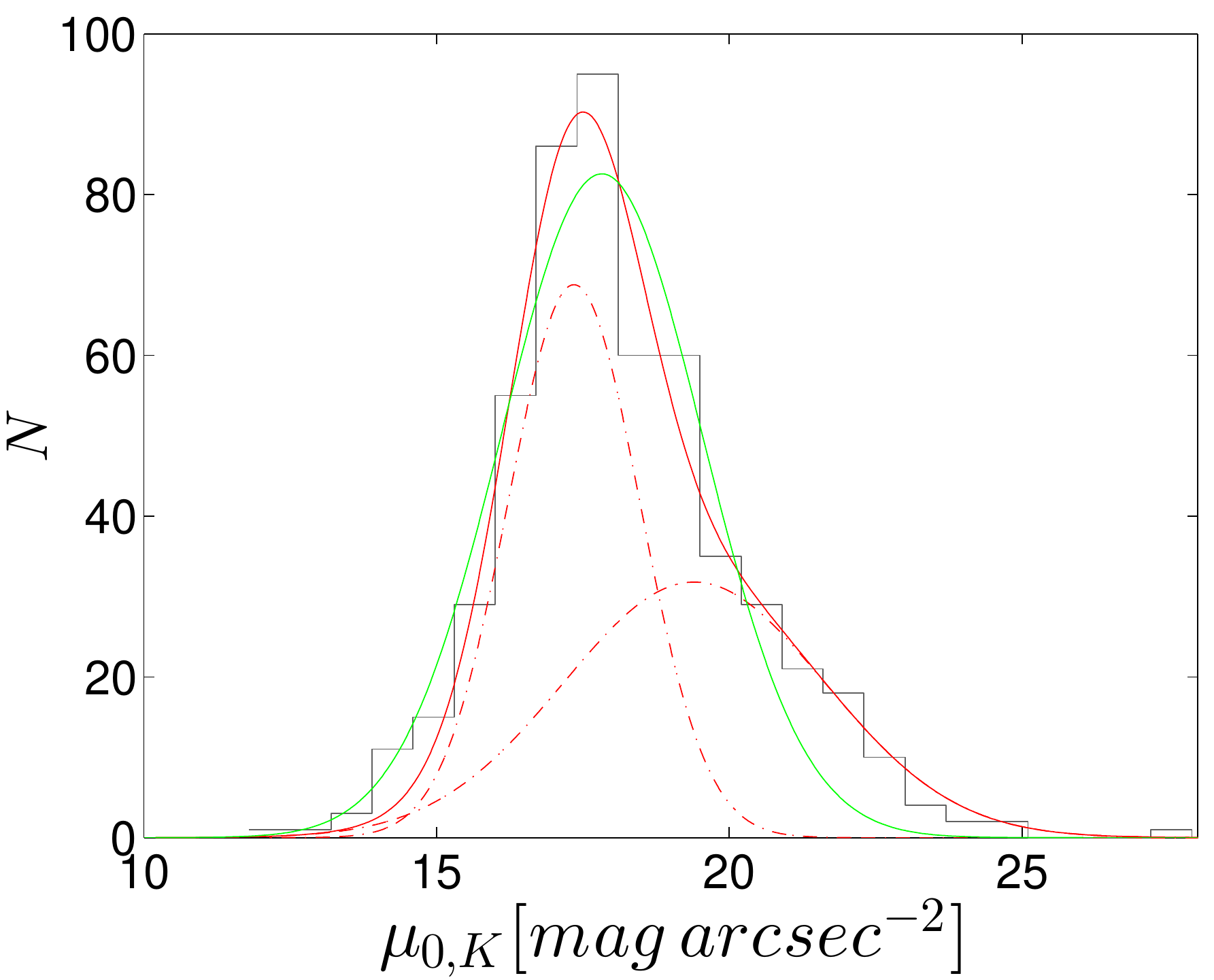}
\includegraphics[width=.24\textwidth]{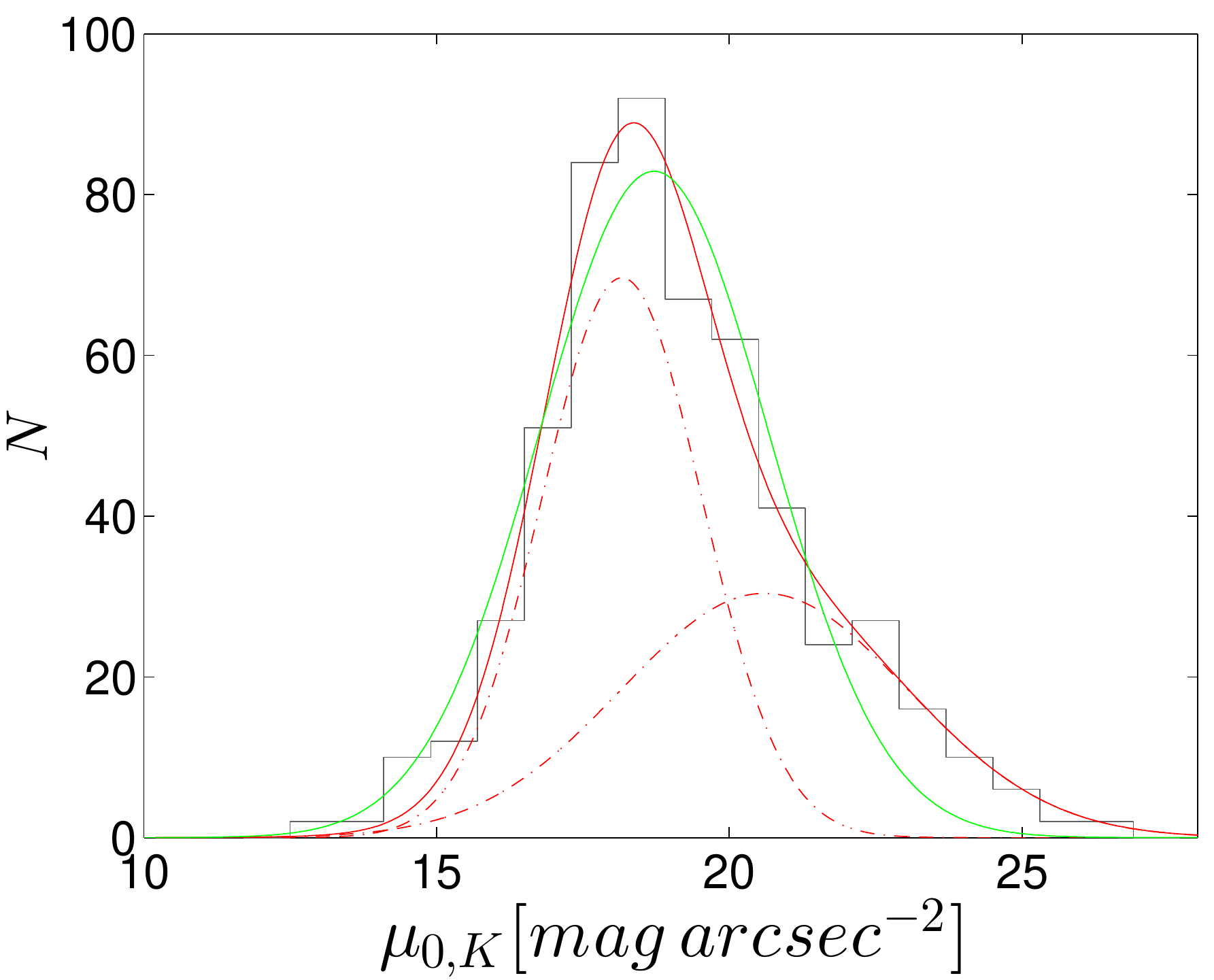}
\includegraphics[width=.24\textwidth]{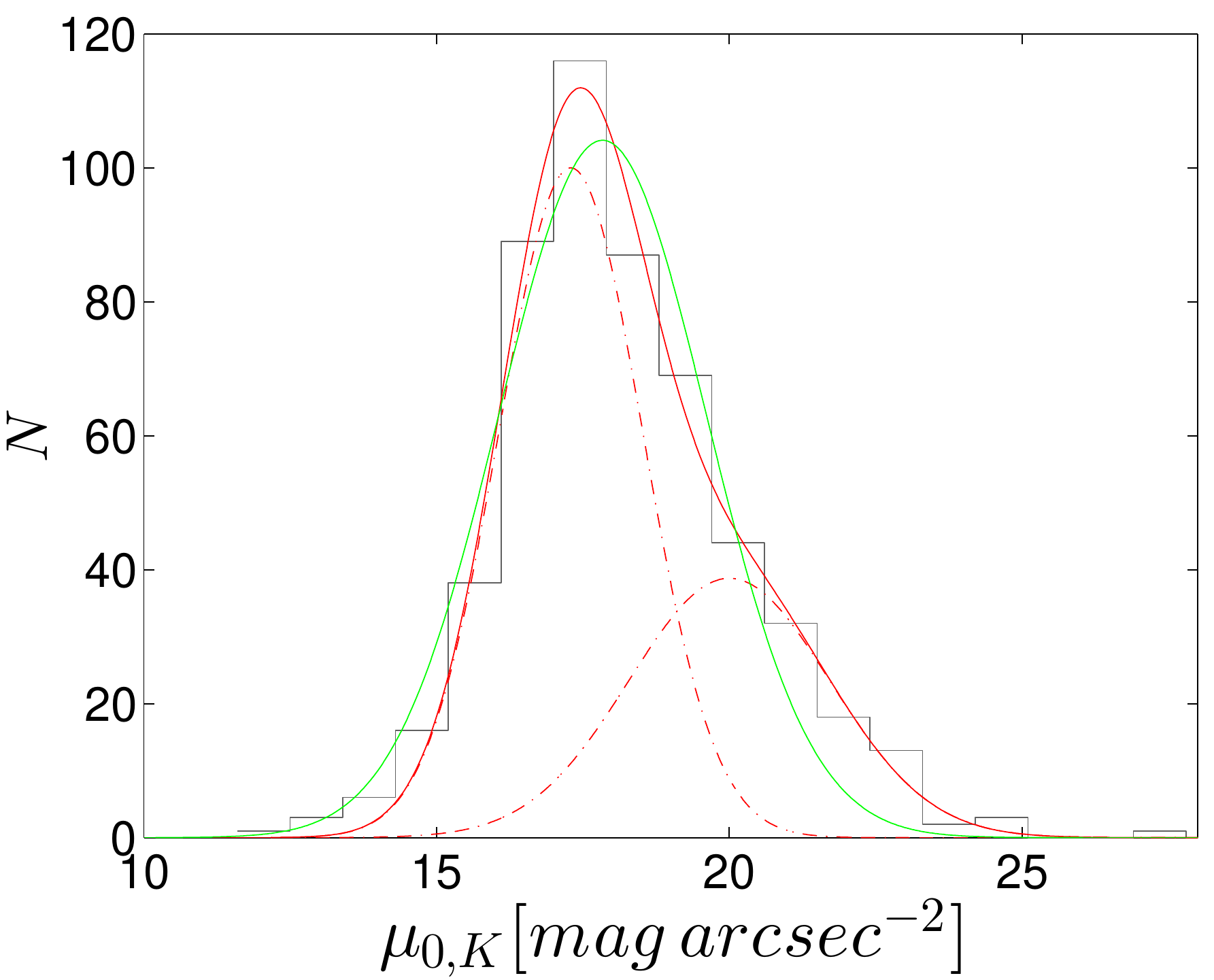}
\end{minipage}
\caption{Histograms of the $\mu_0$ distribution in $K$ band with multi bin sizes ranging from 0.2 to 0.9 $mag$ $arcsec^{-2}$.
Single Gaussian fitting to the distribution is described by green lines, while a sum of double Gaussian profiles is described
using red solid lines and two separate Gaussian profiles are described using red dotted lines.}
\label{fig.surface brightness02-09}
\end{figure*}

Considering that the distribution of $\mu_0$ may be influenced by bin size, 
we change the bin size from 0.2 to 0.9 $mag$ $arcsec^{-2}$ and compare the $\mu_0$ distributions in $K$ band, which is shown in Fig~.\ref{fig.surface brightness02-09}. 
It reveals that a double Gaussian profile is still much better than a single Gaussian profile for the $\mu_0$ distributions fitting with different bin sizes, which also could be shown from Table~\ref{table.AIC_BIC_bin0.2-0.9} that $\bigtriangleup BIC$ and $\bigtriangleup AICc$ are still larger than 10. The locations of double peak centers for the $\mu_0$ distribution with multi bin sizes are shown in Table~\ref{table.AIC_BIC_bin0.2-0.9}. The standard deviations for the location of lower surface brightness Gaussian peak and higher surface brightness Gaussian peak are 0.32 and 0.56, respectively.  
  
Therefore, it is not the bin size that leads to the double Gaussian profiles for the distribution of $\mu_0$ in $K$ band.

\begin{table*}\small
\centering
\caption{The AIC and BIC values of all the sample galaxies with single and double Gaussian fitting with multi bin sizes ranging from 0.2 to 0.9 $mag$ $arcsec^{-2}$ in $K$ band.}
\label{table.AIC_BIC_bin0.2-0.9}
\begin{tabular}{|ccccccccc|}
\hline
bin size & 0.2 & 0.3 & 0.4 & 0.5 & 0.6 & 0.7 & 0.8 & 0.9 \\
\hline
$AICc_{s}$ &164.3 &129.2 &128.1 &106.9 &96.7 &90.7 &78.2 &74.7  \\
$AICc_{d}$ &132.1 &96.0 &106.7 &83.1 &72.7 &73.0 &61.3  &53.9\\
$BIC_{s}$ &171.0 &134.4 &132.3 &110.3 &99.4 &92.7 &79.4 &75.4 \\
$BIC_{d}$ &145.1 &105.3 &113.2 &88.0 &75.7 &74.0 &59.6 &51.3 \\
$\bigtriangleup AICc$ & 32.2 & 33.2 & 21.4 & 23.8 & 24.0 & 17.7 & 16.9 & 20.8 \\ 
$\bigtriangleup BIC$ & 25.9 & 29.1 & 19.1 & 22.3 & 23.7 & 18.7 & 19.8 & 24.1 \\
$\mu_1$ & 17.3 & 17.3 & 17.3 & 17.3 & 17.3 & 17.3 & 18.2 & 17.3 \\
$\mu_2$ & 19.3 & 19.4 & 19.0 & 19.0 & 19.1 & 19.4 & 20.6 & 20.0 \\
\hline
\end{tabular}
\end{table*}

\subsection{Color}
To confirm that there are double Gaussian components in $K$ band, we classify our sample into two subsamples, those are bluer galaxies and redder galaxies, and fit them with single and double Gaussian profiles. 

\begin{figure*}
\centering
\begin{minipage}{\textwidth}
\includegraphics[width =.48\textwidth]{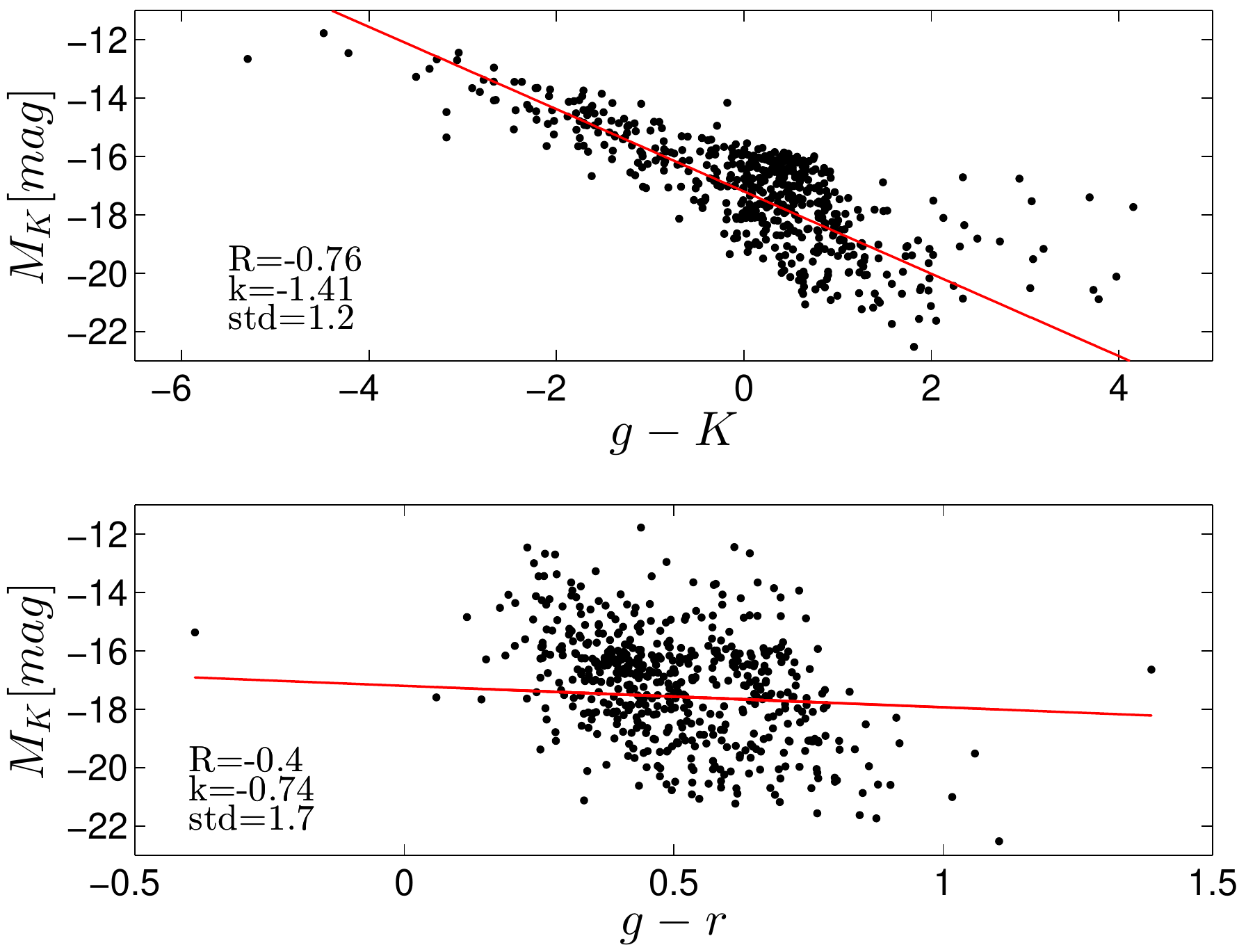}
\includegraphics[width =.44\textwidth]{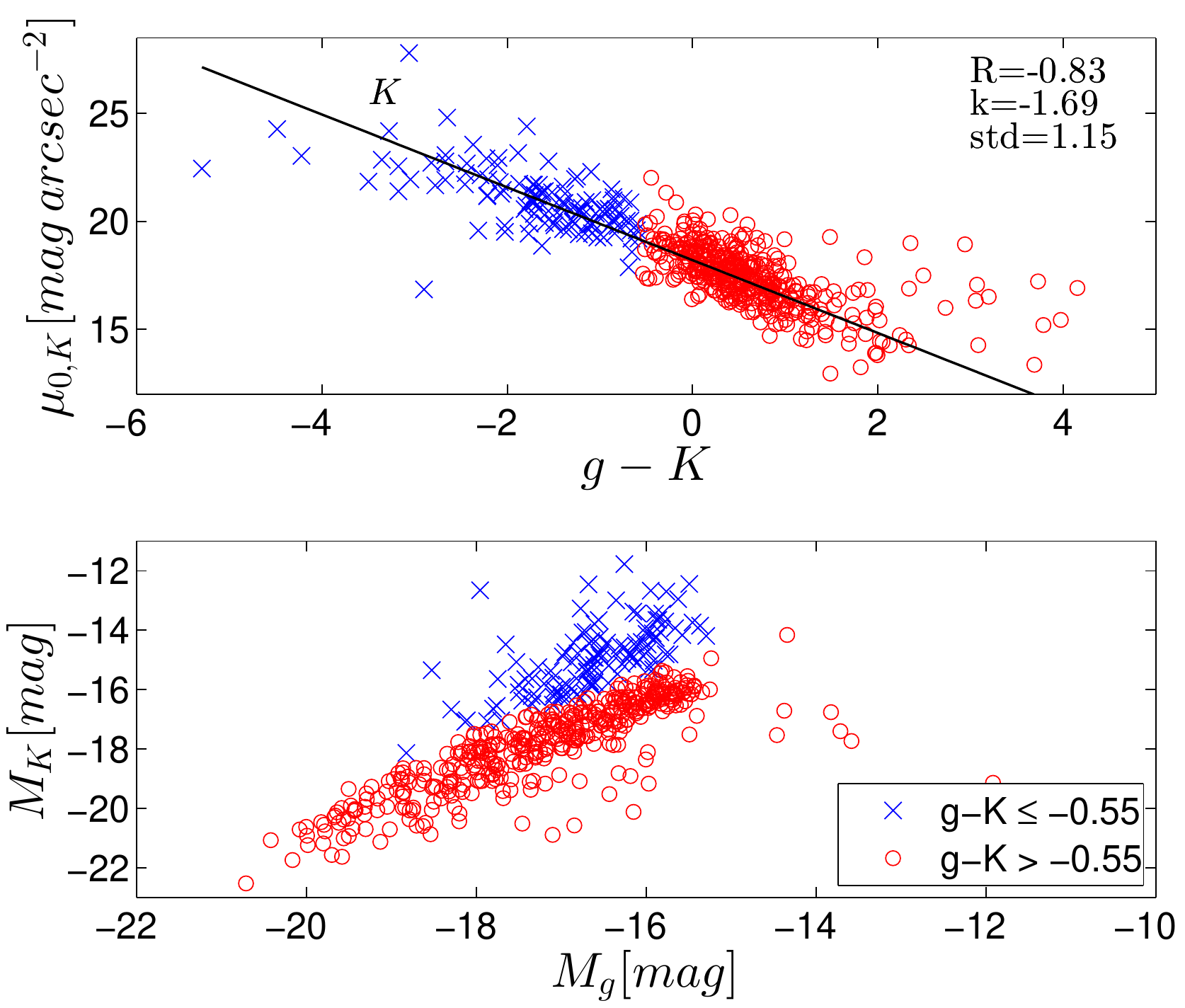}
\end{minipage}
\caption{The left top panel represents the relation between the color ($g$-$K$, which has been translated to AB magnitude) and absolute magnitude in $K$ band ($M_K$, which has been translated to AB magnitude), and the left bottom panel represents the relation between the color $g$-$r$ and $M_K$. The right top panel represents the relation between the color $g$-$K$ and $\mu_0$, the right bottom panel represents the relation between absolute magnitude in $g$ band ($M_g$) and absolute magnitude in $K$ band ($M_K$).}
\label{fig.color_g-K_g-r}
\end{figure*}

The relations between colors and $M_K$ are shown in 
Fig.~\ref{fig.color_g-K_g-r}. The correlation coefficients are -0.76 and -0.40 for relation between $g-K$ and $M_K$ and relation between $g-r$ and $M_K$, respectively. It is much more correlated between $g-K$ and $M_K$. In red bands ($K$ band), bluer galaxies (lower value of $g - K$) tend to have fainter magnitude. There is also a tight correlation between color ($g - K$) and $\mu_0$, which can be shown from the correlation coefficient (-0.83) of the right top panel of Fig.~\ref{fig.color_g-K_g-r}. The right bottom panel of Fig.~\ref{fig.color_g-K_g-r} presents the relation between $M_g$ and $M_K$. The sample could be well separated into two parts by $g - K = -0.55$. The blue crosses in Fig.~\ref{fig.color_g-K_g-r} represent a component of bluer galaxies with $g - K \leq -0.55$ and red circles represent another component of redder galaxies with $g - K > -0.55$. 
 
\begin{figure*}
\centering
\begin{minipage}{\textwidth}
\includegraphics[width=.46\textwidth]{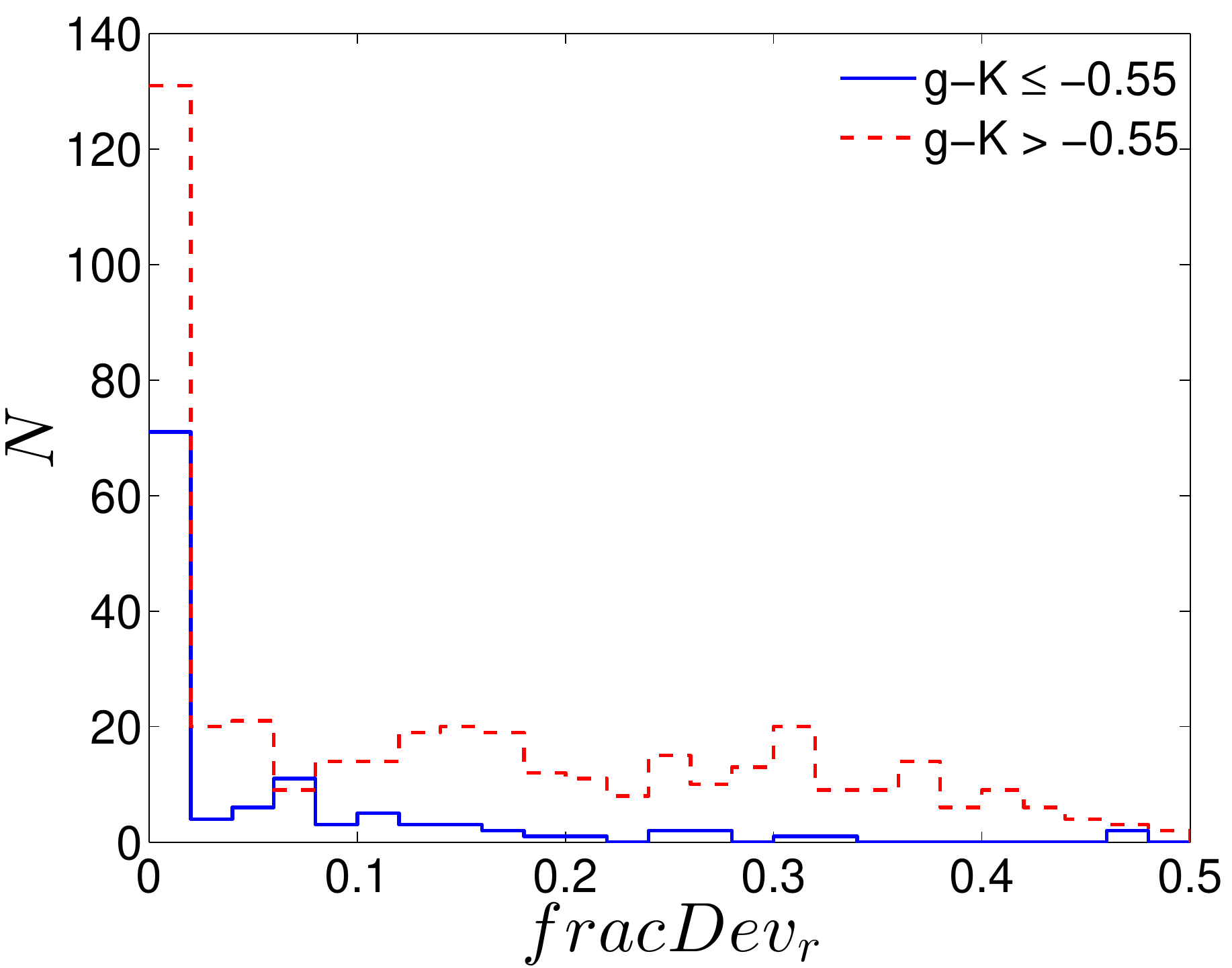}
\includegraphics[width=.44\textwidth]{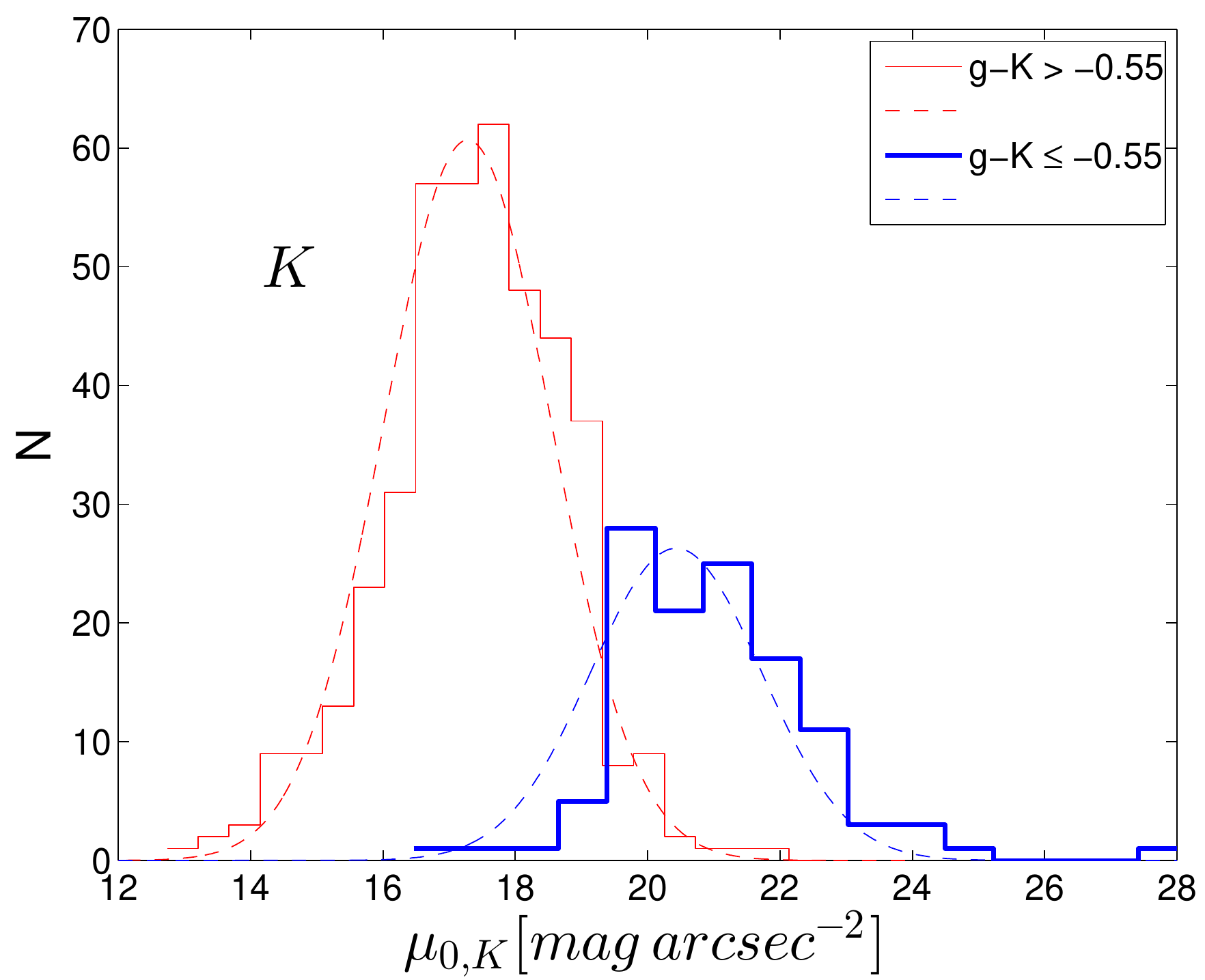}
\end{minipage}
\caption{The left panel shows histograms of the fracDev distribution for galaxies with $g - K \leq -0.55$ (shown with blue solid lines) and $g - K > -0.55$ (shown with red dashed lines), respectively. The right panel presents histograms of the $\mu_0$ distribution with the optimal bin size in $K$ band for galaxies with $g - K \leq -0.55$ (shown with blue thick lines) and $g - K > -0.55$ (shown with red thin lines), respectively.}
\label{fig.fracdev distribution g-K}
\end{figure*}

The left panel of Fig.~\ref{fig.fracdev distribution g-K} shows the $fracDev_r$ distribution for bluer galaxies (with $g - K \leq -0.55$) and redder galaxies (with $g - K > -0.55$). The ratios of galaxies with $fracDev \leq 0.1$ for bluer and redder galaxies are 80.5\% and 46.7\%, respectively. That is to say, bluer galaxies with $g - k \leq -0.55$ contain more galaxies with smaller portion of bulge than redder galaxies.  

From the right panel of Fig.~\ref{fig.fracdev distribution g-K}, for bluer galaxies, the $\mu_0$ distribution is preferred single Gaussian distribution, but it is not clear for redder galaxies. Similar to the Gaussian fitting results for galaxies with larger or smaller fraction of bulge, there is a difference between the peak centers of $\mu_0$ distribution for bluer and redder galaxies. The peak center of $\mu_0$ distribution for bluer galaxies is larger than that for redder galaxies, that is to say, bluer galaxies tend to have lower surface brightness and redder galaxies tend to have higher surface brightness. 

Therefore, there may be some effect for the color (bluer galaxies with $g - k \leq -0.55$ and redder galaxies with $g - K > -0.55$) on the fact of double Gaussian being better than a single
Gaussian fitting for $\mu_0$ distribution in K band.

\begin{table}
\centering
\caption{The AIC and BIC values of our subsamples (bluer galaxies with $g - K \leq -0.55$ and redder galaxies with $g - K > -0.55$) fitting with single and double Gaussian profiles in $K$ band.}
\label{table.color}
\begin{tabular}{|ccc|}
\hline
Value & $g - K \leq -0.55$ & $g - K > -0.55$ \\
\hline
\hline
$AICc_{single}$ &42.2 & 62.1\\\
$AICc_{double}$ &74.1 & 65.4 \\
$BIC_{single}$ &42.6 &63.6 \\
$BIC_{double}$ &73.2 & 64.4\\
$\bigtriangleup AICc$ &-31.9 & -3.3  \\
$\bigtriangleup BIC$  &-30.6 & -0.8 \\
\hline
\hline
\end{tabular}
\end{table}

\subsection{Statistical fluctuations}
The $\mu_0$ distribution of our sample in multi-bands has been presented in the previous Result Section. 
Especially in $K$ band, the evidence of double Gaussian being better than single Gaussian fitting is very strong. Now we discuss 
if the result in these bands results from statistical fluctuations. In order to obtain the likelihood of very strong evidence of double Gaussian being better than a single Gaussian fitting,  
we randomly select the same number of galaxies from our sample, and every single galaxy could be selected any times. 
We repeat 1000 times using both double Gaussian fitting and single Gaussian fitting. Here we also apply AICc and BIC value to 
estimating the goodness of fitting. If AICc and BIC satisfy:
\begin{equation}
   AICc_{single} - AICc_{double} \geq 10, 
\&
   BIC_{single} - BIC_{double} \geq 10,
\end{equation}
the evidence of double Gaussian being better than a single Gaussian fitting for $\mu_0$ distribution is considered to be very strong. In the equation, $AICc_{single}$ and $AICc_{double}$ are AICc values estimated 
using single and double Gaussian fitting, respectively. 
$BIC_{single}$ and $BIC_{double}$ are BIC values estimated using single and double Gaussian fitting, respectively.

\begin{table}
\caption{The likelihood of the very strong evidence of double Gaussian being better than a single Gaussian fitting for $\mu_0$ distribution in multi-bands.}
\label{table.likelihood}
\begin{tabular}{|c|c|c|c|c|c|c|c|c|}
\hline
Band & g & r & i & z & Y & J & H & K \\
Percent & 6.2 & 0 & 0 & 3.0&11.3 &22.1  & 17.2 & 78.3 \\
\hline
\end{tabular}
\end{table}
Table~\ref{table.likelihood} presents the percent of the very strong evidence of double Gaussian being better than a single Gaussian fitting for the $\mu_0$ distribution in the 1000 times test. 
It has the highest probability, 783 out of 1000 tests,  
to obtain a very strong evidence of a double Gaussian profile being better than a single Gaussian profile for the $\mu_0$ distribution in $K$ band, which is similar to fitting the $\mu_0$ distribution of our sample. 
For other bands, the percent is much lower.

The results refute the fact that double Gaussian being better than a single Gaussian fitting could be 
caused by statistical fluctuations.

\subsection{What may cause the double Gaussian components in $K$ band?}

\citealp{2009MNRAS.393..628M} studied the central surface brightness of 65 spiral galaxies in Ursa Major cluster in $K'$ band using bulge and disk decompositions  
and found a bimodal distribution. The sample size is small and the Ursa Major cluster populations include more early-type galaxies. \citealp{2009MNRAS.394.2022M} 
constructed a more complete sample with deep NIR observation of 286 Virgo cluster galaxies, and obtained the same conclusion 
as the UMa cluster that the central surface brightness distribution is bimodal in $H$ band. 
Comparing to \citealp{1997ApJ...484..145T}, \citealp{2009MNRAS.393..628M}, \citealp{2009MNRAS.394.2022M}, 
which found double peaks at about 18 $mag$ $arcsec^{-2}$ and 20 $mag$ $arcsec^{-2}$ and the gap positions at about 19 $mag$ $arcsec^{-2}$ 
for galaxies in Virgo cluster and Ursa Major cluster both in $K'$ and $H$ band, the evidence of double Gaussian distribution of $\mu_0$ for our sample is also strong in 
$H$ and $K$ band, which have double peaks at about 17.8 and 19.2 $mag$ $arcsec^{-2}$ and gap positions at about 18.3 $mag$ $arcsec^{-2}$. 
It is similar between the double peak locations and gap positions in this study and those in the previous studies.
The evidence of double Gaussian being better than single Gaussian fitting is not positive in optical bands, which is 
consistent with \citealp{2009MNRAS.394.2022M}. There are double Gaussian components of $\mu_0$ for our galaxies in near-infrared bands ($Y$, $J$, $H$ and $K$ band), and the evidence of a double Gaussian profile being better than a single Gaussian profile is very strong in $K$ band, especially.

We discuss and find that the sample incompleteness does not change the $\mu_0$ double Gaussian  distribution in $K$ band (More details in Section 5.1.).

To confirm the offset arising from morphological dependencies, 
we classify our sample into disc galaxies with smaller fraction of bulge ($fracDev_r \leq 0.1$) and galaxies with larger fraction of bulge ($0.5 \geq fracDev_r > 0.1$). 
It confirms that the morphology of galaxies may affect the $\mu_0$ distribution, 
that is to say, the fact of double Gaussian components may dependent on the morphology of sample galaxies to some extent. According to Fig.~\ref{fig.plot_u0_fracdev_lt_ht}, disc galaxies with larger fraction of bulge tend to have higher central surface brightness, and 
disc galaxies with smaller fraction of bulge tend to have lower central surface brightness. 
Therefore, there is a possibility that disc galaxies with larger and smaller fraction of bulge
result in higher and lower central surface brightness peaks, respectively. 

According to Fig.~\ref{fig.u0-scalelength_kpc}, there is a weak correlation between disk scalelength and $\mu_0$. 
Galaxies with larger disk scalelength tend to be lower surface brightness galaxies, and galaxies with smaller disk  scalelength tend to be higher surface brightness galaxies. 
That's to say, galaxies with larger and smaller disk scalelength may result in lower and higher central surface brightness peaks, respectively. 

To confirm that the color of galaxies may affect the $\mu_0$ distribution in $K$ band, we classify our sample into two subsamples, those are bluer galaxies with $g - K \leq -0.55$ and redder galaxies with $g - K > -0.55$, and fit them with single and double Gaussian profiles. From Fig.~\ref{fig.color_g-K_g-r} to Fig.~\ref{fig.fracdev distribution g-K}, bluer galaxies with $g - K \leq -0.55$ contain more galaxies with smaller portion of bulge than redder galaxies. Bluer galaxies tend to have lower surface brightness and redder galaxies tend to have higher surface brightness.

To exclude the bias of statistical fluctuations arising by small-number statistics, \citealp{2013MNRAS.433..751S} studied the central surface brightness in 3.6 $\mu m$ of 438 galaxies from 
$S^4G$. Also, a bimodal distribution is presented. It confirms that the bimodality is independent of statistics. We do 1000 times test and also confirm that the fact of double Gaussian distribution for $\mu_0$ distribution in $K$ band is not caused by statistical fluctuations. 
   
Conclusively, the double Gaussian distribution of $\mu_0$ in $K$ band for our sample
may depend on bulge-to-disk ratio, color and disk scalelength, rather than the inclination of sample galaxies, bin size and statistical fluctuations. Higher disc central surface brightness galaxies tend to have larger fraction of bulge and redder color. Lower disc central surface brightness galaxies tend to have smaller fraction of bulge and bluer color. The double Gaussian peak center locations of $\mu_0$ distribution for the final sample are different from those for subsamples with different $fracDev_r$ (galaxies with large fraction of bulge and small fraction of bulge) and subsamples with different color (bluer galaxies with $g - K \leq -0.55$ and redder galaxies with $g - K > -0.55$). Therefore, the double Gaussian distribution of $\mu_0$ in $K$ band for our sample may result from the factors of bulge-to-disk ratio, color and disk scalelength, together. 

As explained in \citealp{1997ApJ...484..145T} that the source of the bimodality may be an instability for galaxies when the baryons and dark matter are co-dominant in the centers. Galaxies in the high surface brightness part in our sample should be dominated by baryons in the galaxies centers while galaxies in the low surface brightness part in our sample should be dominated by dark matter in the galaxies centers. The galaxies in our sample with intermediate central surface brightness are expected to be instable because it is an instable state that baryons and dark matter are co-dominant in the galaxies centers. We intend to center on the work of exploring more possible reasons for bimodality in our future work.

\section{Conclusion}
We analyze 538 galaxies from SDSS DR7 main galaxy catalogue and UKIDSS DR10 LAS to study 
the disc central surface brightness distributions. 
It is representative for our sample galaxies within 57 Mpc and absolute magnitude in $r$ band limited to -16 mag. 
The final sample galaxies lie in clusters and the field. 
Then GALFIT is used to do exponential profile fittings. 
When estimating the $\mu_0$, no correction for dust extinction is applied in this study.   
The results are concluded as follows:

(1) Among the eight bands in optical and near-infrared, the evidence of double Gaussian being better than a single Gaussian fitting for $\mu_0$ distributions is positive in all near-infrared bands, especially it is very strong in $K$ band, but it is not positive in optical bands. 
It reveals higher probability of the very strong evidence of double Gaussian being better than a single Gaussian fitting for $\mu_0$ distributions in $K$ band when we repeat 1000 times test, which refutes the
hypothesis that the double Gaussian distribution could be caused by the statistical fluctuations. 

(2) Although the final sample is not absolutely complete, the sample incompleteness does not change the double Gaussian distribution of $\mu_0$ in $K$ band for our sample. By analyzing a series of subsamples selected from the final sample, 
the fact that double Gaussian being much better than a single Gaussian fitting of $\mu_0$ distributions for our sample in $K$ band may be not caused by inclination, bin size and statistical fluctuations.

(3) There is a probability that the fact of double Gaussian being much better than a single Gaussian fitting for $\mu_0$ distributions in $K$ band is caused by the morphology, color and disk scalelength of galaxies.  
Galaxies with larger fraction of bulge, redder color and smaller disk scalelength  
may result in higher central surface brightness peak. 
Galaxies with smaller fraction of bulge, bluer color and larger disk scalelength may result in lower central surface brightness peak. Therefore, the fact of double Gaussian components for $\mu_0$ distribution may be mainly caused by a combination effect of the morphology, color and disk scalelength of sample galaxies. 

\section*{Acknowledgments}
We appreciate the referee who provided very 
constructive and helpful comments and suggestions, which
helped to improve very well our work. We also thank Fengshan Liu, Chao Liu, Yanbin Yang and Bo Zhang very much for the helpful discussions and comments. 
This project is supported by the National Natural Science Foundation of China (Grant Nos.11733006, U1531245, 11225316, U1631105).
This project is supported by the National Key R\&D Program of China (Grant No.2017YFA0402704).

\appendix
\section{AKAIKE INFORMATION CRITERION AND BAYESIAN INFORMATION CRITERION}

We use Akaike information criterion (AIC) and Bayesian information criterion (BIC) to select a better model from single and double Gaussian fitting.
In the process of model fitting, adding more parameters may increase the likelihood, but it could also bring over fitting problems. 
To solve this problem, both AIC and BIC introduce a penalty term, which is ln(n) for BIC and 2 for AIC, for the number of parameters in the model.

The definition of AIC is 
\begin{equation}
   $AIC = 2$k$ - 2ln(\^{L})$
\end{equation}
In this equation, $k$ is the number of estimated parameters in the model, which is 3 for a single Gaussian profile and 6 for a double Gaussian profile.
\^{L} is the maximum value of the likelihood function for the model. It is a relative value of the set of models which means differences between AICs 
rather than the absolute size of AIC value. This is based on the likelihood theory. In the special case of Least Squares (LS) estimation, AIC can be translated into:
 \begin{equation}
   $AIC = 2$k$ + $n$ln(${\sigma}^2$)$
\end{equation}
In this equation, ${\sigma}^2$ is the estimated variance of a candidate model and n is the number of observed data, which can be estimated using:
\begin{equation}
  $ $\sigma^2$ = $s^2/n$ $,
\end{equation}
where $s^2$ is the residual sum of squares.
If the sample size is small compared with the number of parameters, AIC may perform
poorly and overfit (\citealp{su78}, \citealp{sa86}).  
Then AICc is developed with a second-order variant correction of AIC.
\begin{equation}
  $AICc = AIC + (2$k^2$ + 2$k$)/($n$ - $k$ -1)$.
\end{equation}
Here we use AICc rather than AIC to select models.

The Bayesian information criterion (BIC) is a model selection criteria and based on likelihood function. 
The definition of BIC is:
\begin{equation}
   $BIC = ln($n$)$k$ - 2ln(\^{L})$
\end{equation}
In Gaussian special case, if the model errors or disturbances are assumed to be independent and identically distributed according to a normal distribution, and the boundary condition that the derivative of the log likelihood with respect to the true variance is zero, this becomes:
 \begin{equation}
   $BIC = ln($n$)$k$ + $n$ln(${\sigma}^2$)$
\end{equation}
The model with the lowest BIC is preferred when selecting models. 
With the increasing of the error variance ${\sigma}^2$ and the explained variable $k$, the value of BIC will increase. 
Therefore, a model with lower BIC value implies that it may have either fewer explanatory variables, better fit, or both. 

The strength of the evidence against the model with higher BIC (or AIC) value can be summarized as Table~\ref{table.evidence} (\citealp{ra95}).

\begin{table}
\caption{The strength of the evidence against the model with higher BIC (or AIC) value.}
\label{table.evidence}
\begin{tabular}{|l|l|}
\hline
$\bigtriangleup$ BIC (or $\bigtriangleup$ AICc) & Evidence against higher BIC (or AICc) \\
0 to 2 & Not worth more than a bare mention \\
2 to 6 & Positive \\
6 to 10 & Strong \\
$\geq$ 10 & Very Strong \\
\hline
\end{tabular}
\end{table}
\bsp	
\label{lastpage}
\end{document}